\documentclass[apj,numberedappendix,tighten]{emulateapj}

\pdfoutput=1

\usepackage{amsmath}

\usepackage[T1]{fontenc}

\PassOptionsToPackage{breaklinks=true}{hyperref}

\usepackage[lf,mathtabular]{MinionPro}
\makeatletter
 \def\Mn@Text@Family{MinionPro-TLF}
 
\makeatother

\newcommand{\ergcm}{erg~s$^{-1}$~cm$^{-2}$}
\newcommand{\ergs}{erg~s$^{-1}$}
\renewcommand{\deg}{\ensuremath{^\circ}}

\newcommand{\Min}{_{\text{min}}}
\newcommand{\fmin}{f_{\text{min}}}
\newcommand{\Psel}{P_{\text{sel}}}

\newcommand{\Tspec}{\ensuremath{T_X}}
\newcommand{\TX}{\ensuremath{T_{\!X}}}
\def\Tx2{\ensuremath{T_{\!X,2}}}
\newcommand{\YX}{\ensuremath{Y_{\mkern -1mu X}}}
\newcommand{\LX}{\ensuremath{L_{X}}}
\def\M500{\ensuremath{M_{500}}}
\def\R500{\ensuremath{r_{500}}}
\def\rin{\ensuremath{r_{\rm in}}}
\def\rout{\ensuremath{r_{\rm out}}}
\newcommand{\Mtot}{\ensuremath{M_{\text{tot}}}}
\newcommand{\Mgas}{\ensuremath{M_{\text{gas}}}}

\newcommand{\Msun}{\ensuremath{M_\odot}}
\newcommand{\LCDM}{$\Lambda$CDM}
\def\400d{400d}

\newcommand\Mtrue{\ensuremath{M^{\text{true}}}}
\newcommand\Mest{\ensuremath{M^{\text{est}}}}
\newcommand\sigmaest{\ensuremath{\sigma^{\text{est}}}}
\newcommand\Mmin{\ensuremath{M_{\text{min}}}}
\newcommand\Mmax{\ensuremath{M_{\text{max}}}}
\newcommand\zmin{\ensuremath{z_{\text{min}}}}
\newcommand\zmax{\ensuremath{z_{\text{max}}}}
\newcommand\btheta{\theta}

\newcommand\Asoft{\ensuremath{A_{\text{soft}}}}
\newcommand\dTcal{\ensuremath{\delta T_{\text{cal}}}}

\slugcomment{Submitted 5/12/2008;
  Revised 10/3/2008; Accepted 10/29/2008}
\journalinfo{The Astrophysical Journal, in press (692, 2009 February
  10); arXiv:0805.2207}

\shorttitle{\emph{CHANDRA}\/ CLUSTER COSMOLOGY PROJECT.\quad II.}
\shortauthors{VIKHLININ ET AL.} 

\begin{document}

\title{\emph{Chandra}\/ Cluster Cosmology Project II: SAMPLES AND X-RAY DATA REDUCTION}

\author{
A.~Vikhlinin\altaffilmark{1,2},
R.~A.~Burenin\altaffilmark{2},
H.~Ebeling\altaffilmark{3},
W.~R.~Forman\altaffilmark{1},
A.~Hornstrup\altaffilmark{4},
C.~Jones\altaffilmark{1},
A.~V.~Kravtsov\altaffilmark{5},
S.~S.~Murray\altaffilmark{1},
D.~Nagai\altaffilmark{6},
H.~Quintana\altaffilmark{7},
A.~Voevodkin\altaffilmark{2,8}}

\altaffiltext{1}{Harvard-Smithsonian Center for Astrophysics, 
                 60 Garden Street, Cambridge, MA 02138}
\altaffiltext{2}{Space Research Institute (IKI), Profsoyuznaya 84/32,
                 Moscow, Russia}
\altaffiltext{3}{Institute for Astronomy, University of Hawaii, 2680
Woodlawn Drive, Honolulu, HI 96822}
\altaffiltext{4}{Danish National Space Center, Juliane Maries Vej 30, 
                 Copenhagen 0, DK-2100, Denmark}
\altaffiltext{5}{Dept.\ of Astronomy and Astrophysics,
  Kavli Institute for Cosmological Physics, Enrico Fermi Institute,
  University of Chicago, Chicago, IL 60637}
\altaffiltext{6}{Department of Physics and Yale Center for Astronomy \&
  Astrophysics, Yale University, New Haven, CT 06520}
\altaffiltext{7}{Departamento de Astronomia y Astrofisica, 
                 Pontificia Universidad Catolica de Chile, Casilla 306, 
                 Santiago, 22, Chile}
\altaffiltext{8}{Los Alamos National Laboratory, Los Alamos, NM 87545}

\begin{abstract}

  We discuss the measurements of the galaxy cluster mass functions at
  $z\approx0.05$ and $z\approx0.5$ using high-quality \emph{Chandra}
  observations of samples derived from the \emph{ROSAT} PSPC All-Sky and
  400~deg$^2$ surveys. We provide a full reference for the data analysis
  procedures, present updated calibration of relations between the total
  cluster mass and its X-ray indicators ($\TX$, $\Mgas$, and $\YX$) based on
  a subsample of low-$z$ relaxed clusters, and present a first measurement
  of the evolving $\LX-\Mtot$ relation (with $\Mtot$ estimated from $\YX$)
  obtained from a well-defined statistically complete cluster sample and
  with appropriate corrections for the Malmquist bias applied.  Finally, we
  present the derived cluster mass functions, estimate the systematic
  uncertainties in this measurement, and discuss the calculation of the
  likelihood function.  We confidently measure the evolution in the cluster
  comoving number density at a fixed mass threshold, e.g., by a factor of
  $5.0\pm1.2$ at $M_{500}=2.5\times10^{14}\,h^{-1}\,M_\odot$ between $z=0$
  and $0.5$. This evolution reflects the growth of density perturbations and
  can be used for the cosmological constraints complementing those from the
  distance-redshift relation.
\end{abstract}

\section{Introduction} 
\label{sec:intro} 

This work continues a series of papers in which we present the data for a
new X-ray selected sample of galaxy clusters --- the \400d{} survey ---
based on the data from the \emph{ROSAT} PSPC pointed observations. In the
first paper \citep[][Paper~I~hereafter]{2007ApJS..172..561B}, we presented
the cluster catalog and described the survey's statistical calibration
(selection function, effective area and so on). A complete high-redshift
subsample of the \400d{} clusters, 36 objects at $z=0.35-0.9$ with $\langle
z\rangle=0.5$, has been observed with \emph{Chandra}.  The goal of this
program was to provide X-ray data of sufficient quality for reliable
estimates of the high-redshift ($z\sim0.5$) cluster mass function.

\emph{Chandra} exposures were designed to yield at least 1500--2000 photons
from each cluster. This is sufficient to measure several high-quality total
mass proxies --- average temperature excluding the center, integrated gas
mass, and the $\YX$ parameter (the product of tempeature and gas mass
derived from X-ray data). The resulting mass estimates are much more
reliable than what was achievable in many previous studies where the only
available mass indicator was the X-ray flux \cite{2006ApJ...650..128K}.
Using several mass proxies also allows us to control the systematics by
checking the consistency of results obtained by different methods.

Observations of the high-redshift 400d clusters are complemented by
\emph{Chandra} archival data for a complete, flux-limited sample of nearby
clusters detected in the \emph{ROSAT} All-Sky survey (49 objects at present,
expected to grow by a factor of 1.5 in the near future as the completeness
of the \emph{Chandra} archive expands to lower fluxes). \emph{Chandra} data
for nearby clusters, combined when necessary with the \emph{ROSAT} PSPC
pointings, allow us to measure the same set of total mass proxies in local
and distant clusters.

The present work is a significant step forward in providing observational
foundations for cosmological work with the cluster mass function. First, it
uses a larger sample of high-$z$ clusters than the previous studies. For
example, the best published measurement of the evolution in the cluster
temperature function \citep{2004ApJ...609..603H} was based on 25 low-$z$
objects and 19~clusters with $\langle z\rangle=0.43$. \emph{Chandra}
provides much higher-quality data for each high-$z$ object than were
available before.  Second, we use a more advanced approach to the X-ray data
analysis, partly because this is called for by the \emph{Chandra} data and
partly because of the experience learned from recent deep observations of
low-$z$ clusters \citep[e.g.,][]{2006ApJ...640..691V}.  Last but not least,
the data for high and low-$z$ samples were obtained with the same instrument
and analyzed uniformly, minimizing the potential for systematic errors ---
the crucial ingredient for precise measurement of the evolution of the
cluster mass function.

In this paper, we present the analysis of the \emph{Chandra} observations of
our cluster sample, describe our approach to the cluster total mass
estimates, derive the evolving $M-L_X$ relation, and describe the
computation of the survey volume as a function of mass. We conclude by
presenting the cluster mass functions estimated in the ``concordant''
$\Lambda$CDM cosmology. The cosmological modeling of the cluster mass
function data is presented in an accompanying paper (Vikhlinin et al.,
Paper~III hereafter). The prime goal of this work is to provide a full
reference of the data reduction procedures and discuss the sources of
systematic uncertainties in the cluster mass function estimates at low and
high redshifts.

All distance-dependent quantities are computed assuming a $\Lambda$CDM
cosmological model with $\Omega_M=0.30$, and $\Omega_\Lambda=0.70$. We
also assume $h=0.72$, unless the explicit $h$-scaling is given. The
luminosities and fluxes are in the 0.5--2~keV energy band.

\defcitealias{2007ApJS..172..561B}{Paper~I}

\section{Cluster samples}

\subsection{High-Redshift Sample}
\label{sec:sample:high-z}

\begin{figure}[t]
\centerline{\includegraphics[width=0.97\linewidth]{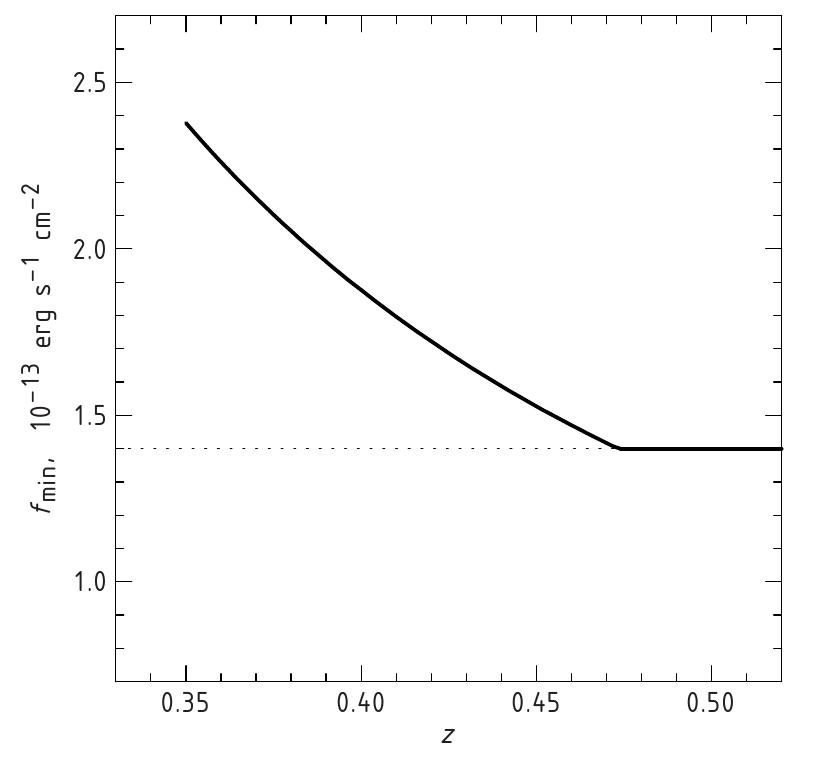}}
\caption{The limiting \emph{ROSAT} flux for selection in the
  \400d-\emph{Chandra} sample, as a function of redshift. At $z>0.473$, the
  limiting flux is $1.4\times10^{-13}$, that in the \400d{} catalog. At
  $0.35<z<0.473$, the flux limit corresponds to the minimum luminosity
  specified in eq.[\ref{eq:fmin:ch}].}
\label{fig:fmin:ch}
\end{figure}

Our high-redshift cluster sample is a well-defined subsample of the $z>0.35$
clusters from the \400d{} survey. The selection was designed to provide a
quasi mass-limited sample at $z\lesssim0.5$ by requiring that the
\emph{ROSAT}-derived luminosity was above a threshold of
\begin{equation}\label{eq:fmin:ch}
L_{X,\text{min}}=4.8\times10^{43}\,\,(1+z)^{1.8}\,\mbox{\ergs}
\end{equation}
in the default \LCDM{} cosmology. This luminosity threshold approximately
corresponds to a mass limit of $10^{14}$~\Msun{} from the low-$z$ $\LX-M$
relation. The redshift factor here corresponds to an early measurement of
the evolution in the $\Mgas-L$ relation \citep{2002ApJ...578L.107V}. The
resulting selection is entirely objective and in fact is formulated as a
redshift-dependent flux limit (shown in Fig.\,\ref{fig:fmin:ch}).  At
$z>0.473$, no additional selection is applied since the minimum flux of the
main \400d{} sample satisfies the luminosity threshold in
eq.~(\ref{eq:fmin:ch}).

Thirty nine objects from the \400d{} catalog satisfy these selection
criteria, and all were observed with \emph{Chandra}. For three clusters,
0216$-$1747, 0521$-$2530, 1117$+$1744, the accurate total X-ray flux
measured by \emph{Chandra} was $<10^{-13}$~\ergcm, significantly below the
target minimum flux in the \400d{} catalog, $1.4\times10^{-13}$~\ergcm. The
existence of such deviations is expected (see, e.g., Fig.~23 in
\citetalias{2007ApJS..172..561B}) because \emph{ROSAT} flux estimates have
large statistical errors. However, the computation of the \400d{} selection
function in this flux regime is less accurate because it depends strongly on
the wings of the distribution of the flux measurement scatter (see \S\,7.1
in \citetalias{2007ApJS..172..561B} for details).  We, therefore, opted not
to use these three clusters in the further analysis. The additional
selection criterion, $f_{\rm true}>10^{-13}$~\ergcm, will be taken into
account in the sample volume computations. The final sample of 36
high-redshift clusters we will use hereafter is presented in
Table~\ref{tab:400d:sample}.

\subsection{Low-Redshift Sample}
\label{sec:samle:low-z}

\begin{deluxetable*}{p{2cm}ccrrrrcrc}
  \tablecaption{High-redshift cluster sample\label{tab:400d:sample}}
  \tablehead{
    \colhead{Name} &
    \colhead{$z$} &
    \colhead{$L_X$,} &
    \colhead{\Tspec} &
    \colhead{$M_Y$} &
    \colhead{$M_G$} &
    \colhead{$M_T$} &
    \colhead{$f_x$,} &
    \colhead{$f_{\it ROSAT}$,} &
    \colhead{Merger?}
    \\
    \colhead{}  &
    \colhead{} & 
    \colhead{erg s$^{-1}$} &
    \colhead{keV} &
    \colhead{$10^{14}\,M_\odot$} &
    \colhead{$10^{14}\,M_\odot$} &
    \colhead{$10^{14}\,M_\odot$} &
    \colhead{$10^{-13}$~cgs}  &
    \colhead{$10^{-13}$~cgs}
    \\
    \colhead{(1)}  &
    \colhead{(2)} &
    \colhead{(3)} &
    \colhead{(4)} &
    \colhead{(5)} &
    \colhead{(6)} &
    \colhead{(7)} &
    \colhead{(8)} &
    \colhead{(9)} &
    \colhead{(10)} 
  }
  \startdata
  $0302$$-$$0423$\dotfill & $0.3501$ & $5.24\times10^{44}$ & $ 4.78\pm 0.75$ & $ 3.72\pm 0.38$& $ 3.58\pm 0.28$& $ 3.26\pm 0.77$& $15.34$& $15.9\pm 1.9$ & \nodata\\
$1212$$+$$2733$\dotfill & $0.3533$ & $3.61\times10^{44}$ & $ 6.62\pm 0.89$ & $ 6.17\pm 0.57$& $ 5.62\pm 0.37$& $ 6.16\pm 1.24$& $10.53$& $12.5\pm 1.7$ & $\checkmark$\\
$0350$$-$$3801$\dotfill & $0.3631$ & $6.80\times10^{43}$ & $ 2.45\pm 0.50$ & $ 1.43\pm 0.19$& $ 1.40\pm 0.18$& $ 1.34\pm 0.41$& $1.68 $& $ 2.9\pm 0.8$ & $\checkmark$\\
$0318$$-$$0302$\dotfill & $0.3700$ & $1.82\times10^{44}$ & $ 4.04\pm 0.63$ & $ 2.82\pm 0.28$& $ 2.44\pm 0.21$& $ 2.86\pm 0.67$& $4.63 $& $ 4.6\pm 0.5$ & $\checkmark$\\
$0159$$+$$0030$\dotfill & $0.3860$ & $1.42\times10^{44}$ & $ 4.25\pm 0.96$ & $ 2.51\pm 0.37$& $ 1.92\pm 0.22$& $ 2.67\pm 0.90$& $3.30 $& $ 3.3\pm 0.4$ & \nodata\\
$0958$$+$$4702$\dotfill & $0.3900$ & $1.04\times10^{44}$ & $ 3.57\pm 0.73$ & $ 1.84\pm 0.25$& $ 1.34\pm 0.15$& $ 2.03\pm 0.63$& $2.22 $& $ 2.8\pm 0.6$ & \nodata\\
$0809$$+$$2811$\dotfill & $0.3990$ & $2.50\times10^{44}$ & $ 4.17\pm 0.73$ & $ 3.69\pm 0.42$& $ 3.98\pm 0.35$& $ 2.96\pm 0.78$& $5.40 $& $ 5.5\pm 0.8$ & $\checkmark$\\
$1416$$+$$4446$\dotfill & $0.4000$ & $1.94\times10^{44}$ & $ 3.26\pm 0.46$ & $ 2.52\pm 0.24$& $ 3.10\pm 0.24$& $ 1.76\pm 0.37$& $4.01 $& $ 4.0\pm 0.5$ & \nodata\\
$1312$$+$$3900$\dotfill & $0.4037$ & $1.37\times10^{44}$ & $ 3.72\pm 1.06$ & $ 2.75\pm 0.57$& $ 2.62\pm 0.42$& $ 2.47\pm 1.06$& $2.71 $& $ 2.6\pm 0.4$ & $\checkmark$\\
$1003$$+$$3253$\dotfill & $0.4161$ & $1.53\times10^{44}$ & $ 5.44\pm 1.40$ & $ 2.80\pm 0.49$& $ 1.57\pm 0.20$& $ 3.83\pm 1.47$& $3.04 $& $ 3.5\pm 0.4$ & \nodata\\
$0141$$-$$3034$\dotfill & $0.4423$ & $1.32\times10^{44}$ & $ 2.13\pm 0.38$ & $ 1.22\pm 0.17$& $ 1.30\pm 0.24$& $ 1.03\pm 0.27$& $2.06 $& $ 3.1\pm 0.9$ & $\checkmark$\\
$1701$$+$$6414$\dotfill & $0.4530$ & $2.39\times10^{44}$ & $ 4.36\pm 0.46$ & $ 3.28\pm 0.24$& $ 3.20\pm 0.20$& $ 2.66\pm 0.42$& $3.91 $& $ 3.9\pm 0.4$ & \nodata\\
$1641$$+$$4001$\dotfill & $0.4640$ & $9.46\times10^{43}$ & $ 3.31\pm 0.62$ & $ 1.70\pm 0.20$& $ 1.34\pm 0.13$& $ 1.73\pm 0.49$& $1.43 $& $ 2.9\pm 0.8$ & \nodata\\
$0522$$-$$3624$\dotfill & $0.4720$ & $1.04\times10^{44}$ & $ 3.46\pm 0.48$ & $ 2.18\pm 0.21$& $ 1.82\pm 0.15$& $ 2.12\pm 0.45$& $1.47 $& $ 1.8\pm 0.3$ & $\checkmark$\\
$1222$$+$$2709$\dotfill & $0.4720$ & $9.88\times10^{43}$ & $ 3.74\pm 0.61$ & $ 2.09\pm 0.24$& $ 1.59\pm 0.16$& $ 2.08\pm 0.51$& $1.39 $& $ 1.9\pm 0.4$ & \nodata\\
$0355$$-$$3741$\dotfill & $0.4730$ & $1.76\times10^{44}$ & $ 4.61\pm 0.82$ & $ 3.02\pm 0.35$& $ 2.44\pm 0.22$& $ 2.87\pm 0.76$& $2.48 $& $ 2.9\pm 0.7$ & \nodata\\
$0853$$+$$5759$\dotfill & $0.4750$ & $8.43\times10^{43}$ & $ 3.42\pm 0.67$ & $ 2.05\pm 0.27$& $ 1.63\pm 0.17$& $ 2.09\pm 0.61$& $1.22 $& $ 2.0\pm 0.5$ & $\checkmark$\\
$0333$$-$$2456$\dotfill & $0.4751$ & $9.79\times10^{43}$ & $ 3.16\pm 0.58$ & $ 1.90\pm 0.22$& $ 1.64\pm 0.17$& $ 1.85\pm 0.51$& $1.33 $& $ 2.4\pm 0.5$ & $\checkmark$\\
$0926$$+$$1242$\dotfill & $0.4890$ & $1.50\times10^{44}$ & $ 4.74\pm 0.71$ & $ 3.00\pm 0.30$& $ 2.02\pm 0.16$& $ 3.42\pm 0.77$& $2.04 $& $ 1.7\pm 0.3$ & $\checkmark$\\
$0030$$+$$2618$\dotfill & $0.5000$ & $1.57\times10^{44}$ & $ 5.63\pm 1.13$ & $ 3.43\pm 0.41$& $ 2.04\pm 0.19$& $ 4.41\pm 1.33$& $2.09 $& $ 2.4\pm 0.3$ & $\checkmark$\\
$1002$$+$$6858$\dotfill & $0.5000$ & $1.71\times10^{44}$ & $ 4.04\pm 0.83$ & $ 2.80\pm 0.40$& $ 2.34\pm 0.27$& $ 2.65\pm 0.81$& $2.19 $& $ 2.0\pm 0.4$ & $\checkmark$\\
$1524$$+$$0957$\dotfill & $0.5160$ & $2.07\times10^{44}$ & $ 4.23\pm 0.51$ & $ 3.24\pm 0.27$& $ 3.08\pm 0.21$& $ 2.82\pm 0.51$& $2.45 $& $ 3.0\pm 0.4$ & $\checkmark$\\
$1357$$+$$6232$\dotfill & $0.5250$ & $1.63\times10^{44}$ & $ 4.60\pm 0.69$ & $ 2.96\pm 0.29$& $ 2.40\pm 0.18$& $ 2.78\pm 0.62$& $1.90 $& $ 2.0\pm 0.3$ & \nodata\\
$1354$$-$$0221$\dotfill & $0.5460$ & $1.40\times10^{44}$ & $ 3.77\pm 0.53$ & $ 2.31\pm 0.23$& $ 1.69\pm 0.16$& $ 2.32\pm 0.48$& $1.45 $& $ 1.5\pm 0.2$ & $\checkmark$\\
$1120$$+$$2326$\dotfill & $0.5620$ & $1.79\times10^{44}$ & $ 3.58\pm 0.44$ & $ 2.50\pm 0.21$& $ 2.32\pm 0.16$& $ 2.13\pm 0.39$& $1.68 $& $ 2.1\pm 0.4$ & $\checkmark$\\
$0956$$+$$4107$\dotfill & $0.5870$ & $1.85\times10^{44}$ & $ 4.40\pm 0.50$ & $ 2.93\pm 0.22$& $ 2.44\pm 0.14$& $ 2.87\pm 0.49$& $1.64 $& $ 1.6\pm 0.3$ & $\checkmark$\\
$0328$$-$$2140$\dotfill & $0.5901$ & $2.30\times10^{44}$ & $ 5.14\pm 1.47$ & $ 3.42\pm 0.66$& $ 2.92\pm 0.38$& $ 3.17\pm 1.36$& $2.09 $& $ 2.1\pm 0.6$ & \nodata\\
$1120$$+$$4318$\dotfill & $0.6000$ & $3.75\times10^{44}$ & $ 4.99\pm 0.30$ & $ 3.92\pm 0.17$& $ 4.20\pm 0.24$& $ 3.00\pm 0.27$& $3.24 $& $ 3.0\pm 0.3$ & \nodata\\
$1334$$+$$5031$\dotfill & $0.6200$ & $2.22\times10^{44}$ & $ 4.31\pm 0.28$ & $ 2.62\pm 0.17$& $ 1.88\pm 0.22$& $ 2.73\pm 0.27$& $1.76 $& $ 1.8\pm 0.3$ & $\checkmark$\\
$0542$$-$$4100$\dotfill & $0.6420$ & $2.91\times10^{44}$ & $ 5.45\pm 0.77$ & $ 4.07\pm 0.39$& $ 3.70\pm 0.25$& $ 3.86\pm 0.82$& $2.21 $& $ 2.2\pm 0.3$ & $\checkmark$\\
$1202$$+$$5751$\dotfill & $0.6775$ & $2.22\times10^{44}$ & $ 4.08\pm 0.72$ & $ 2.90\pm 0.37$& $ 2.85\pm 0.29$& $ 2.42\pm 0.64$& $1.34 $& $ 1.5\pm 0.4$ & $\checkmark$\\
$0405$$-$$4100$\dotfill & $0.6861$ & $2.23\times10^{44}$ & $ 3.98\pm 0.48$ & $ 2.51\pm 0.20$& $ 2.17\pm 0.16$& $ 2.32\pm 0.42$& $1.33 $& $ 1.5\pm 0.4$ & $\checkmark$\\
$1221$$+$$4918$\dotfill & $0.7000$ & $3.35\times10^{44}$ & $ 6.63\pm 0.75$ & $ 4.88\pm 0.38$& $ 4.16\pm 0.23$& $ 5.04\pm 0.86$& $2.06 $& $ 2.1\pm 0.5$ & $\checkmark$\\
$0230$$+$$1836$\dotfill & $0.7990$ & $2.55\times10^{44}$ & $ 5.50\pm 1.02$ & $ 3.46\pm 0.46$& $ 2.70\pm 0.27$& $ 3.57\pm 0.99$& $1.09 $& $ 2.2\pm 0.6$ & $\checkmark$\\
$0152$$-$$1358$\dotfill & $0.8325$ & $5.46\times10^{44}$ & $ 5.40\pm 0.97$ & $ 3.91\pm 0.52$& $ 3.94\pm 0.40$& $ 3.40\pm 0.91$& $2.24 $& $ 1.8\pm 0.3$ & $\checkmark$\\
$1226$$+$$3332$\dotfill & $0.8880$ & $8.42\times10^{44}$ & $11.08\pm 1.39$ & $ 7.59\pm 0.61$& $ 5.75\pm 0.28$& $ 9.91\pm 1.86$& $3.27 $& $ 2.9\pm 0.3$ & $\checkmark$

  \enddata
  \tablecomments{Column (2) --- cluster redshift. Column (3) --- total X-ray
    luminosity (0.5--2~keV band, object frame) measured from accurate
    \emph{Chandra} flux. \emph{Chandra} fluxes and luminosities have
    $\approx2\%$ statistical uncertainties. Column (4) --- average
    temperature from the spectrum integrated in the $[0.15-1]\,r_{500}$
    annulus.  Column (5) --- total mass estimated from $\YX$ parameter
    (\S\,\ref{sec:YX}). Column (6) --- $\Mtot$ estimated from integrated gas
    mass (\S\,\ref{sec:fgas}).  Column (7) --- mass estimated from the
    $\Mtot-\TX$ relation (\S\,\ref{sec:M-T}). Column (8) --- total X-ray
    flux measured by \emph{Chandra} (0.5--2~keV, observer's frame). Column
    (9) --- total X-ray flux (0.5--2~keV, observer's frame) reported in the
    400d catalog from \emph{ROSAT} PSPC data.  Column (10) --- approximate
    classification into mergers and relaxed clusters
    (\S\,\ref{sec:M-T:unrelaxed}). }
\end{deluxetable*}

\begin{deluxetable*}{p{2cm}cccrrrrc}
  \tablecaption{Low-redshift sample\label{tab:cat}}
  \tablehead{
    \colhead{Name} &
    \colhead{$f_x$,} &
    \colhead{$z$\tablenotemark{a}} &
    \colhead{$L_X$,} &
    \colhead{\Tspec} &
    \colhead{$M_Y$} &
    \colhead{$M_G$} &
    \colhead{$M_T$} &
    \colhead{Merger?}
    \\
    \colhead{}  &
    \colhead{$10^{-11}$~cgs}  &
    \colhead{} & 
    \colhead{erg s$^{-1}$} &
    \colhead{keV} &
    \colhead{$10^{14}\,M_\odot$} &
    \colhead{$10^{14}\,M_\odot$} &
    \colhead{$10^{14}\,M_\odot$}
    \\
    \colhead{(1)}  &
    \colhead{(2)} &
    \colhead{(3)} &
    \colhead{(4)} &
    \colhead{(5)} &
    \colhead{(6)} &
    \colhead{(7)} &
    \colhead{(8)} &
    \colhead{(9)}
  }
  \startdata
  A\,3571\dotfill  & 7.42 & 0.0386 & $2.37\times10^{44}$ & $ 6.81\pm 0.10$ & $ 5.90\pm 0.06$& $ 5.30\pm 0.07$& $ 6.61\pm 0.15$& \nodata\\
A\,2199\dotfill  & 6.43 & 0.0304 & $1.27\times10^{44}$ & $ 3.99\pm 0.10$ & $ 2.77\pm 0.05$& $ 2.80\pm 0.04$& $ 2.92\pm 0.11$& \nodata\\
2A~0335\dotfill  & 6.24 & 0.0346 & $1.60\times10^{44}$ & $ 3.43\pm 0.10$ & $ 2.33\pm 0.05$& $ 2.53\pm 0.05$& $ 2.32\pm 0.11$& \nodata\\
A\,496\dotfill   & 5.33 & 0.0328 & $1.23\times10^{44}$ & $ 4.12\pm 0.07$ & $ 2.96\pm 0.04$& $ 3.02\pm 0.04$& $ 3.07\pm 0.07$& \nodata\\
A\,3667\dotfill  & 4.64 & 0.0557 & $3.14\times10^{44}$ & $ 6.33\pm 0.06$ & $ 7.35\pm 0.07$& $ 8.62\pm 0.15$& $ 6.74\pm 0.09$& $\checkmark$\\
A\,754\dotfill   & 4.35 & 0.0542 & $2.78\times10^{44}$ & $ 8.73\pm 0.00$ & $ 8.47\pm 0.13$& $ 6.68\pm 0.12$& $11.05\pm 0.00$& $\checkmark$\\
A\,85\dotfill    & 4.30 & 0.0557 & $2.91\times10^{44}$ & $ 6.45\pm 0.10$ & $ 5.98\pm 0.07$& $ 5.91\pm 0.10$& $ 6.03\pm 0.14$& \nodata\\
A\,2029\dotfill  & 4.23 & 0.0779 & $5.72\times10^{44}$ & $ 8.22\pm 0.16$ & $ 8.64\pm 0.14$& $ 8.35\pm 0.20$& $ 8.66\pm 0.25$& \nodata\\
A\,478\dotfill   & 4.16 & 0.0881 & $7.24\times10^{44}$ & $ 7.96\pm 0.27$ & $ 8.15\pm 0.17$& $ 7.82\pm 0.12$& $ 8.20\pm 0.42$& \nodata\\
A\,1795\dotfill  & 4.14 & 0.0622 & $3.52\times10^{44}$ & $ 6.14\pm 0.10$ & $ 5.46\pm 0.06$& $ 5.34\pm 0.06$& $ 5.58\pm 0.14$& \nodata\\
A\,3558\dotfill  & 4.11 & 0.0469 & $1.96\times10^{44}$ & $ 4.88\pm 0.10$ & $ 4.78\pm 0.07$& $ 5.43\pm 0.09$& $ 4.54\pm 0.15$& $\checkmark$\\
A\,2142\dotfill  & 3.94 & 0.0904 & $7.20\times10^{44}$ & $10.04\pm 0.26$ & $11.96\pm 0.20$& $11.91\pm 0.16$& $11.70\pm 0.45$& \nodata\\
A\,2256\dotfill  & 3.61 & 0.0581 & $2.66\times10^{44}$ & $ 8.37\pm 0.24$ & $ 7.84\pm 0.15$& $ 6.14\pm 0.09$& $10.33\pm 0.45$& $\checkmark$\\
A\,4038\dotfill  & 3.48 & 0.0288 & $6.18\times10^{43}$ & $ 2.61\pm 0.05$ & $ 1.65\pm 0.02$& $ 2.03\pm 0.05$& $ 1.52\pm 0.04$& \nodata\\
A\,2147\dotfill  & 3.47 & 0.0355 & $9.40\times10^{43}$ & $ 3.83\pm 0.12$ & $ 3.10\pm 0.08$& $ 3.52\pm 0.14$& $ 3.15\pm 0.15$& $\checkmark$\\
A\,3266\dotfill  & 3.39 & 0.0602 & $2.69\times10^{44}$ & $ 8.63\pm 0.18$ & $ 9.00\pm 0.13$& $ 7.66\pm 0.12$& $10.82\pm 0.34$& $\checkmark$\\
A\,401\dotfill   & 3.19 & 0.0743 & $3.90\times10^{44}$ & $ 7.72\pm 0.30$ & $ 8.63\pm 0.24$& $ 9.27\pm 0.20$& $ 7.88\pm 0.46$& \nodata\\
A\,2052\dotfill  & 2.93 & 0.0345 & $7.47\times10^{43}$ & $ 3.03\pm 0.07$ & $ 1.84\pm 0.03$& $ 1.95\pm 0.04$& $ 1.91\pm 0.07$& \nodata\\
Hydra-A\dotfill  & 2.91 & 0.0549 & $1.93\times10^{44}$ & $ 3.64\pm 0.06$ & $ 2.83\pm 0.03$& $ 3.34\pm 0.04$& $ 2.51\pm 0.06$& \nodata\\
A\,119\dotfill   & 2.47 & 0.0445 & $1.06\times10^{44}$ & $ 5.72\pm 0.00$ & $ 4.50\pm 0.03$& $ 3.61\pm 0.06$& $ 5.80\pm 0.00$& $\checkmark$\\
A\,2063\dotfill  & 2.39 & 0.0342 & $5.98\times10^{43}$ & $ 3.57\pm 0.19$ & $ 2.21\pm 0.08$& $ 2.13\pm 0.07$& $ 2.46\pm 0.19$& \nodata\\
A\,1644\dotfill  & 2.33 & 0.0475 & $1.14\times10^{44}$ & $ 4.61\pm 0.14$ & $ 4.21\pm 0.09$& $ 4.66\pm 0.11$& $ 4.16\pm 0.19$& $\checkmark$\\
A\,3158\dotfill  & 2.30 & 0.0583 & $1.72\times10^{44}$ & $ 4.67\pm 0.07$ & $ 4.13\pm 0.05$& $ 4.74\pm 0.09$& $ 3.67\pm 0.09$& \nodata\\
MKW\,3s\dotfill  & 2.08 & 0.0453 & $9.28\times10^{43}$ & $ 3.03\pm 0.05$ & $ 2.09\pm 0.03$& $ 2.52\pm 0.05$& $ 1.90\pm 0.05$& \nodata\\
A\,1736\dotfill  & 2.04 & 0.0449 & $8.94\times10^{43}$ & $ 2.95\pm 0.09$ & $ 2.10\pm 0.06$& $ 2.10\pm 0.13$& $ 2.10\pm 0.09$& $\checkmark$\\
EXO~0422\dotfill & 2.01 & 0.0382 & $6.35\times10^{43}$ & $ 2.84\pm 0.09$ & $ 1.51\pm 0.04$& $ 1.46\pm 0.06$& $ 1.73\pm 0.09$& \nodata\\
A\,4059\dotfill  & 2.00 & 0.0491 & $1.05\times10^{44}$ & $ 4.25\pm 0.08$ & $ 2.81\pm 0.04$& $ 2.58\pm 0.04$& $ 3.19\pm 0.09$& \nodata\\
A\,3395\dotfill  & 1.95 & 0.0506 & $1.09\times10^{44}$ & $ 5.10\pm 0.17$ & $ 6.74\pm 0.18$& $ 6.74\pm 0.20$& $ 6.74\pm 0.34$& $\checkmark$\\
A\,2589\dotfill  & 1.94 & 0.0411 & $7.09\times10^{43}$ & $ 3.17\pm 0.27$ & $ 1.94\pm 0.11$& $ 2.01\pm 0.10$& $ 2.04\pm 0.26$& \nodata\\
A\,3112\dotfill  & 1.89 & 0.0759 & $2.43\times10^{44}$ & $ 5.19\pm 0.21$ & $ 4.20\pm 0.11$& $ 4.12\pm 0.09$& $ 4.28\pm 0.26$& \nodata\\
A\,3562\dotfill  & 1.84 & 0.0489 & $9.58\times10^{43}$ & $ 4.31\pm 0.12$ & $ 3.28\pm 0.07$& $ 3.48\pm 0.09$& $ 3.26\pm 0.14$& \nodata\\
A\,1651\dotfill  & 1.80 & 0.0853 & $2.93\times10^{44}$ & $ 6.41\pm 0.25$ & $ 5.78\pm 0.15$& $ 5.55\pm 0.12$& $ 5.89\pm 0.35$& \nodata\\
A\,399\dotfill   & 1.78 & 0.0713 & $2.01\times10^{44}$ & $ 6.49\pm 0.17$ & $ 6.18\pm 0.11$& $ 5.66\pm 0.12$& $ 6.95\pm 0.27$& $\checkmark$\\
A\,2204\dotfill  & 1.74 & 0.1511 & $9.35\times10^{44}$ & $ 8.55\pm 0.58$ & $ 9.40\pm 0.43$& $ 9.32\pm 0.28$& $ 8.87\pm 0.90$& \nodata\\
A\,576\dotfill   & 1.72 & 0.0401 & $5.99\times10^{43}$ & $ 3.68\pm 0.11$ & $ 2.34\pm 0.05$& $ 2.27\pm 0.06$& $ 2.57\pm 0.12$& \nodata\\
A\,2657\dotfill  & 1.62 & 0.0402 & $5.66\times10^{43}$ & $ 3.62\pm 0.15$ & $ 2.24\pm 0.06$& $ 2.14\pm 0.05$& $ 2.51\pm 0.16$& \nodata\\
A\,2634\dotfill  & 1.61 & 0.0305 & $3.20\times10^{43}$ & $ 2.96\pm 0.09$ & $ 1.74\pm 0.04$& $ 1.83\pm 0.04$& $ 1.85\pm 0.08$& \nodata\\
A\,3391\dotfill  & 1.58 & 0.0551 & $1.05\times10^{44}$ & $ 5.39\pm 0.19$ & $ 4.06\pm 0.10$& $ 3.58\pm 0.09$& $ 4.58\pm 0.24$& \nodata\\
A\,2065\dotfill  & 1.56 & 0.0723 & $1.82\times10^{44}$ & $ 5.44\pm 0.09$ & $ 4.98\pm 0.07$& $ 4.90\pm 0.09$& $ 5.31\pm 0.14$& $\checkmark$\\
A\,1650\dotfill  & 1.53 & 0.0823 & $2.33\times10^{44}$ & $ 5.29\pm 0.17$ & $ 4.59\pm 0.11$& $ 4.78\pm 0.10$& $ 4.39\pm 0.21$& \nodata\\
A\,3822\dotfill  & 1.48 & 0.0760 & $1.91\times10^{44}$ & $ 5.23\pm 0.30$ & $ 4.63\pm 0.18$& $ 4.50\pm 3.91$& $ 4.98\pm 0.43$& $\checkmark$\\
S\,1101\dotfill  & 1.46 & 0.0564 & $1.03\times10^{44}$ & $ 2.44\pm 0.08$ & $ 1.57\pm 0.03$& $ 1.99\pm 0.05$& $ 1.36\pm 0.07$& \nodata\\
A\,2163\dotfill  & 1.38 & 0.2030 & $1.37\times10^{45}$ & $14.72\pm 0.31$ & $21.98\pm 0.31$& $24.17\pm 0.34$& $22.83\pm 0.72$& $\checkmark$\\
Zw\,Cl1215\dotfill & 1.38 & 0.0767 & $1.80\times10^{44}$ & $ 6.54\pm 0.21$ & $ 5.75\pm 0.12$& $ 5.32\pm 0.10$& $ 6.10\pm 0.29$& \nodata\\
RX~J1504\dotfill & 1.35 & 0.2169 & $1.56\times10^{45}$ & $ 9.89\pm 0.53$ & $10.07\pm 0.35$& $ 9.01\pm 0.20$& $10.70\pm 0.86$& \nodata\\
A\,2597\dotfill  & 1.35 & 0.0830 & $2.09\times10^{44}$ & $ 3.87\pm 0.11$ & $ 2.84\pm 0.06$& $ 3.03\pm 0.06$& $ 2.72\pm 0.12$& \nodata\\
A\,133\dotfill   & 1.35 & 0.0569 & $9.60\times10^{43}$ & $ 4.01\pm 0.11$ & $ 2.57\pm 0.05$& $ 2.37\pm 0.04$& $ 2.91\pm 0.12$& \nodata\\
A\,2244\dotfill  & 1.34 & 0.0989 & $2.98\times10^{44}$ & $ 5.37\pm 0.12$ & $ 5.11\pm 0.08$& $ 5.80\pm 0.10$& $ 4.46\pm 0.15$& \nodata\\
A\,3376\dotfill  & 1.31 & 0.0455 & $5.89\times10^{43}$ & $ 4.37\pm 0.13$ & $ 3.01\pm 0.07$& $ 2.53\pm 0.06$& $ 3.84\pm 0.17$& $\checkmark$

  \enddata
  \tablenotetext{a}{Redshifts were converted to the CMB reference
    frame.}  
  \tablecomments{Columns (3)--(9) have the same meaning as in
    Table~\ref{tab:400d:sample}. Column (2) gives the total flux
    (0.5--2~keV) from the best source available (\emph{Chandra} if
    cluster the cluster is at sufficiently high redshift to fit the
    field of view, \emph{ROSAT} PSPC pointing, and re-measurement from
    the All-Sky survey data as a last resort).}
\end{deluxetable*}

The low-redshift cluster sample was selected, similarly to the procedure
described in \cite{2004ApJ...601..610V}, from several samples based on
the \emph{ROSAT} All-Sky Survey (RASS) data
(\citealt{2000MNRAS.318..333E} --- BCS; \citealt{1999ApJ...514..148D};
\citealt{2004A&A...425..367B} --- REFLEX; \citealt{2002ApJ...567..716R}
--- HIGFLUGCS). Overlaps between the catalogs were removed. The objects
at Galactic latitude $|b|<20\deg$, as well as those around LMC, SMC, and
the Virgo cluster were excluded (the exclusion regions were adopted from
\citealt{2002ApJ...567..716R}). The total area covered by these catalogs
is 8.14~sr. The X-ray fluxes were remeasured (starting from a list of
objects with cataloged fluxes $f>5.3\times10^{-12}$~\ergcm{} in the
0.5--2~keV band), using the data from pointed \emph{ROSAT} PSPC
observations, when available. Our final sample consists of 49 clusters
(Table~\ref{tab:cat}) with the re-measured flux
$f>1.3\times10^{-11}$~\ergcm{} in the 0.5--2~keV band, well above the
sensitivity limit of all initial RASS cluster catalogs, and $z>0.025$
(the lower redshift cut was used to ensure that a large fraction of the
cluster virial radius fits inside the \emph{Chandra} field of view). All
objects in this sample have archival \emph{Chandra} observations,
providing accurate X-ray spectral data.

\begin{figure}
\centerline{\includegraphics[width=0.97\linewidth]{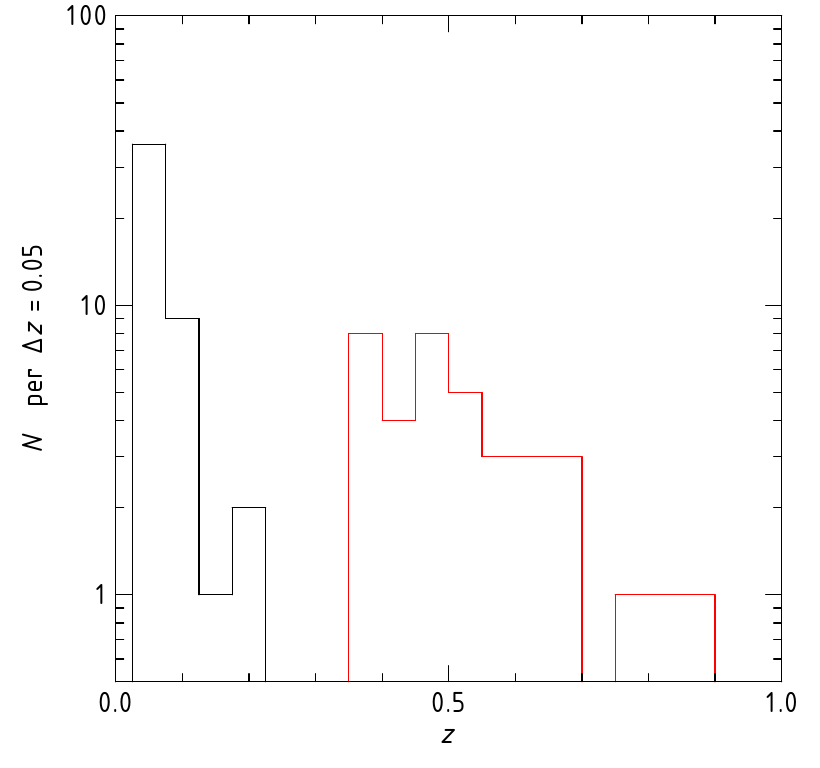}}
\caption{Redshift distribution of clusters in our low and high-$z$
samples.} 
\label{fig:z:distr}
\end{figure}

\subsection{General Characteristics of the Cluster Samples}

The combined cluster sample is a unique, uniformly observed dataset. The
volume coverage and effective mass limits in the low and high-redshift
subsamples are similar ($M_{\rm min}\simeq
(1-2)\times10^{14}\,h^{-1}\,\Msun$). The median mass at all redshifts is
near $\M500=2.5\times10^{14}\,h^{-1}\,\Msun$, which corresponds to
$T=4.5$~keV clusters at $z=0$. Observations suggest that 
clusters of such and larger mass exhibit scalings between
their observables and mass close to the expectations of self-similar
model \citep{2007ApJ...668....1N}, which makes our sample particularly useful for cosmological
applications. 

Our cluster sample is selected essentially using only the X-ray flux.
Cluster detection efficiency is, in principle, also depends on the object
surface brightness. However, the surface brightness effects are minimal for
our objects. For the low-$z$ sample, this is achieved by selecting objects
with fluxes a factor of $>5$ higher than the detection threshold in the
parent \emph{ROSAT} All-Sky Survey samples. For the 400d clusters, we used a
highly sensitive detection method tailored for finding extended sources. The
resulting sensitivity of the detection efficiency to the cluster angular
size has been extensively studied \citepalias{2007ApJS..172..561B} and found
to be small. Furthermore, optical identifications also played no role in
selecting the sample --- essentially all X-ray candidates at both low and
high redshifts were identified as galaxy clusters. Therefore, we do not miss
objects because of misclassification caused by the presence of central or
background AGNs.

The redshift histograms for the low and high-redshift samples are shown in
Fig.\,\ref{fig:z:distr}. The depth of the low-redshift sample is $z\sim
0.15$; there are only 3 clusters beyond this $z$.  Therefore, the
low-redshift sample is effectively ``local'' and it gives us a snapshot of
the cluster population at $z\approx0$. The high-redshift sample starts at
$z=0.35$ and extends to $z=0.9$. The median redshift of the distant sample
is $\langle z\rangle=0.5$.

Data of sufficient quality are available for utilizing three different
X-ray total mass proxies for all our clusters. These observations
provide us with a reliable measure of the evolution of the cluster mass
function between $z\approx 0.5$ and 0, or over $\approx37\%$ of the
present age of the Universe.

\section{\emph{Chandra} and \emph{ROSAT} Data Reduction}

\emph{Chandra} observations provide the basis for our X-ray analysis of
both high and low-redshift clusters. We also make use of the
\emph{ROSAT} PSPC data for the low-$z$ objects (pointed observations
when available and All-Sky Survey data for 8 objects). In low-$z$
clusters, the statistical accuracy of the X-ray surface brightness
determination at large radii is limited mostly by the \emph{Chandra}
field of view. The analysis in such cases benefits from using the
\emph{ROSAT} data that cover a much larger region although with a lower
sensitivity.  Below, we discuss the issues related to the initial data
preparation, spectral analysis, and producing the ``calibrated'' X-ray
images. How these data are used to derive the basic ICM parameters and
the cluster $\Mtot$ is discussed in \S\,\ref{sec:spectral:analysis},
\ref{sec:icm:pars}, and~\ref{sec:M:estimates}.

\subsection{Initial Data Reduction and Calibration Uncertainties}

\begin{figure*}[p]
\centerline{\includegraphics[width=0.4\linewidth]{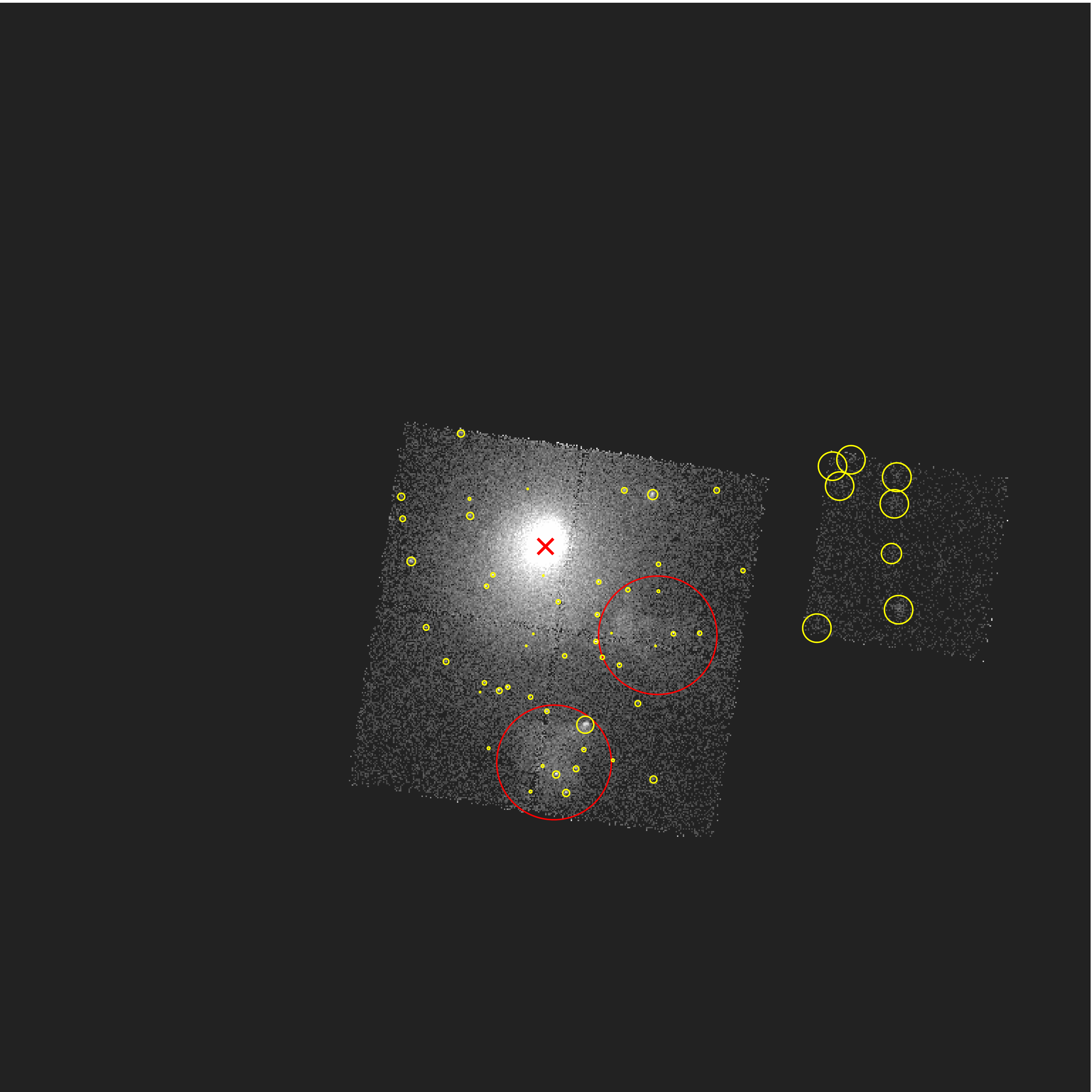}\quad\includegraphics[width=0.4\linewidth]{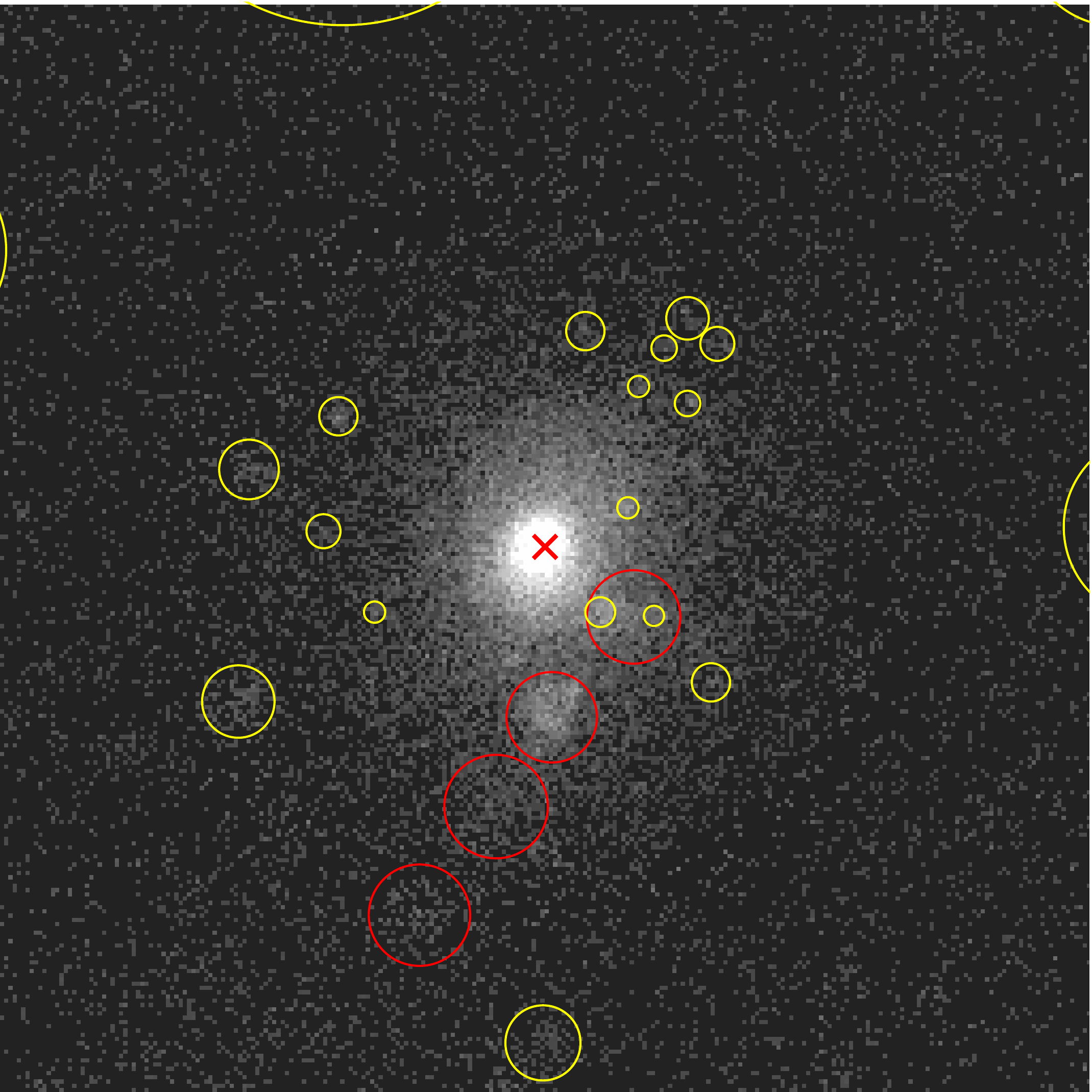}}
\medskip
\centerline{\includegraphics[width=0.4\linewidth]{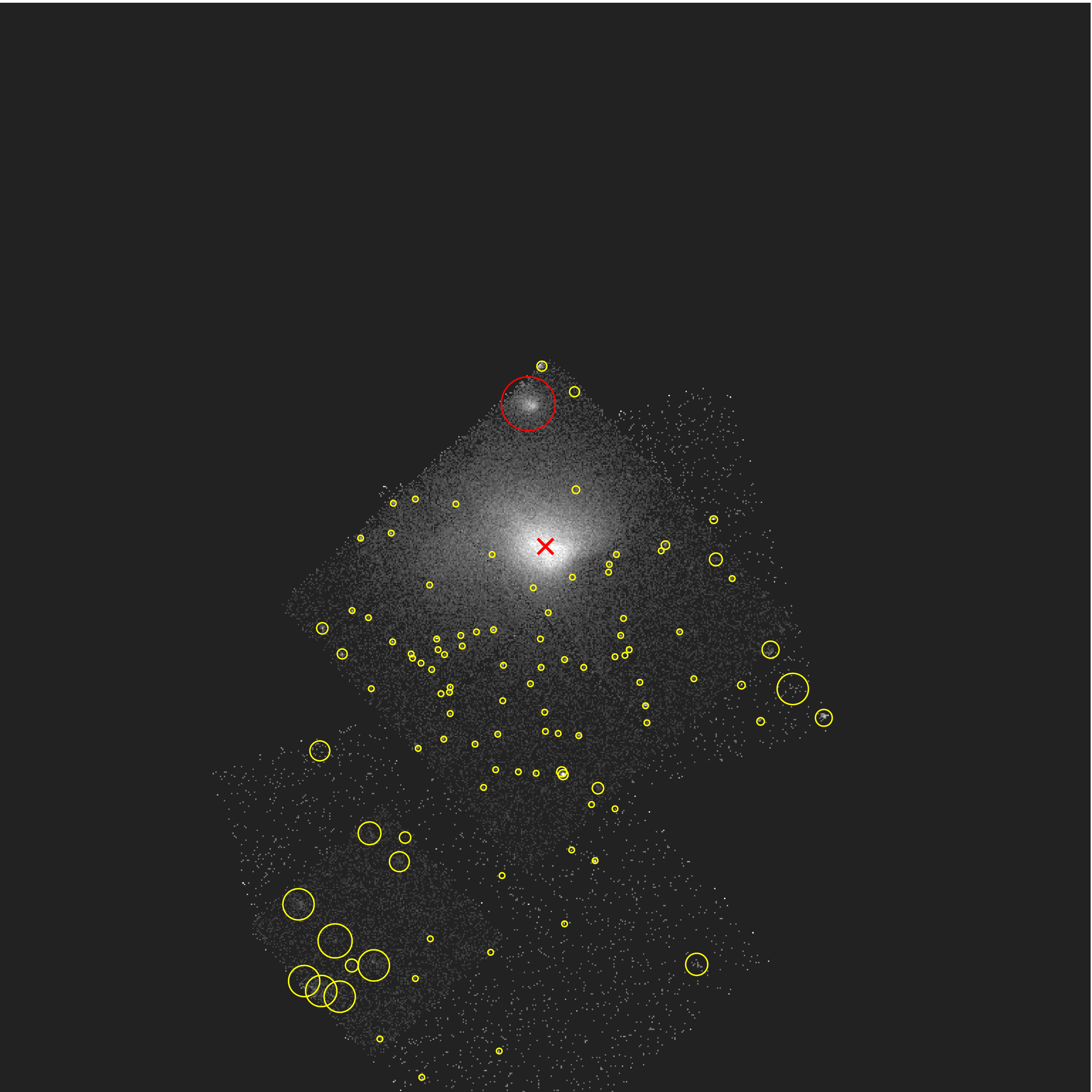}\quad\includegraphics[width=0.4\linewidth]{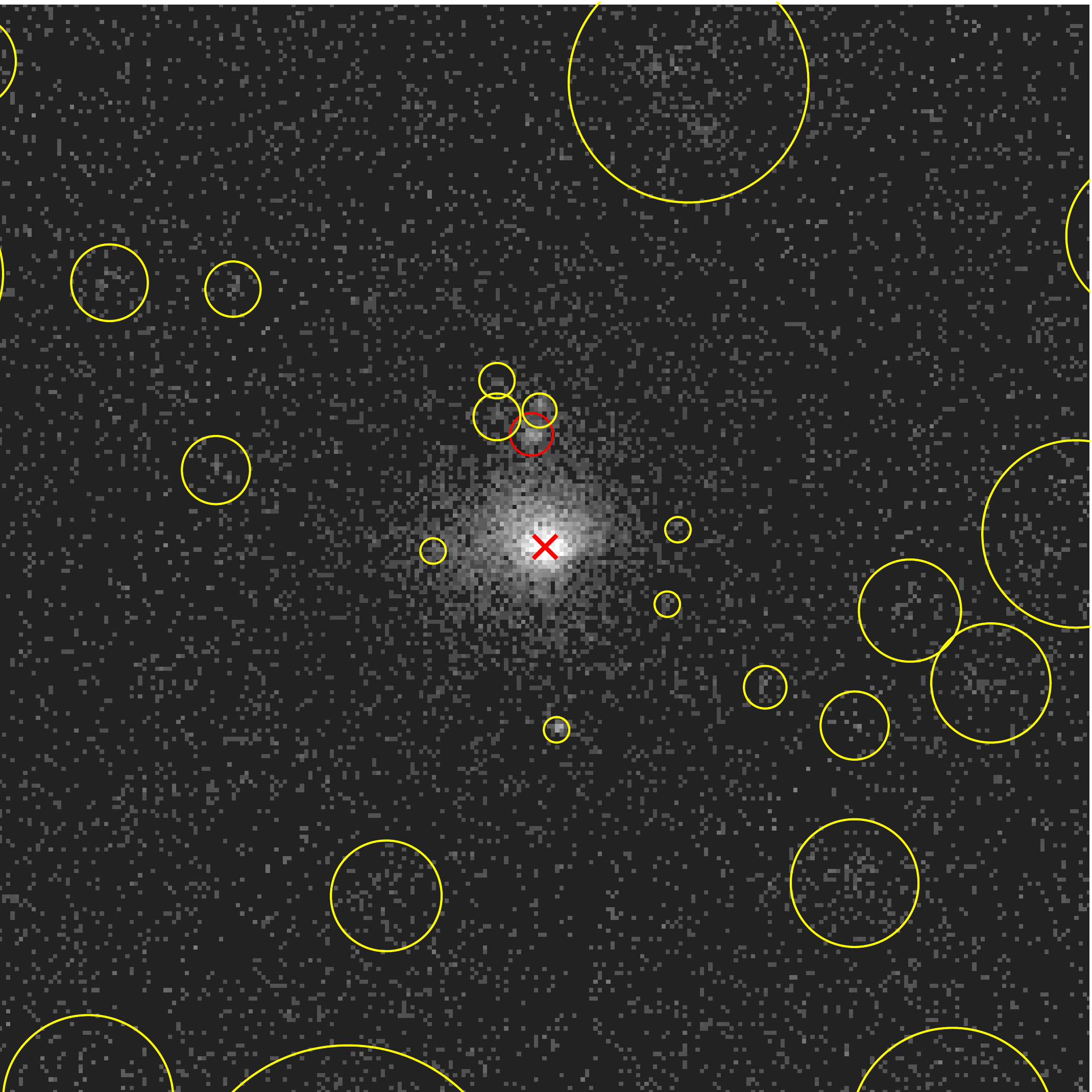}}
\medskip
\centerline{\includegraphics[width=0.4\linewidth]{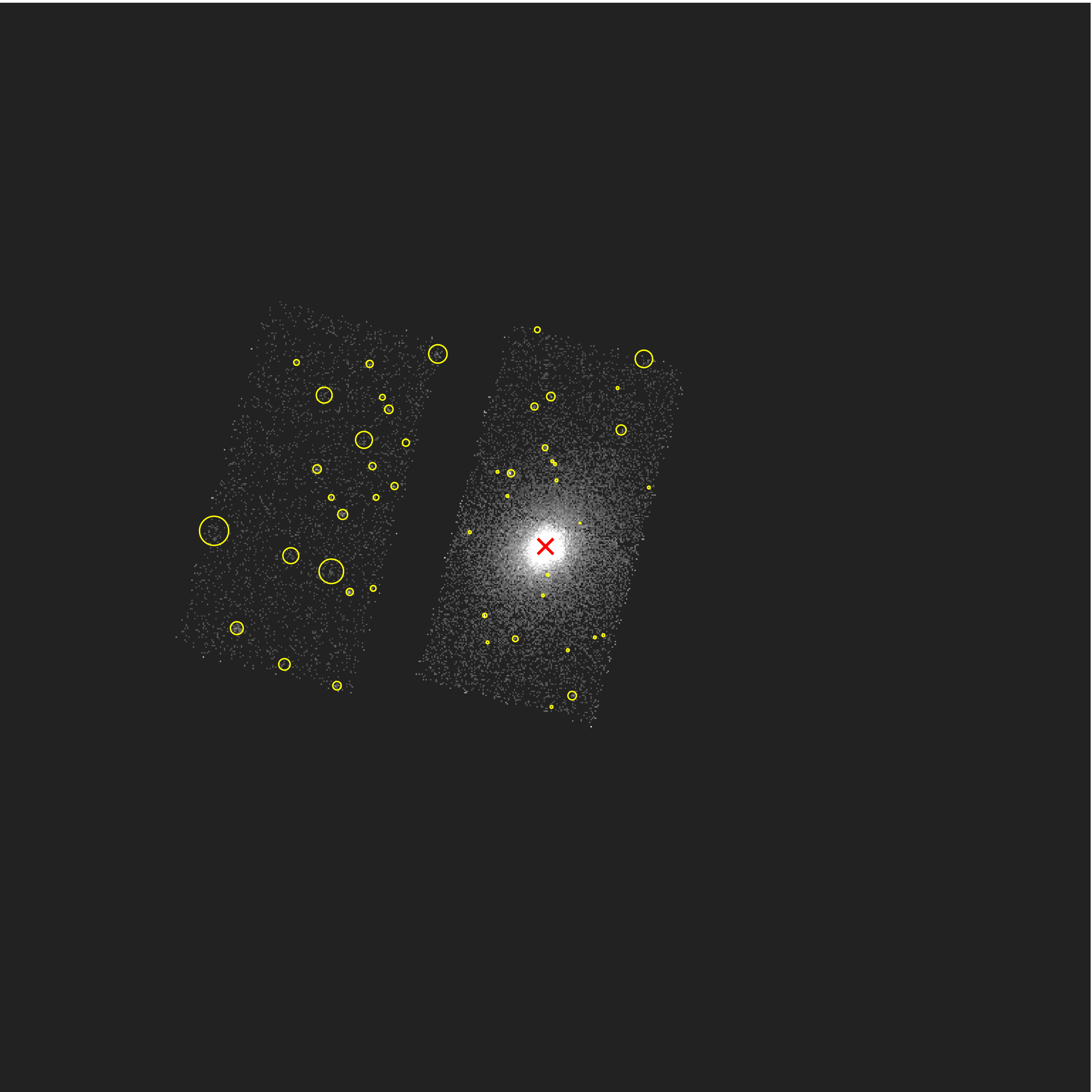}\quad\includegraphics[width=0.4\linewidth]{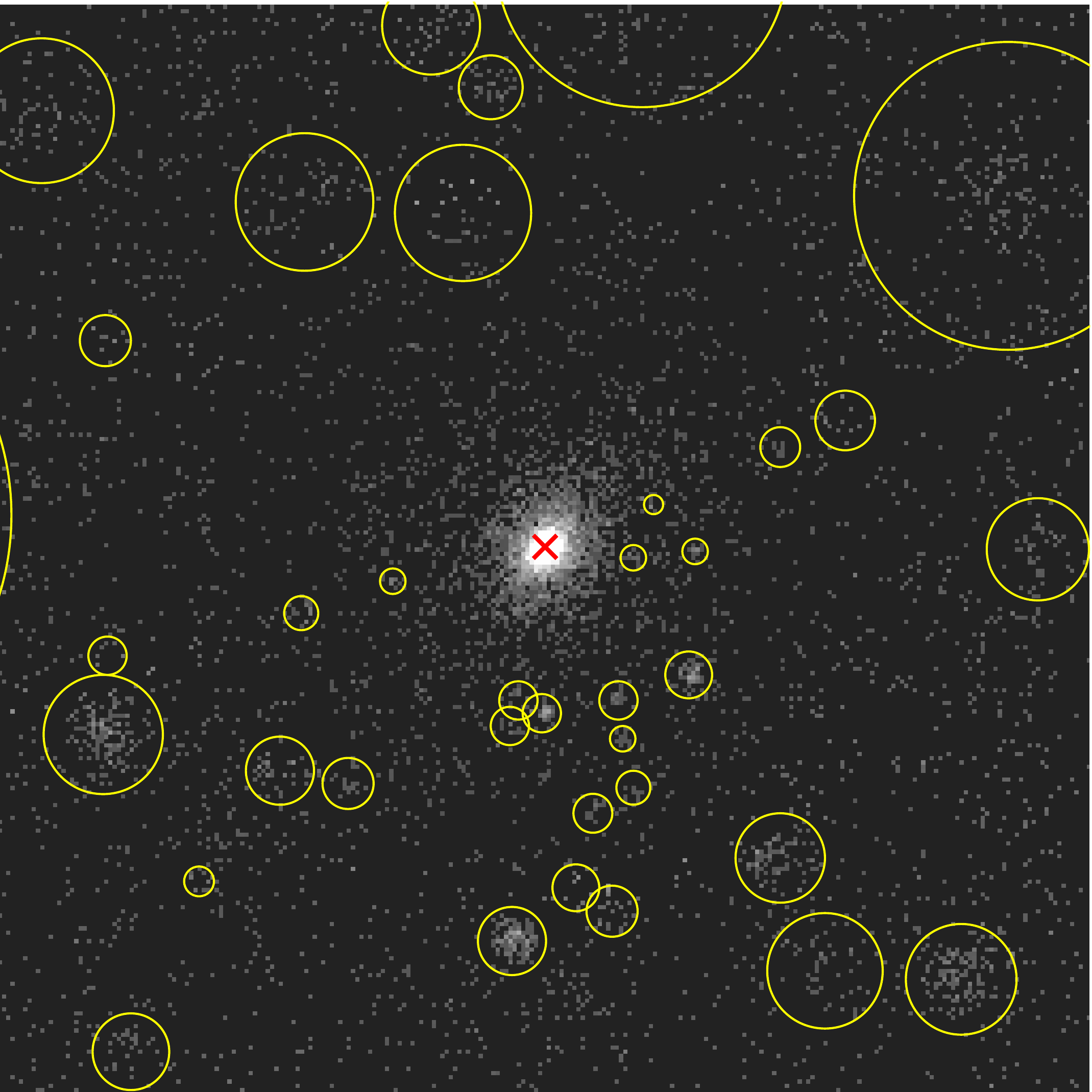}}
\medskip
\caption{Typical examples of X-ray images for the low-redshift clusters
  (A85, A2163, and A2597 top to bottom). Left panel show the
  \emph{Chandra} images (each panel is $50'\times50'$).  \emph{ROSAT}
  PSPC images ($64'\times64'$) are shown on the right.  Yellow circles
  show detected sources unrelated to the clusters; the general increase
  of their radius at large off-cluster distances reflects the
  degradation of the telescope PSF. The red circles indicate the cluster
  substructures that were removed from the profile analysis
  (\S\,\ref{sec:substr}).  The red crosses mark the location of the
  adopted cluster centroid (\S\,\ref{sec:center}).}
\label{fig:img:examples}
\end{figure*}

\subsubsection{Chandra}

For \emph{Chandra}, our data reduction procedure is adopted with no
changes from \citet[V05 hereafter]{2005ApJ...628..655V}. This includes
careful filtering for high background periods and applying all the
latest calibration corrections to the detected X-ray photons, and
determination of the background intensity in each observation.

\defcitealias{2005ApJ...628..655V}{V05}

The quiescent \emph{Chandra} background is dominated by the events
induced by charged particles. This component can be subtracted
exquisitely accurately \citep[with a $\lesssim 2\%$ scatter,
see][]{2006ApJ...645...95H}. A much smaller contribution is provided by
a fraction of the cosmic X-ray background not resolved into discrete
sources.  This component is modeled adequately by using the
``blank-sky'' background datasets which include both the
particle-induced and unresolved sky components. Finally, there is a
non-negligible diffuse soft component attributable to the Galactic ISM
emission \citep{2003ApJ...583...70M} and in some cases, to the
geocoronal charge exchange \citep{2004ApJ...607..596W}.  The soft
background component is the hardest to model because its intensity
depends on the pointing direction, and can even be variable in the case
of charge exchange emission. Fortunately, the soft component can still
be subtracted sufficiently accurately because it is separated spectrally
from the cluster emission (since it is dominated by emission lines near
0.6~keV, see \S\,2.3.2 in \citetalias{2005ApJ...628..655V}).

\begin{figure*}[t]
\centerline{\includegraphics[width=0.4\linewidth]{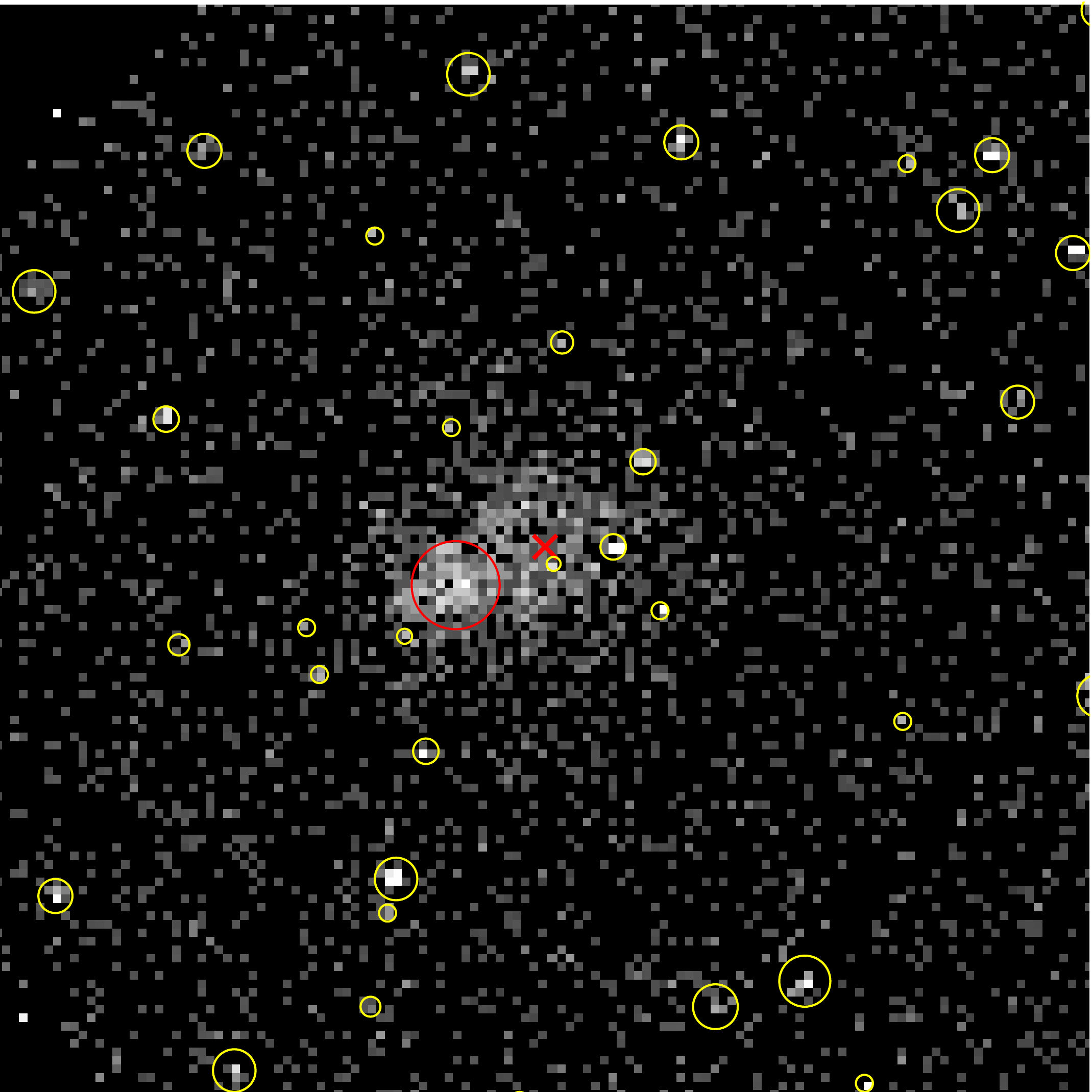}\quad\includegraphics[width=0.4\linewidth]{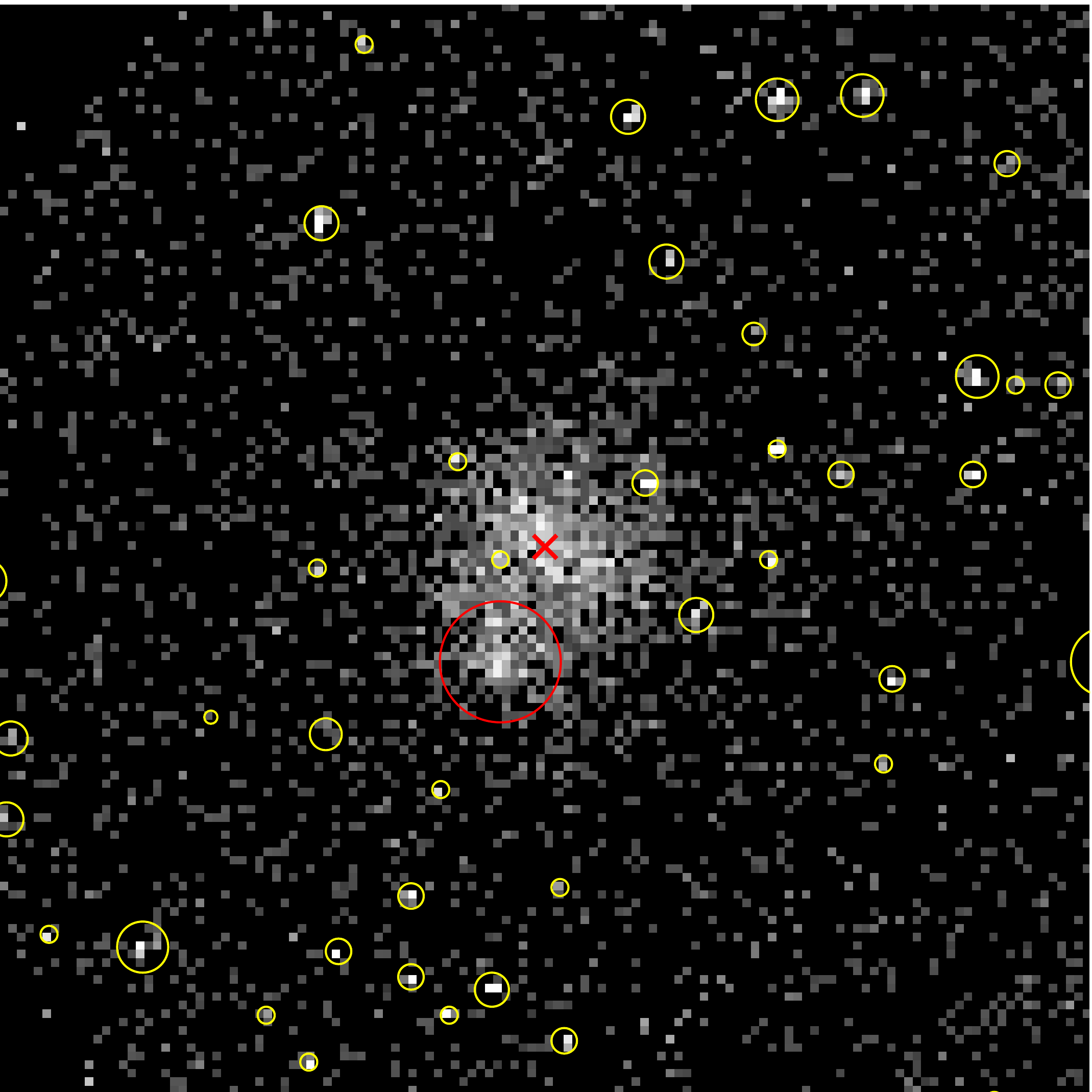}}
\medskip
\caption{Typical examples of \emph{Chandra} images for the high-redshift
  clusters (0230+1836, $z=0.80$ and cl1120+2326, $z=0.56$). Each panel
  is $8.4'\times8.4'$.  The meaning of the region marks is the same as
  in Fig.\,\ref{fig:img:examples}.} 
\label{fig:img:examples:high-z}
\end{figure*}

Uncertainties in determining each of the background components were
propagated in the further analysis. Their impact on the analysis of the
\emph{Chandra} cluster observations is extensively discussed in
\citetalias{2005ApJ...628..655V}. Here we only note that this source of
uncertainty is negligible for the measurements of the average cluster
temperatures dominated by the bright inner region; similarly, the gas mass
measurements are based on the surface brightness profiles in the soft band
where the background is lower relative to the cluster flux.

Conversion of the observed X-ray fluxes to physical quantities such as the
temperature and density of the intracluster gas relies on accurate
calibration of the spectral response. An extensive pre-flight calibration
program was designed to provide \emph{absolute} calibration of the effective
area of the mirror+ACIS system to within 2\% at all locations and across the
entire energy band. The in-flight performance was degraded somewhat but by
2005--2006, the calibration accuracy was restored to near-preflight levels.
Currently, the uncertainty in relative (position and time-dependent)
variations of the effective area is $<3\%$ within the energy band we use in
the present work\footnote{The current status of the \emph{Chandra}
  calibration is summarized on the WWW page http://cxc.harvard.edu/cal. See
  also \citetalias{2005ApJ...628..655V} for discussion relevant to the
  cluster data analysis.}. The estimated uncertainties in the absolute
effective area are $\lesssim5\%$ at all energies. The systematic effect of
such uncertainties on the estimated cluster mass function is small, as
summarized in \S\,\ref{sec:error:budget:X-ray:cal}. We also note that the
calibration uncertainties in the measurement of the \emph{evolution} of the
mass function are nearly canceled because we use the same telescope and
uniform analysis of both low- and high-$z$ samples.

\subsubsection{ROSAT}
\label{sec:data:rosat}

The \emph{ROSAT} PSPC data were reduced as described in
\cite{1999ApJ...525...47V}. The reduction pipeline was based on
S.~Snowden's software \citep{1994ApJ...424..714S}. This software
eliminates periods of high particle and scattered solar backgrounds as
well as those intervals when the detector may be unstable. Exposure maps
in several energy bands are then created using detector maps obtained
during the \emph{ROSAT} All-Sky Survey. The exposure maps include
vignetting and all detector artifacts. The unvignetted particle
background is estimated and subtracted from the data, even though the
PSPC particle background is low compared to the cosmic X-ray background.
The scattered solar X-ray background also should be subtracted
separately, because, depending on the viewing angle, it can introduce a
constant background gradient across the image.  Most Solar X-rays were
eliminated by simply excluding time intervals when this emission was
high, but the remaining contribution was also modeled and subtracted. If
the cluster was observed in several pointings, each pointing was reduced
individually and the resulting images were merged.

The energy resolution of the \emph{ROSAT} PSPC is insufficient to
separate the soft background components spectrally, which was possible
in the case of \emph{Chandra}. However, the \emph{ROSAT} field of view
is much larger and usually we can reliably measure the uniform
background level from the cluster observations themselves. Our procedure
for the background determination was to fit the observed surface
brightness profile at large radii, $r\gtrsim0.7\,r_{500}$, to a power
law plus constant model \citep[as discussed in][]{1999ApJ...525...47V}.
The additional power law component is required since at lower $z$, the
contribution of the cluster brightness is small but non-negligible even
near the edge of the \emph{ROSAT} PSPC field of view. The tests show
that this procedure provides a relative uncertainty in the background
determination of $\sim 5\%$ \citep{1999ApJ...525...47V}. This
uncertainty was propagated into the further analysis.

The limited bandpass of the \emph{ROSAT} PSPC (limited to $E<2$~keV)
does not allow one to measure the cluster temperatures with an accuracy
useful for our purposes. However, the observed count rate can be
converted to a broad-band flux very reliably, as confirmed directly by
excellent agreement with the \emph{Chandra}-derived flux from the same
region (see \S\,\ref{sec:icm:pars}).

\subsection{Removal of Substructures and Identification of the Cluster
  Center}
\label{sec:substr}

After the initial data preparation, we have flat-fielded and
background-subtracted images in the 0.7--2~keV energy band\footnote{The
  0.7--2~keV band is chosen to maximize the ratio of the cluster and
  typical background brightness.}. These images contain only the cluster
emission and other X-ray sources. Our next step is to remove all
point-like sources, as well as substructures within the cluster. The
point source removal is the most straightforward step. Our detection
routine is based on the wavelet decomposition technique documented in
\cite{1998ApJ...502..558V}. The point sources are identified using the
small scales of the wavelet decomposition and the corresponding regions
are masked out from all further analysis. The exclusion radius takes
into account the variation of the PSF size with the offaxis angle (this
is especially important in the case of \emph{ROSAT} PSPC pointed
observations).

\defcitealias{2006ApJ...640..691V}{V06}
\begin{figure*}
\centerline{%
\includegraphics[width=0.99\columnwidth]{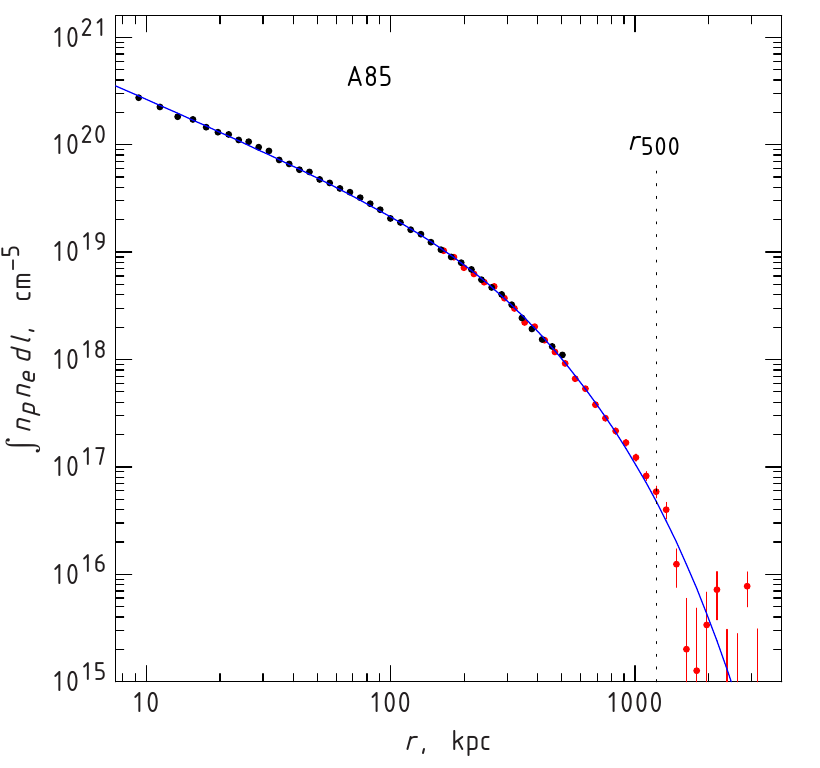}~~%
\includegraphics[width=0.99\columnwidth]{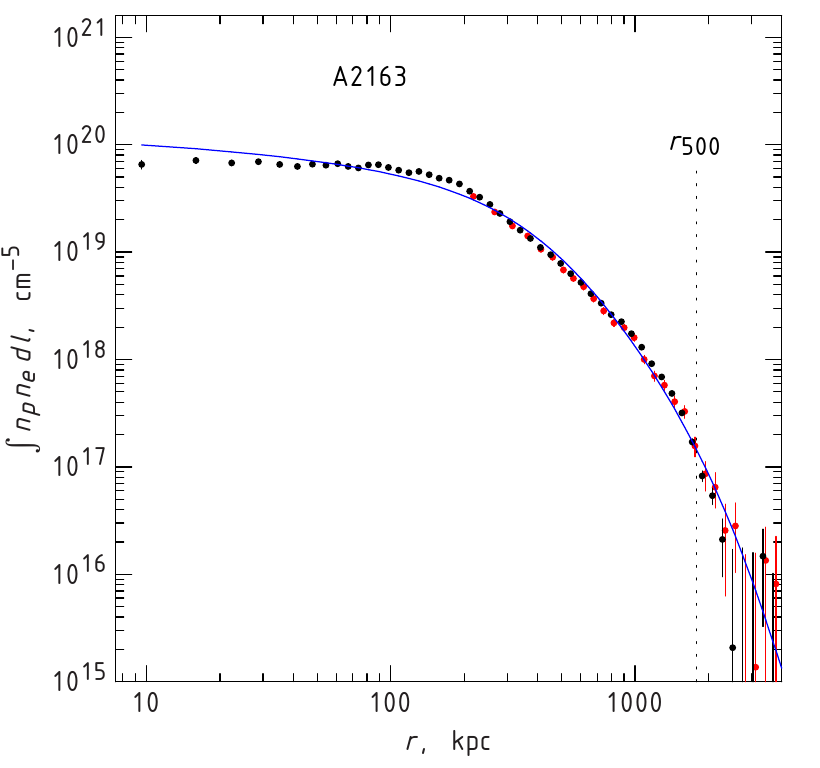}%
}
\medskip
\centerline{%
\includegraphics[width=0.99\columnwidth]{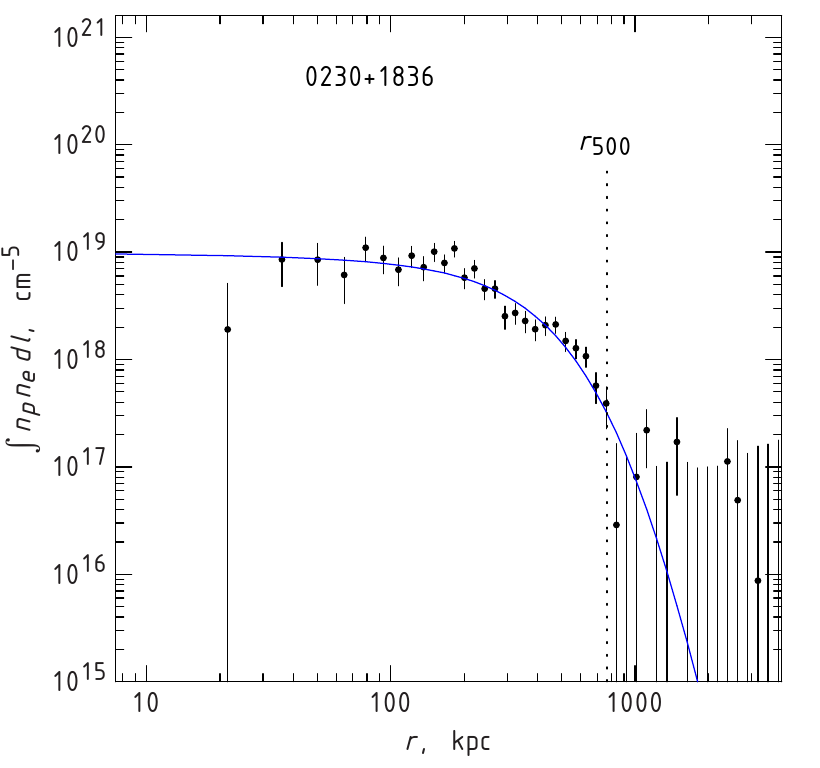}~~%
\includegraphics[width=0.99\columnwidth]{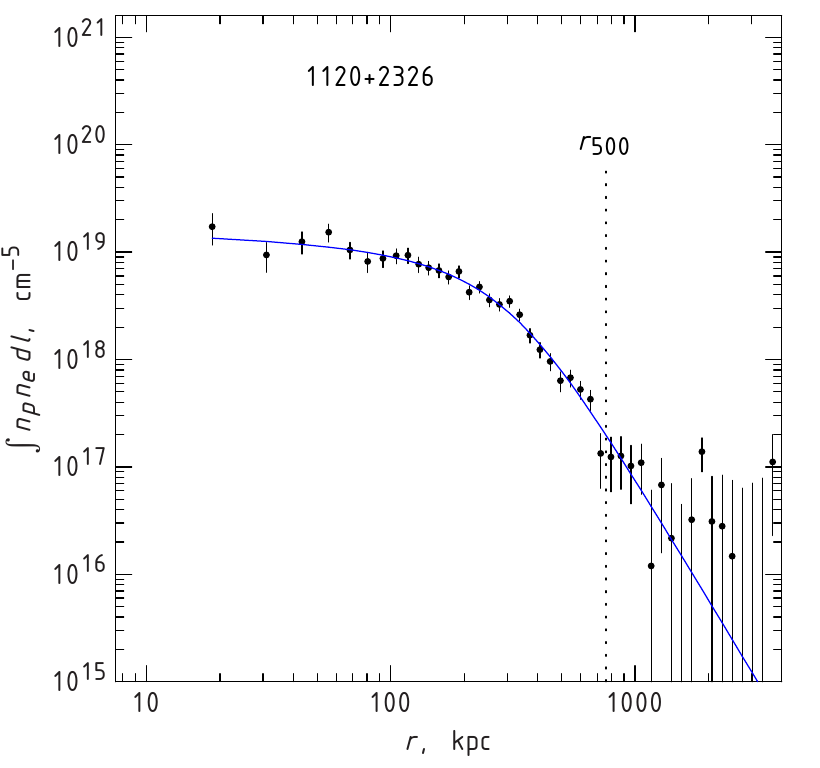}%
}
\caption{Examples of the surface brightness profile modeling for
  clusters shown in Fig.\,\ref{fig:img:examples} and
  \ref{fig:img:examples:high-z}. The observed X-ray count rates are
  converted to the projected emission measure integral (see
  \S\,\ref{sec:icm:pars} and \citetalias{2006ApJ...640..691V}). The
  black and red data points show the \emph{Chandra} and \emph{ROSAT}
  measurements, respectively. The best fit models (the projected
  emission measure integral for the three-dimensional distribution given
  by eq.~\ref{eq:density:model}) are shown by solid lines. The dashed
  lines indicate the estimated $r_{500}$ radii (the $\YX$-based value,
  see \S\,\ref{sec:M:estimates}), for reference. Note that in all cases,
  the surface brightness is traced accurately to $\R500$.  In relaxed
  clusters such as A85, the model describes the data very accurately. In
  strong mergers such as A2163, we see systematic deviations from the
  fit. The effect of such deviations on the cluster mass proxies was
  studied in \citet[][see also
  \S\,\ref{sec:nagai:summary}]{2007ApJ...655...98N}. }
\label{fig:prof:examples}
\end{figure*}

We also mask out any detectable, well-defined substructures within the
cluster (they are included only in the total X-ray luminosity). The
detection of substructures was fully automatic and based on the analysis
of large scales of the wavelet decomposition process. We masked out only
the regions associated with the prominent secondary maxima in the X-ray
surface brightness, keeping the weaker components such as filamentary
structures.  Examples are shown in Fig.\,\ref{fig:img:examples}
and~\ref{fig:img:examples:high-z}. Removal of obvious substructure
reduces the scatter in the relation between the total mass and X-ray
proxies, although the effect is small in most cases because we exclude
only a small fraction ($<20\%$) of the total flux. We note that removal
of substructures was included in the mock \emph{Chandra} analysis
\citep{2007ApJ...655...98N} which we use to assess the uncertainties in
the calibrations of the $\Mtot$ vs.\ proxy relations.

The only quantity we measure without removing the large
scale-substructures is the total X-ray luminosity. The luminosity
determines the detectability of the cluster in shallow surveys. These
surveys usually lack sensitivity and angular resolution to remove the
substructure and detect the clusters on the basis of its total flux.

\label{sec:center}

The further steps in the X-ray data reduction are based on the analysis of
the azimuthally averaged profiles. For this, we need to define the cluster
center in each case. In the case of relaxed clusters with cooling flows, the
center is defined to be simply at the location of the X-ray peak. The
situation is less straightforward for the non-cooling flow clusters or those
with substructure. Instead of using the maximum in the X-ray brightness map,
we center the profiles at the ``center of gravity'' for the main cluster
body. This is done by computing the mean emission-weighted coordinates using
the X-ray brightness in the annulus $r=[250-500]$\,kpc, and iterating this
procedure 2--3 times. The selection of the centroids is illustrated in
Fig.\,\ref{fig:img:examples}.

\subsection{Chandra Spectral Analysis} 
\label{sec:spectral:analysis}

There is an important difference in the approach for determination of
the average temperature from the \emph{Chandra} data for high- and
low-$z$ clusters. The procedure is straightforward for the high-$z$
objects that fall entirely inside the \emph{Chandra} field of view. In
this case, we can measure the average temperature simply by fitting a
single-$T$ model to the X-ray spectrum in the 0.6--10~keV band
integrated in the radial range of interest, e.g., $r=(0.15-1)\,r_{500}$.
This is a common, straightforward analysis; the interested reader can
find all the details of our approach in \cite{2005ApJ...628..655V}.

The situation is more complicated for low-$z$ clusters where typically
not all position angles fall inside the ACIS field of view at large
radii (Fig.\,\ref{fig:img:examples}). If we simply fit the integrated
spectrum within the ACIS field of view, the contribution of the central
region to the total flux will be higher than it should be in the case of
complete coverage.  This introduces a bias if the ICM temperature
distribution is not uniform.  Usually, $T$ is overestimated because the
observed $T(r)$ decreases at large radii
\citep{1998ApJ...503...77M,2002ApJ...567..163D,2005ApJ...628..655V,2007A&A...461...71P}.
Our solution is to measure the temperatures independently in several
annuli (we use annuli of equal logarithmic width, $r_{\rm out}/r_{\rm
  in}=1.5$, within which the overall gradient of $T(r)$ can be
neglected) and then average the obtained temperature profile weighting
each bin not with observed counts but with the total flux expected in
the given annulus if it were completely covered with the field of view.
The surface brightness profile needed to compute this weighting function
can always be derived from \emph{ROSAT} data that always cover the
radial range of interest. In principle this is not an exact method since
this weighting function is proportional essentially to the emission
measure integral, while the weighting corresponding to the spectroscopic
mean is different \citep{2004MNRAS.354...10M,2006ApJ...640..710V}. In
practice, however, this makes a negligible difference for our clusters,
as was verified using the clusters from the \cite{2005ApJ...628..655V}
sample that have the adequate radial coverage for exact computation of
$\langle T\rangle$.

The X-ray spectral model we fit to the observed \emph{Chandra} spectra
includes foreground absorption in the Galactic ISM. In most cases, the
absorbing column density, $N_H$, was fixed at the value provided by
radio surveys \citep{1990ARA&A..28..215D} but we always checked that it
is consistent with the observed spectrum. In a few cases (2A\,0335,
A2634, A478, A2390) the X-ray spectrum indicated a significantly higher
absorption than suggested by the radio data, most likely due to the
presence of molecular gas and dust along the line of sight.  In these
cases, $N_H$ was derived directly from the X-ray spectrum.  A cautionary
note is that small variations of $N_H$, of order $\pm
2\times10^{20}$~cm$^{-2}$, cannot be detected in the \emph{Chandra}
spectra\footnote{Nor in the combined \emph{Chandra} \& ROSAT spectrum in
  the 0.2--10~keV band if the nominal $N_H$ is greater than
  approximately $5\times10^{20}$~cm$^{-2}$. We note, however, that the
  \emph{ROSAT} data were checked for consistency with the nominal $N_H$
  for all clusters in our low-redshift sample.} because they are
indistinguishable from variations of the temperature. For the typical
values $N_H=4\times10^{20}$~cm$^{-2}$ and $T=5$~keV, the variation of
$N_H$ by $\pm 2\times10^{20}$~cm$^{-2}$ changes the best-fit temperature
by $\pm 7\%$, and also changes the derived gas mass by $\pm 3.5\%$,
anticorrelated with $T$. Such variations are smaller than the scatter of
these quantities for a fixed mass \citep*{2006ApJ...650..128K} but still
should be kept in mind. In this regard we note that $\YX=\Mgas\times
\TX$ is less sensitive to variations of $N_H$ because they have the
opposite effect on $\TX$ and $\Mgas$.

The last issue that should be discussed in relation with the X-ray
spectral analysis is the treatment of the ICM metallicity. In low-$z$
clusters, the statistical quality is sufficient to measure the metal
abundance simultaneously with the temperature. This is impossible for
most of our high-$z$ clusters. In these cases, we fixed the metallicity
at $Z=0.3\,Z_\odot$, the typical value at both low and high redshifts
\citep{1997ApJ...481L..63M,2003ApJ...593..705T}. We verified that
variations of $Z$ in the range $0.1-0.5$ (conservative bracket) have a
small effect on the derived parameters --- for a $T=5$~keV cluster at
$z=0.5$ the temperature changes by $\pm 5\%$ and $\Mgas$ changes by $\pm
2\%$, correlated with $T$.

The instrumental uncertainties in $\TX$ measurement are systematic and
uniform (do not introduce object-to-obejct scatter or any significant
redshift-dependent trends). They are considered separately in
\S\,\ref{sec:error:budget:X-ray:cal}. 

%
%
%

\subsection{Gas Mass Measurements}
\label{sec:icm:pars}

Two of the mass proxies we utilize for the $\Mtot$ estimates
(\S\,\ref{sec:M:estimates}) use the gas mass within $r<r_{500}$.
Derivation of the gas mass from the X-ray imaging data is relatively
straightforward, but a few points are still worth noting here. Our
procedure for the $\Mgas$ measurements follows that used for a more
detailed analysis of a smaller sample of low-redshift clusters described
in \citet[V06 hereafter]{2006ApJ...640..691V}, and the main steps are
outlined here for completeness.

\defcitealias{2006ApJ...640..691V}{V06}

The X-ray flux in the 0.7--2~keV energy band is very insensitive to the
plasma temperature, as long as $T\gtrsim2$~keV
\citep{1980ApJ...241..552F}. The observed brightness gives essentially
the integral of $\rho_g^2$ along the line of sight. This is why the ICM
mass is robustly derived from the X-ray data even if the detector has
almost no energy resolution and a limited bandpass. Even though the
effects of the temperature and metallicity are very weak, we applied the
appropriate corrections to the observed surface brightness profiles as
detailed in \citetalias{2006ApJ...640..691V}; this correction also
removes the effects of spatial variations of the telescope effective
area. The corrected profiles are expressed in units of the projected
emission measure integral, $\int n_e\,n_p\,dl$. They are deprojected to
reconstruct the 3-dimensional profile of $\rho_g(r)$. This is done by
fitting the projected data to an analytical model,
\begin{equation}\label{eq:density:model}
  \begin{split}
    n_p\,n_e = n_0^2\;\frac{(r/r_c)^{-\alpha}}{(1+r^2/r_c^2)^{3\beta-\alpha/2}}
                \;&
                \frac{1}{(1+r^\gamma/r_s{}^\gamma)^{\varepsilon/\gamma}}\\
                &+ \frac{n_{02}^2}{(1+r^2/r_{c2}^2)^{3\beta_2}}. 
  \end{split}
\end{equation}
that represents all main features observed in real clusters --- the
$\beta$-model \citep{1978A&A....70..677C} profile
\citep{1984ApJ...276...38J} that may steepen at large radii, and also
show a power-law cusp and possibly a separate component in the center. 
These modifications of the $\beta$-model greatly enhance the functional
freedom and improve the reliability of the X-ray modeling at large radii
(see discussion in \citetalias{2006ApJ...640..691V}). 

The parameters of the 3-dimensional model~(\ref{eq:density:model}) are
obtained by numerically projecting it along the line of sight and
fitting to the observed profile. The best fit directly gives us the
analytic expression for the 3-dimensional profile of $\rho_g(r)$ which
can be integrated to determine $\Mgas$ in the given range of radii.
Several examples of this analysis are shown in
Fig.\,\ref{fig:prof:examples}. Note the excellent agreement between the
\emph{Chandra} and \emph{ROSAT} measurements in the same regions
indicating an accurate cross-calibration between the two instruments.
The uncertainties of $\rho_g(r)$ and $\Mgas$ are derived via Monte-Carlo
simulations~\citepalias[see][]{2006ApJ...640..691V}.

\subsection{Verification by Mock Observations of\\ Cosmological Simulations}
\label{sec:nagai:summary}

We note that our approach to the measurements of the ICM mass and
average temperature has been fully tested by the analysis of the mock
\emph{Chandra} observations of the clusters from high-resolution
cosmological simulations \citep{2007ApJ...655...98N}. The cosmological
cluster simulations used in this work should correctly reproduce the
main aspects of the ICM structure in real clusters, including the
large-scale deviations of the main cluster body from spherical symmetry
and intermediate-scale nonuniformities of the ICM density and
temperature. In fact, the simulations reproduce the detailed X-ray
properties of the ICM in the cluster outskirts ($r\gtrsim 0.2r_{500}$)
quite well \citep{2007ApJ...668....1N} and are therefore sufficiently
realistic for our purposes. In constructing the mock observations of
these simulations, we carefully reproduced the essential observational
effects such as the \emph{Chandra} sensitivity to plasma of different
temperatures, the background level and photon statistics found in
typical observations for both low and high-redshift clusters. The mock
data were reduced by the same software that we use for the analysis of
real cluster observations.

The mock data analysis thus tests the combined effect of inaccuracies in
all steps of our analysis, including removal of substructures,
temperature measurements, and modeling of the X-ray brightness profile.
The mock analysis shows that we recover $\Mgas$ and average temperatures
very accurately. For example, the bias in $\Mgas$ within $r=\R500$ due
to small-scale non-uniformities of the ICM is only $+3\%$, independent
of redshift. The unrelaxed clusters are not significantly different from
the relaxed ones, except for a small number of outliers where the
$\Mgas$ measurement can be biased by 10--15\%. This is significantly
smaller than the biases reported in the earlier work by
\citet*{1999ApJ...520L..21M}. The improvement can be explained by
advances in the data analysis (in particular, the relaxation of the
assumption that the ICM density follows the $\beta$-model) and inclusion
in our mock analysis of the effect of substructure removal which was
always used by observers.

To summarize, we can state that the results from the analysis of mock
observations validate our analysis methods. The expected residual biases
have almost no effect on the derivation of the cluster mass function.

\section{Total mass estimates}
\label{sec:M:estimates}

The cluster mass is not a well-defined quantity and can be defined in a
variety of ways \citep[see, e.g.,][]{2001A&A...367...27W}. We choose to define mass
within the radius corresponding to a fixed
mean overdensity, $\Delta$, with respect to the critical density at the cluster
redshift, $\rho_c\equiv 3H^2(z)/8\pi G$:
\begin{equation}
  M_\Delta=M(<r_\Delta):\quad \frac{M_\Delta}{4/3\,\pi\,r_\Delta^3}=\Delta\times\rho_c.
\end{equation}
The choice of the overdensity threshold is driven by practical
considerations. The ultimate goal of these measurements is to compare the
observed mass function with the theoretical predictions. The mass function
models, which are calibrated by numerical simulations
\citep[e.g.,][]{2001MNRAS.321..372J}, are more robust for low values of
$\Delta$, where the role of numerical resolution in the simulations and
non-gravitational effects within clusters is minimal. On the contrary, the
masses derived from X-ray data are more robust for high values of $\Delta$,
where the statistical quality is higher, hydrostatic equilibrium assumption
is more accurate, etc. We need, therefore, to choose a compromise between
conflicting theoretical and observational requirements. We choose
$\Delta=500$ --- the radius within which the clusters are relatively relaxed
\citep{1996ApJ...469..494E} and good measurements of gas mass and
temperature can be obtained with our \emph{Chandra} observations
\citep{2006ApJ...640..691V,2007ApJ...655...98N}. This is, effectively, the
largest radius at which the ICM temperature can be reliably measured with
\emph{Chandra} and \emph{XMM-Newton}
\citep[e.g.][]{2005ApJ...628..655V,2007A&A...461...71P}.  Using
significantly lower $\Delta$ dramatically increases observational
uncertainties; at significantly higher values of $\Delta$, the theoretical
uncertainties start to increase while there is no crucial gain on the
observational side.

The total cluster masses, $\M500$, are estimated from observed ICM
parameters. We employ the three X-ray proxies for $\Mtot$ discussed in
\citet[][KVN hereafter]{2006ApJ...650..128K} --- the core-excised average
temperature, $\TX$; the hot gas mass, $\Mgas$; and the estimated total
thermal energy, $\YX=\TX\times\Mgas$. We rely on the existence of
low-scatter scaling relations between these parameters and $\Mtot$, as
predicted by self-similar theory and confirmed by high-resolution
cosmological simulations.

\defcitealias{2006ApJ...650..128K}{KVN}

The mass vs.\ proxy relations are calibrated using the hydrostatic
$\Mtot$ estimates in a sample of well-observed, low-redshift, relaxed
clusters, 10 clusters from \citetalias{2006ApJ...640..691V} plus seven
additional objects (A2717, A3112, A1835, A1650, A2107, A4059,
RXJ\,1504--0248) whose deep \emph{Chandra} observations appeared in the
archive since 2006\footnote{These data were reduced completely
  identically to \citet{2005ApJ...628..655V} and
  \citetalias{2006ApJ...640..691V}. All primary conclusions of these
  papers hold for these additional objects. The only effect is to
  improve the accuracy of the $\Mtot$ vs.\ proxy relations.}. In
principle, the hydrostatic method can underestimate the total mass due
to non-thermal pressure components. For example, the analysis of mock
observations presented in \cite{2007ApJ...655...98N} suggests that
$M_{500}$ can be underestimated by $\sim 15\%$, and this effect can be
attributed to the bulk motions of the gas at large radii.  We do not
correct the normalization of the mass vs.\ proxy relation for any such
effects because there are theoretical uncertainties in their magnitudes
(e.g., the ICM viscosity can affect the average velocity of small-scale
bulk motions). We simply account for the possible $\Mtot$ biases in the
total systematic error budget (see \S\,\ref{sec:sys:err} below).
Ultimately, a reliable calibration of the mass vs.\ proxy relation can
be obtained through a stacked weak lensing analysis
\citep[e.g.,][]{2001ApJ...554..881S} of a representative sample of
clusters with high-quality X-ray data. Such data are only starting to
become available now
\citep{2007MNRAS.379..317H,2008MNRAS.384.1567M,2008A&A...482..451Z} and
we in fact use them to place limits on systematic errors in our
calibration of the $\Mtot$ measurements (\S\,\ref{sec:Xray:vs:WL}).

We do apply, however, small first-order corrections to the observed mass
vs.\ proxy relations when they are required to transfer the calibration
from relaxed clusters to the entire population or to account for
expected departures from self-similarity in the evolution of these
relations. In doing this, we try to use only the most robust predictions
from the simulations and to rely on the directly observed properties as
much as possible. The corrections to each proxy are detailed below. The
largest corrections are applied for the $\Mtot-\TX$ relation, while the
$\Mtot-\YX$ relation does not require any corrections (and hence is
potentially the most reliable).

\subsection{$\Mtot-\TX$ Relation}
\label{sec:M-T}

The average X-ray temperature is one of the most widely used cluster
mass indicators. The $M-T$ relation expected in self-similar theory is
given by
\begin{equation}\label{eq:M-T}
  \M500 \propto T^{3/2}\, E(z)^{-1}, \quad {\rm where} \quad
  E(z)\equiv H(z)/H_0.
\end{equation}
This relation arises in a self-similar model simply because the ICM temperature is expected
to scale with the depth of gravitational potential $T\propto M/R$ and mass and
radius in our adopted definition are related ($R\propto M^{1/3}$). 
The relation~(\ref{eq:M-T}) also generally describes the ICM temperatures
found in the cosmological numerical simulations
\citep[][\citetalias{2006ApJ...650..128K}]{1996ApJ...469..494E,2001ApJ...546..100M,2004MNRAS.348.1078B}. 

The average cluster temperature can be defined in different ways but the
most practical, from the observational point of view, is the average
spectral temperature --- the value derived from a single temperature fit
to the total cluster spectrum integrated within a given radial range. We
refer to this temperature as $\TX$.

\subsubsection{Definition and Determination of\/ $\TX$}
\label{sec:Tx:def}

Spatially-resolved X-ray spectroscopy became available with the launch
of the \emph{ASCA} satellite, and since then many studies has indicated
that the cluster scaling relations become tighter if the average
temperature is measured excluding the cluster centeral region which is
often affected by radiative cooling.  This is well illustrated by the
reduction in scatter in the $L_X-T$ relation shown in Fig.1--2 of
\cite{1998ApJ...504...27M}.  The temperature profiles show a large
object-to-object scatter in the centers of even relaxed clusters
\citepalias{2006ApJ...640..691V}.  Clearly, the central cluster region
should be excluded from the measurement of $\TX$.
\cite{1998ApJ...504...27M} has used $\rin=70$~kpc (this inner cutoff
radius was also used in \citetalias{2006ApJ...640..691V} and several
other works).  Perhaps a better motivated choice is to set $\rin$ at a
fixed fraction of $\R500$ \citep{2005A&A...441..893A}. We will use, as
in \citetalias{2006ApJ...650..128K}, $\rin=0.15\,\R500$, because
approximately outside this radius the observed profiles of relaxed
clusters are self-similar \citepalias{2006ApJ...640..691V}.
\emph{Chandra}'s angular resolution is sufficient to resolve
$0.15\,\R500$ even in the highest-redshift objects. An algorithmic
complication is that the cutoff radius is expressed through $\Mtot$
which is itself estimated from, e.g., $\TX$. This is not a big problem
since $\TX$ is not very sensitive to the exact value of $\rin$, and
hence the following iteration scheme converges quickly: (a) measure
$\TX$ including the central region; (b) estimate mass from $M-T$
relation; (c) re-measure $\TX$ using $\rin=0.15\,\hat{r}_{500}$ and
estimate new mass; repeat step (c) until convergence is reached.

We also need to address the issue of the outer radius for integration of
the X-ray spectrum. The cluster properties seem to become progressively
self-similar at large radii \citep*{2007ApJ...668....1N}. Therefore,
ideally, the spectrum should be integrated as far out as possible. The
exact value of $\rout$ is unimportant because the total X-ray flux
converges quickly at $r\rightarrow\infty$. A good practical choice is to
set $\rout=\R500$, because outside approximately this radius, the X-ray
brightness is low compared with the background (e.g.,
Fig.\,\ref{fig:prof:examples}).

\begin{figure}[t]
\centerline{\includegraphics[width=0.97\linewidth]{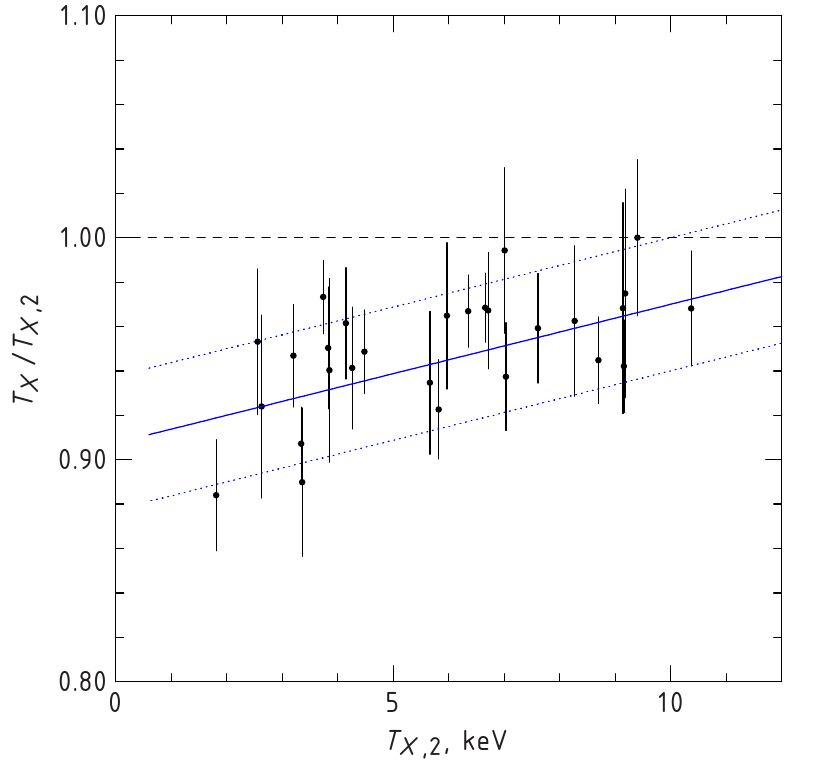}}
\caption{Ratio of the X-ray spectral temperatures measured in the radial
  ranges $(0.15-0.5)\,\R500$ (``$T_{X,2}$'') and $(0.15-1)\,\R500$
  (``$\TX$''), for clusters in the local sample that have a sufficient
  \emph{Chandra} coverage.  The solid line shows the linear
  approximation given by eq.(\ref{eq:t3:t4}), with a 3\% level of
  scatter indicated by dotted lines. The ratio of the temperatures is
  also consistent with a constant value, $\simeq 0.95$ except for a few
  outliers at low $T$.}
\label{fig:t3:t4}
\end{figure}

Given the arguments presented above, we determine $\TX$ in the radial
range $0.15\,\R500-1\,\R500$. This is a straightforward measurement for
our high-$z$ clusters because this region fits completely inside the
\emph{Chandra} field of view and exposures were designed to provide a
sufficient statistical accuracy.  However, for a large fraction of the
low-$z$ clusters, integration to $\R500$ is impossible because of the
limited field of view\footnote{Note that the gas mass can still be
  measured in these clusters out to $\R500$ using the \emph{ROSAT} PSPC
  data.}. A simple investigation shows that we can use a smaller value
of $\rout$ in such cases. First we note that the temperature can be
integrated to $0.5\,\R500$ for all clusters (we refer to this value as
$\Tx2$). For clusters that have sufficient radial coverage, we measured
temperatures both in $(0.15-0.5)\,\R500$ and $(0.15-1)\,\R500$ radial
ranges. The ratio of the two values is shown in Fig.\,\ref{fig:t3:t4}.
It is consistent with a linear relation,
\begin{equation}\label{eq:t3:t4}
  \TX/\Tx2=0.9075 + 0.00625\,\Tx2,
\end{equation}
where the temperatures are in units of keV. The observed scatter around the
linear fit is negligible, $\lesssim 3\%$. The ratio is also consistent with
a constant value, $\simeq 0.95$, except for a few outliers at low
temperatures. Since a tight correlation is observed, we can measure $\Tx2$
and then estimate $\TX$ with a sufficient accuracy using
equation~(\ref{eq:t3:t4}) for those clusters that are not covered by
\emph{Chandra} at large radii.

Finally, we note that even if the trend in $\TX/\Tx2$ is real, this does
not necessarily imply deviations from self-similarity. Because $T$ is
not constant as a function of radius, we have a mixture of spectral
components within any aperture.  A single-tem\-pe\-ra\-ture fit to such
a spectrum gives a weighted average which is different from the
mass-weighted $T$ and weighting itself depends on the typical
temperature in the spectrum
\citep{2004MNRAS.354...10M,2006ApJ...640..710V}. Therefore, we expect
trends in the $\TX/\Tx2$ ratio even if the scaled 3-dimensional
temperature profiles for low and high-$T$ clusters are identical.

\subsubsection{Calibration of $\Mtot-\TX$ Relation at Low Redshifts
  using Relaxed Clusters}
\label{sec:M-Tx:cal}

\begin{figure}[t]
\centerline{\includegraphics[width=0.97\linewidth]{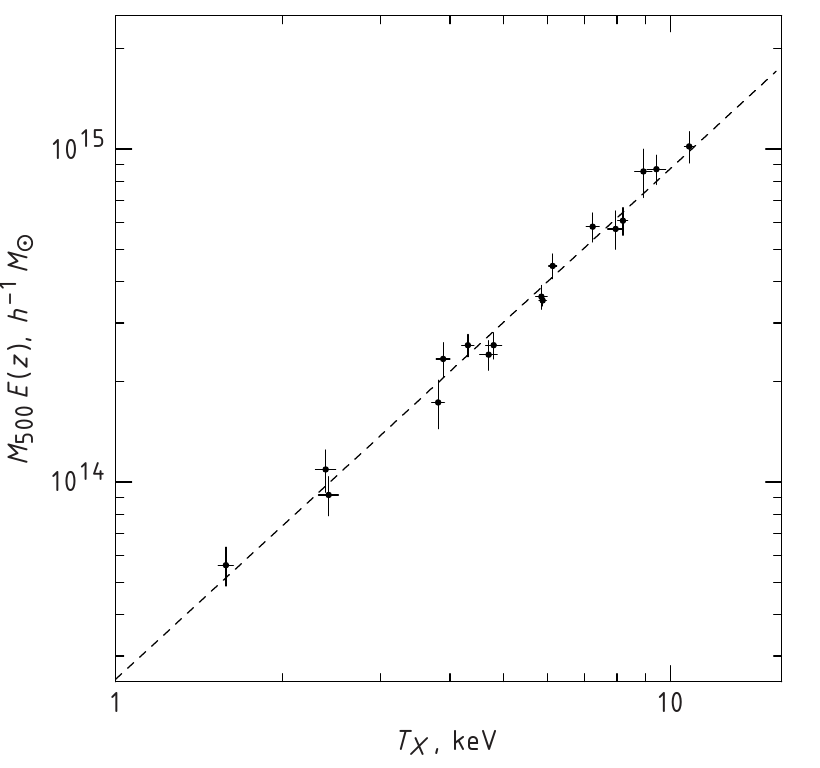}}
\caption{Calibration of the $M-T$ relation using X-ray hydrostatic mass
  measurements for a sample of 10 relaxed \emph{Chandra} clusters with the
  temperature profile measurements extending to $r=\R500$. The mass
  measurements are taken from \citetalias{2006ApJ...640..691V} with 7
  additional clusters (see \S\,\ref{sec:M:estimates}), the temperatures
  match our definition of $\TX$ (see \S~\ref{sec:Tx:def}).  The dashed line
  shows the best-fit power law relation (parameters given in
  Table~\ref{tab:relations:cal}).}
\label{fig:M-T:cal}
\end{figure}

For 17 low-redshift relaxed clusters, there exist very high-quality
\emph{Chandra} observations, providing temperature profiles extending
sufficiently far to permit hydrostatic mass estimates at $r=\R500$ (see
introduction to \S\,\ref{sec:M:estimates}). These observations are a basis
of our calibration of the $\Mtot-\TX$ relation at low redshifts.  The mass
and temperature measurements for these 17 clusters (Fig.\ref{fig:M-T:cal};
note that we symmetrize the error bars for simplicity) are fit to the power
law,
\begin{equation}\label{eq:M-T:fit}
  M = M_0 \, E(z)^{-1}\, (T/5\,\text{keV})^\alpha,
\end{equation}
normalized at $T=5$~keV because this is approximately the median temperature
for this sample and therefore the estimates for $M_5$ and $\alpha$ should be
uncorrelated. The fit is performed using the bisector modification of the
\citet[and references therein]{1996ApJ...470..706A} linear regression
algorithm that allows for intrinsic scatter and nonuniform measurement
errors in both variables. The uncertainties were evaluated by bootstrap
resampling \citep[e.g.,][]{numerical_recipes}, while simultaneously adding
random measurement errors to $M$ and~$T$. The results are shown in
Fig.\,\ref{fig:M-T:cal} and the best-fit parameters of the power law fit are
reported in Table~\ref{tab:relations:cal}. The best-fit slope, $1.53\pm0.08$
is consistent with the expectation of the self-similar theory
(eq.\ref{eq:M-T}). Fixing the power law slope at 1.5 does not significantly
reduce the uncertainty in the normalization (Table~\ref{tab:relations:cal}).
The \emph{XMM-Newton} determination of the $M-T$ relation
\citep{2005A&A...441..893A} is close to our measurement.

Our procedure for hydrostatic $\Mtot$ estimates was fully tested using mock
data from the simulations in \citet{2007ApJ...655...98N}. This work shows
that the inaccuracies introduced by the X-ray data analysis --- e.g., those
related to departures of the cluster body from spherical symmetry --- are
small.  The dominant source of error are departures from equilibrium and
non-thermal pressure components --- the effect fundamentally missed by the
X-ray hydrostatic mass estimates. For example, the residual random gas
motions in ``relaxed'' clusters in the Nagai et al.\ sample seem to result
in a 10--20\% underestimation of $\Mtot$ near $r=\R500$.  Unfortunately,
direct measurements of the ICM turbulence (and other non-thermal pressure
terms) presently are unavailable. We thus face a dilemma: should we use the
theoretical modeling to estimate corrections to the X-ray mass estimates, or
should we rely only on observations? Our choice is to follow the philosophy
outlined in the introduction to \S\ref{sec:M:estimates} and to use the
corrections suggested by simulations as an estimate of the systematic
errors. A better estimate (9\%) for the systematic uncertainties in the
\emph{Chandra} cluster mass scale can be obtained from comparison of X-ray
and weak lensing mass measurements, see \S\,\ref{sec:Xray:vs:WL} below.

\begin{deluxetable*}{p{3.5cm}lll}
  \tablecaption{Calibration of mass-observable relations\label{tab:relations:cal}}
  \tablehead{
      \colhead{Relation} &
      \colhead{Form} &
      \colhead{$M_0$, $f_{g,0}$} &
      \colhead{$\alpha$}
    }
  \startdata
$\M500-\TX$\dotfill & $M_{500} = M_0\,(T/5\,{\rm keV})^{\alpha}\,E(z)^{-1}$ &
  $(3.02\pm0.11)\times10^{14}\,h^{-1}\,M_\odot$ & $1.53\pm0.08$ \\
$\M500-\TX$\dotfill & $M_{500} = M_0\,(T/5\,{\rm keV})^{\alpha}\,E(z)^{-1}$ &
  $(2.95\pm0.10)\times10^{14}\,h^{-1}\,M_\odot$ & $1.5$, fixed \\
$\M500-\Mgas$\dotfill & $f_g = f_{g,0} + \alpha \log M_{15}$ &
$(0.0764\pm0.004)\,h^{-1.5}$ & $0.037\pm0.006$ \\
$\M500-\YX$\dotfill & $M_{500} = M_0\,(\YX/3\times10^{14}\,\Msun\,{\rm keV})^{\alpha}\,E(z)^{-2/5}$ &  $(5.77\pm0.20)\times10^{14}\,h^{1/2}\,M_\odot$ & $0.57\pm0.03$ \\
$\M500-\YX$\dotfill & $M_{500} = M_0\,(\YX/3\times10^{14}\,\Msun\,{\rm keV})^{\alpha}\,E(z)^{-2/5}$ & $(5.78\pm0.30)\times10^{14}\,h^{1/2}\,M_\odot$ & $0.6$, fixed
\enddata
\tablecomments{To apply the relations, measure the mass proxy for your
  $h$ of choice, and scale the normalization factor in column~(2)
  according to the $h$-dependence given in column~(3). $\Mtot-\YX$
  relation should be applied according to eq.(\ref{eq:m-yx:reallife}).
  The $f_g$ trend is used in the $\Mtot-\Mgas$,
  eq.(\ref{eq:Mgas:Mtot:2}) and~(\ref{eq:Mgas:Mtot:1}); the
  $z$-dependence of this relation is discussed in
  \S\,\ref{sec:fgas:corr:evol}.}
\end{deluxetable*}

\begin{figure}[tb]
\centerline{\includegraphics[width=0.97\linewidth]{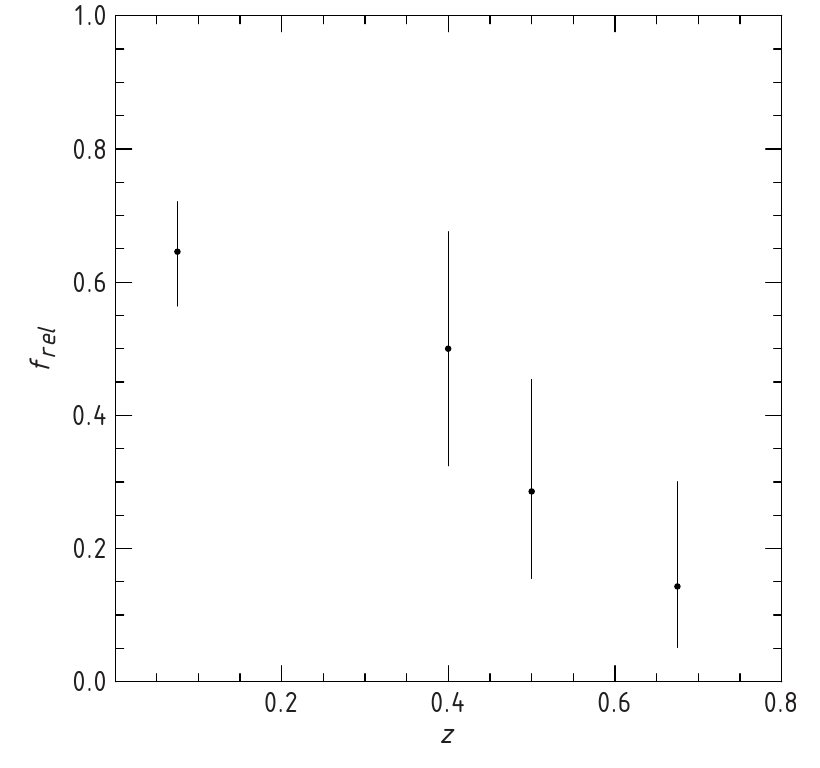}}
\caption{The fraction of clusters which could be classified as ``relaxed''
  based on their observed X-ray morphology (presence of secondary peaks,
  large centroid shift etc., see \S\,\ref{sec:M-T:unrelaxed}), as a function
  of $z$.}
\label{fig:relaxed:frac}
\end{figure}

\subsubsection{Transfer of $\Mtot-\TX$ Calibration to Entire Population}
\label{sec:M-T:unrelaxed}

The simulations suggest a systematic offset in the normalization of the
$\Mtot-\TX$ relation for relaxed and unrelaxed clusters, in the sense that the
merging clusters tend to have lower temperatures for the same mass
\citep[\citetalias{2006ApJ...650..128K}]{2001ApJ...546..100M,ventimiglia08}. Since our
calibration of the $\Mtot-\TX$ is for a subsample of relaxed clusters, we need
a procedure to transfer this calibration for the entire population that
contains both relaxed and merging clusters. This can be achieved using a
simple, first-order correction outlined below. 

First, we note that the systematic offset in the $\Mtot-\TX$ relation
cannot be measured directly using the X-ray data. Ultimately, it can be
measured with a weak lensing analysis of a large sample. The results
presented in \citetalias{2006ApJ...650..128K} (their Table~2) suggest
that the offset is $(17\pm5)\%$ in mass for a fixed $\TX$. There is no
obvious trend of this offset with redshift, or the difference in the
slope of the relations for relaxed and merging clusters.  Most
importantly for our application, this offset can lead to departures from
self-similar evolution in the $\Mtot-\TX$ relation for the entire
cluster population, because the fraction of merging clusters is expected
to increase at high redshifts
\citep[e.g.][]{2001ApJ...546..223G,2005APh....24..316C}, as we indeed
observe in our sample (Fig.\,\ref{fig:relaxed:frac}).

Second, \citetalias{2006ApJ...650..128K} and \citet{2007ApJ...655...98N}
classified the simulated clusters as relaxed and unrelaxed using only
the morphology of their mock X-ray images. We, therefore, can apply the
equivalent classification to the observed clusters in our high- and
low-$z$ samples. ``Unrelaxed'' clusters are those with secondary maxima,
filamentary X-ray structures, or significant isophotal centroid shifts.
\citet{2007ApJ...655...98N} show examples of this classification applied
to simulated data; more examples in the real data can be found in
Fig.~\ref{fig:img:examples}--\ref{fig:img:examples:high-z}. The
$\TX$-based mass estimates for clusters identified as mergers should be
corrected upwards by a factor of 1.17. Formal uncertainties on this
correction factor are $\pm0.05$ \citepalias[Table~2
in][]{2006ApJ...650..128K}; the average uncertainties for the entire
population are futher reduced because masses have to be corrected only
for a fraction of clusters (see below). Applying such a correction for
individual objects automatically takes into account any
redshift-dependent changes in the fraction of mergers, and thus removes
this source of departures from self-similar
evolution. \citet{ventimiglia08} show that the deviations from the mean
$\Mtot-\TX$ relation in the simulated clusters are correlated with the
quantitative substructure measures; using such an approach instead of
our simple classification is potentially more accurate and would be
warranted in samples of larger size.

We note that the underlying source of difference in the $\Mtot-\TX$
normalization between mergers and relaxed clusters is incomplete
relaxation of the intracluster gas. A fraction of energy is contained in
bulk motions of the gas and it is gradually converted into heat as the
cluster relaxes after a merger. This process (unlike, e.g., radiative
cooling in the center) should be reliably reproduced by current
simulations, and so our reliance on the simulations to derive this
correction is justified. Our dichotomical classification is of course
very approximate and a more accurate approach should take into account
the cluster relaxation history.  However, the current X-ray data does
not allow us to quantify the cluster dynamical state with the required
precision. Even rather simple substructure measures
\citep[e.g.,][]{2007arXiv0708.1518J} require more photons than we have
for distant objects. Moreover, the process of relaxation should be
sensitive to effective viscosity in the ICM and it is unclear that the
current simulations which incorporate only low, numerical viscosity can
accurately predict the $\Mtot-\TX$ relation for semi-relaxed clusters.
They, however, should still be reliable for the extreme cases.  In
nearly-relaxed clusters, the turbulent motions are very weak in the
inner regions (which dominate the $\TX$ measurements) even in the
zero-viscosity simulations. In post-merger clusters, most of the
turbulent energy is in the large-scale flows which dissipate on long
time scales even if viscosity is high (Coulomb). Furthermore, the
magnitude of this correction is relatively small. We estimate that the
fraction of non-relaxed clusters in the sample changes from 35\% at
$z=0$ to $\sim 80\%$ at $z=0.6$ (Fig.\,\ref{fig:relaxed:frac}). The
corresponding correction for the $\Mtot-\TX$ normalization for the
entire populations is $+6\%$ and $+13\%$ at $z=0$ and $z=0.6$,
respectively; thus the redshift-dependent correction is only $7\%$ in
mass.
\looseness-1

\subsubsection{Summary of Mass Estimates through $\Mtot-\TX$ Relation}
\label{sec:M-T:summary}

This section provides a summary of how we use the X-ray temperature for
the cluster $\Mtot$ estimates. 

First, an initial value of $\TX$ is obtained from the \emph{Chandra}
spectrum integrated within a wide aperture (not excluding the
center). This $\TX$ is used to estimate $\M500$ using the power law
fit~(\ref{eq:M-T:fit}) and thus $\R500$. The temperature is then
remeasured in the annulus $(0.15-1)\,\R500$ and this procedure is
iterated several times until convergence is reached. If the radius
$\R500$ is well outside the \emph{Chandra} field of view, we use a
smaller aperture, $(0.15-0.5)\,\R500$, and apply corrections detailed
in \S\,\ref{sec:Tx:def}. 

Our $\Mtot-\TX$ relation is calibrated by very high-quality \emph{Chandra}
observations of 17 low-redshift relaxed clusters with a wide range of
masses. The statistical accuracy of this calibration is $\approx 3\%$, so
the dominant source of uncertainty is systematics, mostly related to the
possible presence of non-thermal pressure components in the ICM.

Next, we need to compensate for the expected systematic difference in
the $\Mtot-\TX$ relation for relaxed and unrelaxed clusters.  If the
X-ray morphology shows that the cluster is unrelaxed, the mass estimated
from the $\Mtot-\TX$ relation is multiplied by a factor of 1.17
(\S\,\ref{sec:M-T:unrelaxed}). In doing so, we assume that $\Mtot-\TX$
relations for relaxed and unrelaxed clusters separately evolve precisely
as expected in the self-similar theory ($M$ for fixed $T$ scales as
$E(z)^{-1}$). We cannot verify this assumption independently of the
background cosmology we would like to measure. Instead, we rely on the
simulations to estimate the magnitude of possible departures from the
self-similar scaling. Such departures must be treated as systematic
errors which affect the cosmological constraints. From the results
presented in \citet{2007ApJ...655...98N}, we estimate this uncertainty
to be equivalent to $\approx 7\%$ difference in the normalization of
$\Mtot-\TX$ relations at $z=0$ and $z=0.6$.

To properly compute the likelihood function for the estimated cluster
mass functions, we need to know the intrinsic scatter in the $\TX$-based
mass estimates. The simulations suggest that this scatter is $\simeq
20\%$, and we adopt this value. We later verified that reasonable
variations of the scatter (in the range $15-25\%$) have negligible
effect on fitting the estimated mass function. This range brackets the
scatter observed in the simulations separately for relaxed and unrelaxed
subpopulations, as well as for low and high-redshift clusters
\citepalias[Table~2 in][]{2006ApJ...650..128K}. Therefore, our analysis
is insensitive to realistic trends of the scatter with redshift.

\subsection{$\Mtot-\Mgas$ Relation}
\label{sec:fgas}

Our second method of estimating the cluster total mass uses the X-ray
derived hot gas mass as a proxy. The application of this proxy is extremely
simple in an ideal case in which all cluster baryons are in the ICM, the ICM
strictly follows the distribution of dark matter, and clusters contain
exactly the cosmic mix of baryonic and non-baryonic matter
\citep{2004ApJ...601..610V}. The total mass in this case is given simply by
\begin{equation}\label{eq:Mgas:Mtot:1}
  \Mtot = f_g^{-1} \Mgas,
\end{equation}
where $\Mgas$ is provided by the X-ray data, and $f_g$, to the first
approximation, equals $\Omega_b/\Omega_M$, the ratio which is accurately
given by the CMB measurements. To estimate the mass corresponding to a
given critical overdensity, we need to solve
\begin{equation}\label{eq:Mgas:Mtot:2}
  \frac{\Mgas(r)\,f_g^{-1}}{4/3\,\pi\,r^3\,\rho_c(z)} = \Delta
\end{equation}
for $r$ to find the corresponding overdensity radius,
$r_\Delta$. Equation~(\ref{eq:Mgas:Mtot:1}) with $\Mgas$
evaluated at $r_\Delta$ is then used to find $\Mtot$.

\subsubsection{Corrections for Non-Universality of Gas Fraction}
\label{sec:fgas:corr}

\begin{figure}[t]
\centerline{\includegraphics[width=0.97\linewidth]{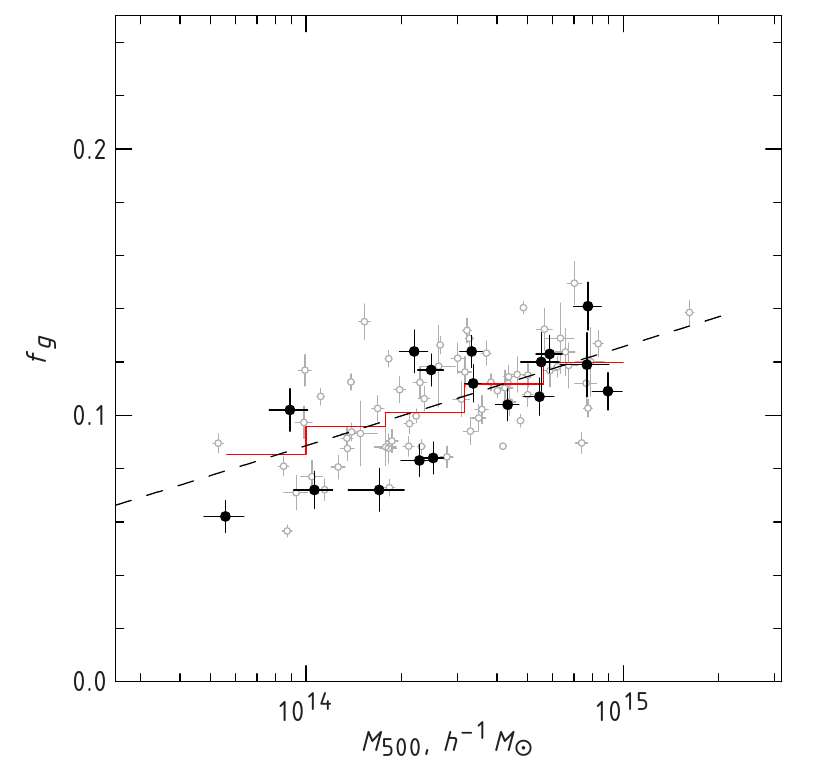}}
\caption{Trend of $f_g$ within $r=r_{500}$ with cluster mass derived from
  X-ray observations. The solid black circles show the results from direct
  hydrostatic mass measurements (\citetalias{2006ApJ...640..691V} with 7
  additional clusters, see \S\,\ref{sec:M:estimates}). Grey circles show
  approximate estimates using the $\Mtot-\TX$ correlation (see text). The
  scatter is consistent with being purely due to mass measurement
  uncertainties, either from hydrostatic estimates
  \citep{2007ApJ...655...98N} or from $\Mtot-\TX$ correlation
  \citepalias{2006ApJ...650..128K}. The error bars indicate only the formal
  measurement uncertainties.}
\label{fig:fgas:trend}
\end{figure}

In reality, the $\Mgas$-based estimate is more complicated because the
observed gas fraction in clusters is significantly lower than the cosmic
average
\citep[e.g.,][]{2003MNRAS.344L..13E,2004MNRAS.353..457A,2006ApJ...652..917L,2007MNRAS.378..293A}
and moreover, there are trends of observed $f_g$ with the cluster mass
(e.g., \citeauthor{1999ApJ...517..627M}~\citeyear{1999ApJ...517..627M};
\citetalias{2006ApJ...640..691V},
\citeauthor{2006A&A...456...55Z}~\citeyear{2006A&A...456...55Z}). This
trend can be related to the baryon cooling and galaxy formation
\citep{2005ApJ...625..588K}, energy feedback from the central AGNs
\citep{2007ApJ...663..139B}, evaporation of supra-thermal protons
\citep{2007JCAP...03..001L} etc.\ --- processes whose theoretical
modeling is highly uncertain at present.  The best approach is therefore
to derive the trend $f_g(M)$ observationally. Once this is done, it can
be straightforwardly taken into account in eq.[\ref{eq:Mgas:Mtot:2}] ---
we just need to use $f_g(4/3\,\pi\,r^3\,\rho_c)$ instead of a
constant.\footnote{We assume that the cluster mass is the only parameter
  controlling systematic trends in $f_g$. This assumption is consistent
  with current observations (see caption to Fig.\ref{fig:fgas:trend}).
  If there are additional parameters, their role would be to introduce
  systematic scatter in the $\Mgas/\Mtot$ ratio for fixed $\Mtot$. The
  observed scatter is consistent with the value we adapt based on the
  simulations. If the scatter can be related to easily measured X-ray
  observables, it would be possible to improve the quality of the
  $\Mgas$ proxy still futher.}

The main problem is that direct X-ray hydrostatic $\Mtot$ measurements
near $\R500$ are feasible only in a small number of clusters,
insufficient to establish the functional form of the $f_g(M)$ trend. We
can, however, follow the approach used in \citet{1999ApJ...517..627M}
--- the total mass (and hence, $\R500$) can be estimated from the
average temperature (see \S\,\ref{sec:M-T} above), and then the gas mass
determined from the X-ray image within that radius. Such estimates of
$f_g$ have substantial uncertainties because of the scatter in the
$\Mtot-\TX$ relation, but this method can be applied virtually to any
cluster. The results for our low-$z$ sample are shown by grey points in
Fig.\,\ref{fig:fgas:trend}. The histogram shows the averages of these
crude estimates in several mass intervals.  Clearly, the data suggest an
approximately linear trend of $f_g$ with $\log M$. The $f_g$ values
obtained from hydrostatic mass measurements closely follow the same
trend (solid black points in Fig.\,\ref{fig:fgas:trend}).  These, more
accurate, values are used to determine the normalization and slope of
the $f_g(M)$ trend,
\begin{equation}\label{eq:fg:M:trend}
  f_g\; (h/0.72)^{1.5} = 0.125 + 0.037\,\log M_{15},
\end{equation}
where $M_{15}$ is the cluster total mass, $\M500$, in units of
$10^{15}\,h^{-1}\,M_\odot$. Extrapolation of this trend to lower masses
described the observed $f_g$ for galaxy groups \citep{msun08}. The
uncertainties of the coefficients are such that $f_g$ is determined to
$4-5\%$ across the useful mass range, $10^{14}-10^{15}\,h^{-1}\,\Msun$,
resulting in the same systematic uncertainty in the $\Mtot$ estimates
because of the $f_g(M)$ trend. The systematic uncertainties are, however,
dominated by those of the hydrostatic mass estimates (discussed in
\S\,\ref{sec:M-Tx:cal} and \ref{sec:Xray:vs:WL}) and so the overall
calibration of the absolute mass scale with the $\Mgas$ method is the same
as that in the $\Mtot-\TX$ or $M-\YX$ relations.

The observed $f_g$ within for the highest-mass clusters is $\sim 25\%$
lower than the cosmic baryon fraction, $\Omega_b/\Omega_M=0.165\pm0.005$
\citep{2008arXiv0803.0547K}. Partly, the remaing baryons can be in the
form of stars. The observed star-to-gas ratios for massive clusters are
in the range of 0.05--0.1 \citep{2007ApJ...666..147G} but the stellar
masses are derived from population synthesis models and thus can
uncertain by factors of order 2. The tension is reduced still futher if
the Hubble constant value is lower than we assume. For example, for
$h=0.685$ \citep[the lower $1\sigma$ bound for the combined constraints
in]{2008arXiv0803.0547K}, the X-ray derived $f_g$ values are 8\% higher
than we quote in eq.(\ref{eq:fg:M:trend}).

\begin{figure}[t]
\centerline{\includegraphics[width=0.97\linewidth]{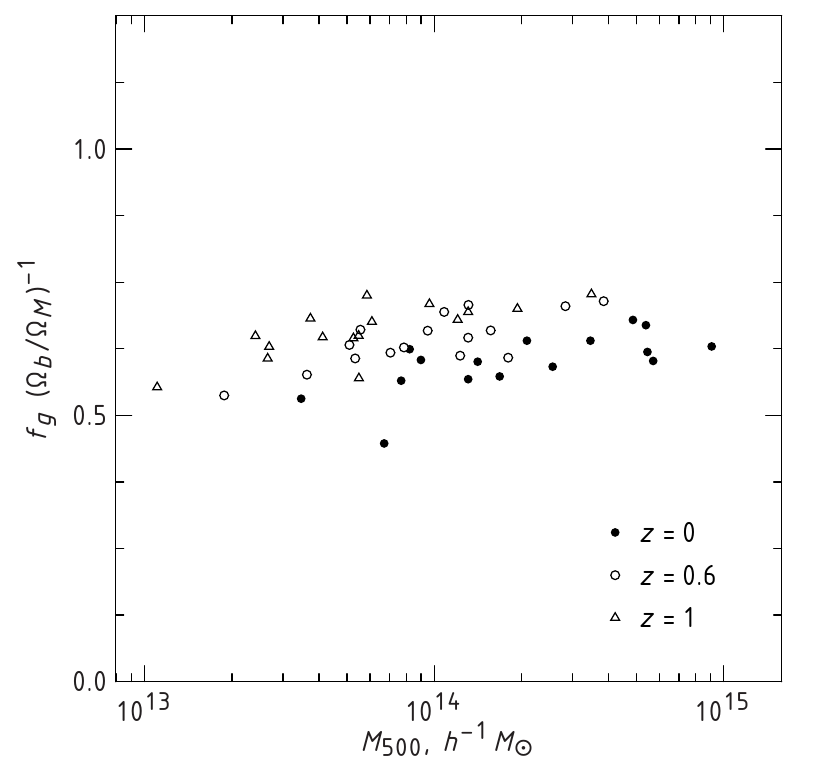}}
\smallskip
\centerline{\includegraphics[width=0.97\linewidth]{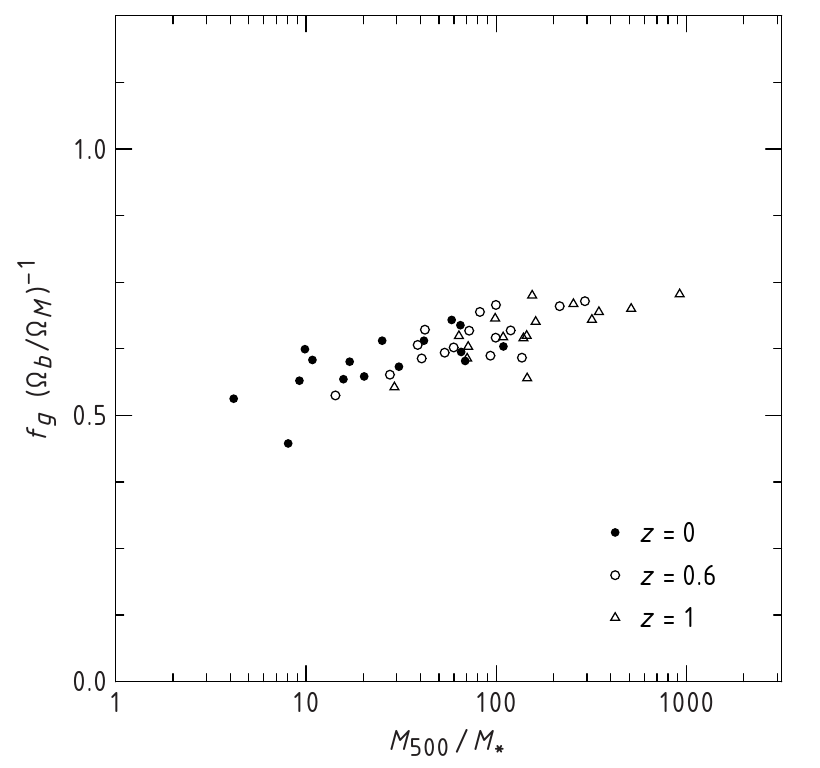}}
\caption{The dependence of $f_g$ within $\R500$ on the cluster mass
  observed in high-resolution cosmological simulations with cooling,
  star formation, and feedback
  \citep{2005ApJ...625..588K,2007ApJ...655...98N}. The simulated
  clusters show qualitatively the same trend as that observed at $z=0$
  (Fig.\,\ref{fig:fgas:trend}),  but there is a clear evolution of $f_g$
  for the given $M$. The $z$-dependence is almost completely removed if
  we scale the cluster masses by the characteristic non-linear mass
  scale, $M_*$ (lower panel).}
\label{fig:fgas:trend:sim}
\end{figure}

\subsubsection{Evolutionary Corrections}
\label{sec:fgas:corr:evol}

Unfortunately, we cannot observationally establish the $f_g(M)$ trend for
high-$z$ clusters independent of the underlying cosmology.  Therefore, we
have to rely on the theoretical models that explain the observed trend at
$z=0$ and can predict its evolution at least for the cosmologies close to
the ``concordance'' \LCDM. Since, unfortunately, no completely satisfactory
model currently exists, this step is a major source of systematic
uncertainties.

One such model can be based on the numerical simulations presented in
\citet{2005ApJ...625..588K}. The simulated clusters show the trend in
$f_g(M)$ which is very close to that observed at $z=0$, both in terms of
slope and magnitude of the deviation from the global baryon fraction,
$\Omega_b/\Omega_m$ (Fig.\,\ref{fig:fgas:trend:sim}\,\emph{a}). The
``missing'' baryon component in the \citeauthor{2005ApJ...625..588K}\
clusters is converted into stellar material, so that $(M_{\rm
  stars}+\Mgas)/\Mtot$ is within 10\% of the value $\Omega_b/\Omega_m$
specified in the simulation. Observational support for this model is
provided by the recent work of \citet{2007ApJ...666..147G} who show that the
trend in stellar mass fraction, $f_{\rm stars}(M)$, is such that it roughly
compensates for decreasing $f_g$ in low-mass clusters, making $f_{\rm
  stars}+f_g$ nearly constant at $\M500\gtrsim10^{14}\,h^{-1}\Msun$ clusters,
although not quite bringing it to the WMAP value of $\Omega_b/\Omega_M$
(see discussion at the end of \S\,\ref{sec:fgas:corr}).

The trend in the \citeauthor{2005ApJ...625..588K}\ simulations show a
clear dependence on the redshift
(Fig.\,\ref{fig:fgas:trend:sim}\,\emph{a}) in the sense that a given
value of $f_g$ corresponds to a systematically decreasing $M_{500}$,
although at each $z$, $f_g(M)$ seems to follow lines with the same
slope.  Empirically, we find that the dependence on the redshift is
almost completely removed (Fig.\,\ref{fig:fgas:trend:sim}\,\emph{b}), if
we scale the cluster masses by $M_*$, the mass scale corresponding to a
linear fluctuations amplitude of 1.686: $\sigma(M_*)=1.686$, where
$\sigma(M)$ is the \emph{rms} fluctuation of density field smoothed with
a top hat filter containing mass $M$. In other words, the simulations
indicate that $f_g(M/M_*)$ is almost independent of redshift, at least
at $z\le1$. A simple explanation of such a scaling can be related to the
mass distribution of the cluster progenitors at high
redshifts\footnote{Most of the stellar mass in the red and old galaxies
  of observed clusters, which contain the bulk of the cluster stellar
  mass, and in cosmological simulations is in place by $z\sim 1-2$, so
  the stellar fraction within each lower-redshift cluster is effectively
  ``pre-set'' at high redshifts.  Efficiency of star formation within
  each galaxy-sized dark matter halo depends on the halo mass.
  Therefore, the stellar fraction within clusters is probably defined by
  the mass function of its progenitors near the redshift of the peak
  star formation ($z=2-3$).  Indeed, the calculations of the progenitor
  mass functions using extended Press-Schechter theory
  \citep{1993MNRAS.262..627L} show that they are much more similar for
  clusters with the same $M_{500}/M_*$ than for those with the same
  $M_{500}$ at different $z$.}. We also note that qualitatively similar
scaling ($f_g$ for a fixed mass increases at high $z$) is expected if
the gas distribution in a cluster potential well does not evolve at all
(e.g., inner regions of a cluster remain in equilibrium and do not
evolve significantly) and gas fraction is constant, but mass $M_{500}$
changes simply due to evolution of the background critical density,
$\rho_c(z)$, to which it is tied by the definition.

In the spirit of our general approach of using the theoretical results in
the cluster mass estimates as minimally as possible, we use the observed
dependence of $f_g$ on mass for low-redshift clusters
(eq.[\ref{eq:fg:M:trend}]), and take a suggestion from simulation that the
same trend should hold at all redshifts, if masses are scaled by $M_*$
computed for the given cosmological model. This gives us $f_g(M,z)$,
necessary to estimate the total cluster mass from the observed $\Mgas$
(eq.[\ref{eq:Mgas:Mtot:2}]). Although this adopted $z$-dependence is
motivated only qualitatively, the predicted overall correction is small. For
example, for the cosmological model with $\Omega_M\approx0.28$,
$\sigma_8\approx0.78$ (close to the best-fit to our cluster data),
$M_*=3.6\times10^{12}\,\Msun$ and $8.1\times10^{11}\,\Msun$ at $z=0.05$ and
0.5, respectively. The median masses of clusters in our sample are
$4.8\times10^{14}\,\Msun$ at $z=0.05$ and $2.3\times10^{14}\,\Msun$ (see
below). The ratio $\M500/M_*$, therefore, varies from $\sim 130$ to $\sim
280$, corresponding to a predicted change in $f_g$ for the median mass
clusters of 11\% (eq.[\ref{eq:fg:M:trend}]). A reasonable estimate for the
systematic error is around 50\% of this overall correction, or 5--6\% in
terms of mass between redshifts of 0 and 0.5.

\subsubsection{Summary for $\Mtot-\Mgas$ Relation}

To summarize, our approach to the $\Mgas$-based estimates of the total
cluster mass is based on using eq.[\ref{eq:Mgas:Mtot:2}] to find
$\R500$, and hence $\M500$, for each cluster. In this equation,
$\Mgas(r)$ is the observed gas mass profile derived from the X-ray
image, and $f_g$ is the estimated gas fraction as a function of mass and
redshift\footnote{Formally, we can write the gas fraction to be a
  function of radius, as
  $f_g=f_g(M/M_*)=f_g(500\,\rho_c(z)\,4/3\,\pi\,r^3/M_*(z))$, and
  then~[\ref{eq:Mgas:Mtot:2}] becomes an implicit equation, which can be
  solved for $\R500$ numerically.}. The
dependence $f_g(M)$ is determined empirically at $z\approx0$
(\S\,\ref{sec:fgas:corr}, eq.[\ref{eq:fg:M:trend}]). It is assumed that
this trend evolves with redshift such that $f_g$ remains constant for
clusters with a fixed $M/M_*$ (this is justified in
\S\,\ref{sec:fgas:corr:evol}).

The systematic uncertainties of this $\Mtot$ estimate are dominated by those
of $f_g$. The latter can be factorized into two components, the
uncertainties of the empirical measurements at $z\approx0$, and the
uncertainties of the assumed evolution with redshift. The low-redshift
uncertainties are essentially those of the X-ray total mass estimates,
discussed above in connection with the $\Mtot-\TX$ relation. More important
for cosmological constraints is the redshift-dependent uncertainty. Within
our redshift range, it can be estimated as 5--6\%
(\S\,\ref{sec:fgas:corr:evol}).

The object-to-object scatter in the $\Mgas$-based total mass estimates can
be easily derived from the analysis of mock X-ray data for simulated
clusters. This was done in \citet{2006ApJ...650..128K} and
\citet{2007ApJ...655...98N}, who find that the scatter in the $\Mtot-\Mgas$
relation is approximately 11\% in $\Mtot$ for a given $\Mgas$. Most of this
scatter results from the X-ray analysis, as intrinsic scatter of the gas
mass for a fixed total mass in simulated clusters is $<5\%$.

\subsection{$\Mtot-\YX$ Relation}
\label{sec:YX}

The final $\Mtot$ proxy we use is the most robust X-ray mass estimator proposed
by \citetalias{2006ApJ...650..128K}. The quantity, $\YX$, is defined as
\begin{equation}
  \label{eq:YX:def}
  \YX = T_X \times M_{{\rm gas},X},
\end{equation}
where $T_X$ is the temperature derived from fitting the cluster X-ray
spectrum integrated within the projected radii
\mbox{$0.15\,\R500-1\,\R500$}, and $M_{{\rm gas},X}$ is the hot gas
mass within the sphere $\R500$, \emph{derived from the X-ray image}.

The quantity that $Y_X$ approximates is the total thermal energy of
the ICM within $\R500$, and also the integrated low-frequency
Sunyaev-Zeldovich flux \citep{1972CoASP...4..173S}. The total thermal
energy, $Y$, was found in the simulations to be a very good indicator
of the total cluster mass
\citep{2004MNRAS.348.1401D,2005ApJ...623L..63M,2006ApJ...648..852H,2006ApJ...650..538N}.
In the simplest self-similar model
\citep{1986MNRAS.222..323K,1991ApJ...383..104K}, $Y$ scales with the
cluster mass as
\begin{equation}\label{eq:Y:evol}
  \Mtot \propto Y^{3/5}\,E(z)^{-2/5}
\end{equation}
(e.g., \citetalias{2006ApJ...650..128K}). This scaling is a
consequence of the expected evolution in the $\Mtot-T$ relation
(eq.[\ref{eq:M-T}]) and the assumption of the self-similar model that
$f_g$ is independent of cluster mass. Hydrodynamic simulations show
that the expected scaling~[\ref{eq:Y:evol}] is indeed valid, and
moreover, the relation shows a smaller scatter in $M$ for fixed $Y$
than, e.g., the $M-\TX$ relation. The primary reason is that the total
thermal energy of the ICM is not strongly disturbed by cluster mergers
\citep{2007astro.ph..1586P}, unlike $\TX$ or X-ray luminosity
\citep{2001ApJ...561..621R}.

It is reassuring that the $\Mtot-Y$ scaling also appears to be not very
sensitive to the effects of gas cooling, star formation, and energy feedback
\citep{2006ApJ...650..538N} --- these effects do not affect the power slope
or the evolution law, although change somewhat the overall normalization.
The stability of $Y$ is primarily explained by the fact that gas cooling
tends to remove from the ICM the lowest-entropy gas
\citep{2001Natur.414..425V}, increasing the average temperature of the
remaining gas and thus affecting $\TX$ and $\Mgas$ in opposite ways. Direct
hydrodynamic simulations of \citet{2007ApJ...668....1N} confirm this
expectation.

\begin{figure}[t]
\centerline{\includegraphics[width=0.97\linewidth]{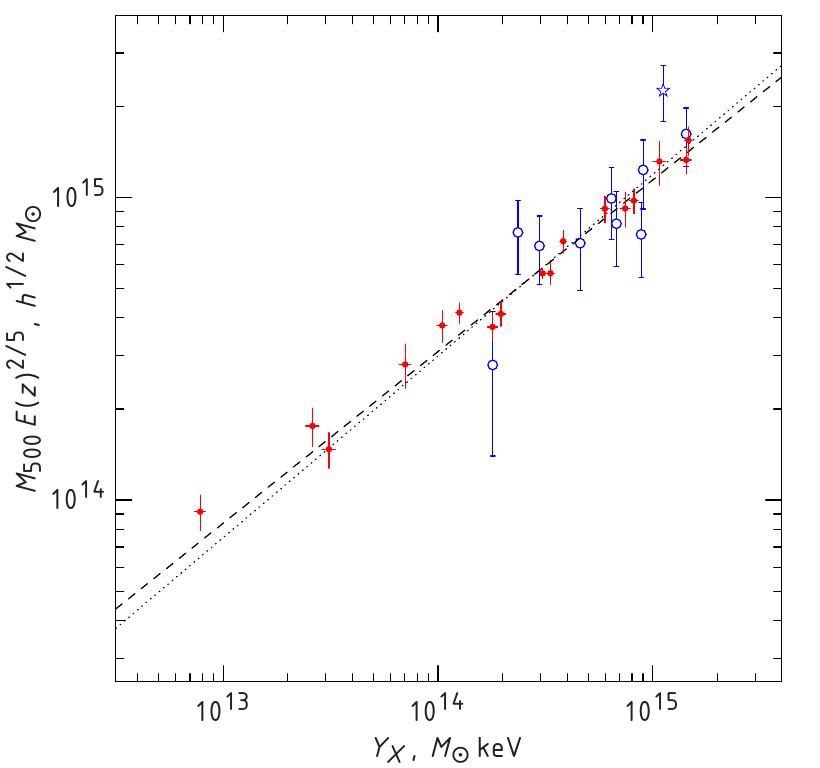}}
\caption{Calibration of the $\Mtot-\YX$ relation. Points with errorbars
  show \emph{Chandra} results from \citet{2006ApJ...640..691V} with 7
  additional clusters (\S\,\ref{sec:M:estimates}). Dashed line shows a
  power law fit (excluding the lowest-mass cluster) with the free slope.
  Dotted line shows the fit with the slope fixed at the self-similar
  value, $3/5$ (parameters for both cases are given in
  Table~\ref{tab:relations:cal}). Open points show weak lensing
  measurements from \citet{2007MNRAS.379..317H} (these data are not used
  in the fit); the strongest outlier is A1689 (open star), a known case
  of large scale structures superposed along the line of sight.  }
\label{fig:M-Y:cal}
\end{figure}

As discussed in \citetalias{2006ApJ...650..128K}, the X-ray proxy, $\YX$, is
potentially even more stable with respect to cluster mergers than the
``true'' $Y$. In the post-merger state, for example, the temperature and
thus $Y$ is biased somewhat low because of incomplete dissipation of bulk
ICM motions. The same bulk motions, however, cause the gas density
fluctuations, which leads to an overestimation of $\Mgas$ from the X-ray
analysis \citep{1999ApJ...520L..21M}. Therefore, the merger-induced
deviations of the average temperature and derived $\Mgas$ are
anti-correlated and hence partially canceled out in $\YX$. Even the
strongest mergers in the simulated cluster sample used in
\citetalias{2006ApJ...650..128K} do not lead to large deviations of $\YX$
from the mean scaling. There is also no detectable systematic offset in the
normalization of the $\Mtot-\YX$ relations for relaxed and unrelaxed
clusters. The upper limit for the difference in $\Mtot$ for fixed $\YX$
within the \citetalias{2006ApJ...650..128K} simulated sample is 4\% (see
their Table~2).

Since $\YX$ is so insensitive to the cluster dynamical state, it is
straightforward to calibrate the $\Mtot-\YX$ relation using the sample of
\emph{Chandra} clusters from \citetalias{2006ApJ...640..691V}, and then it
is reasonable to assume that the same relation is also valid for unrelaxed
clusters. The observed $\Mtot-\YX$ relation does follow very closely the
expected self-similar scaling of eq.~\ref{eq:Y:evol}
\citep[Fig.\,\ref{fig:M-Y:cal}; see also][]{2007A&A...474L..37A}. The
best-fit power law is
\begin{equation}
  M\,E(z)^{2/5} \propto \YX^{0.53\pm0.04}
\end{equation}
when all clusters are included. The marginal deviation of the slope from
a self-similar value of $3/5$ is driven primarily by the
lowest-temperature cluster (MKW4), for which both the total mass and
$\YX$ measurements are most uncertain. Excluding this cluster (its
$\Mtot$ is in any case smaller than the lower mass threshold in the
cluster mass functions in our samples), the power law fit becomes
\begin{equation}
  M\,E(z)^{2/5} \propto \YX^{0.57\pm0.05},
\end{equation}
fully consistent with the self-similar relation (shown by a dashed line in
Fig.\,\ref{fig:M-Y:cal}). We use the latter fit for the $\YX$-based cluster
mass estimates. Note that \cite{msun08} find a slope of 0.57 when they fit
jointly their galaxy group sample with the \citetalias{2006ApJ...640..691V}
clusters, supporting the notion that the MKW4 measurement can be ignored.
The normalization constant is provided in Table~\ref{tab:relations:cal}
\citep[it is consistent with the \emph{XMM-Newton} results
of][]{2007A&A...474L..37A}. Note that the $h$-dependence of the
normalization constant in the $\Mtot-\YX$ relation is $\propto h^{1/2}$,
different from the usual $h^{-1}$ in, e.g., the $\Mtot-\TX$ relation. This
is the consequence of the $h$-dependence of the X-ray $\Mtot$ and $\Mgas$
estimates, see \citetalias{2006ApJ...650..128K} for details.

The overall uncertainties of the calibration of the $\Mtot-\YX$ are
identical to those for the $\Mtot-\TX$ relation (see~\S\,\ref{sec:M-T}),
with the exception that we do not expect an additional source of
uncertainty related to the transfer of calibration from relaxed clusters
to the entire population. As for the $\Mtot-\TX$ relation, we also have
to rely on the simulations for an estimate of redshift-dependent
departures from the expected self-similar scaling. The results of
\citetalias{2006ApJ...650..128K} provide an upper limit of $<5\%$ for
the evolution of the amplitude of the relation at a fixed $\YX$ between
$z=0$ and $0.6$. The expected level of scatter in the $\Mtot-\YX$
relation (7\% in $\Mtot$, see \citetalias{2006ApJ...650..128K}) is below
the uncertainties of \emph{Chandra} hydrostatic mass estimates for
individual clusters.  Indeed, the intrinsic scatter is undetectable in
the data. Since the scatter is expected to be small, its exact value is
unimportant for modeling the mass function, and thus can be safely
adopted from the simulations.

\subsubsection{Systematic Error of \emph{Chandra} Mass Measurements}
\label{sec:Xray:vs:WL}

Using the $\Mtot-\YX$ relation, we can address the question of absolute
calibration of the \emph{Chandra} mass estimates through comparison with
recent weak lensing mass measurements in representative samples of
clusters.  Weak lensing measurements of $M_{500}$ in individual objects
still have $\sim 30\%$ uncertainties, and are expected to have a similar
intrinsic uncertainty due to projection of structures along the line of
sight \citep{2001ApJ...547..560M}. However, as the current weak lensing
samples start to include more than 10 objects, the average normalization
of $\Mtot$ vs.\ proxy relations can be measured to better than 10\%.
The two useful recent studies are those of \citet{2007MNRAS.379..317H}
and \citet{2008A&A...482..451Z}. In Fig.\,\ref{fig:M-Y:cal}, we compare
the \emph{Chandra} $\Mtot-\YX$ relation with that for low-$z$ clusters
in the \citet{2007MNRAS.379..317H} sample. The $\YX$ for all
\citeauthor{2007MNRAS.379..317H} clusters in this plot were derived from
\emph{Chandra} data using the procedure applied to our cosmological
samples.  With the exception of a single outlier \citep[A1689, a known
case of of large scale structures superposed along the line of sight,
e.g.,][]{2006MNRAS.366L..26L}, the weak lensing masses for given $\YX$
are in good agreement with the \emph{Chandra} values. Fitting the ratio
of normalizations of the $\Mtot-\YX$ relations obtained from bisector
fits to the two dataset with the slope fixed at 0.57, we find $M^{({\rm
    wl})}/M^{(\it Chandra)} = 1.01\pm0.11$. A similar agreement is found
for the weak lensing masses in \citet{2008A&A...482..451Z}. The
normalizations relevant for our case are presented in their Table~3.
After correcting their $\YX$ by +7\% to compensate for a systematic
difference currently observed between \emph{Chandra} vs.\
\emph{XMM-Newton} temperatures\footnote{Here, we are interested only in
  bringing all the measurements to the \emph{Chandra} temperature scale
  because we use \emph{Chandra} data. It may well be that the \emph{XMM}
  temperatures are in fact correct. The temperature calibration
  uncertainties should be treated as an additional source of systematic
  errors (see \S\,\ref{sec:error:budget:X-ray:cal} for more details).
  Fortunately, the estimated uncertainties are within the bounds
  suggested by comparison of the \emph{Chandra} and weak lensing mass
  measurements.  }, we find $M^{({\rm wl})}/M^{(\it Chandra)} =
1.05\pm0.07$.  The weighted average for the two samples is $M^{({\rm
    wl})}/M^{(\it Chandra)} = 1.04\pm0.06$.  The integrated probability
within the $M^{({\rm wl})}/M^{(\it Chandra)} =[0.91-1.09]$ interval is
0.7, thus $\pm9\%$ is a good estimate for $1\sigma$ systematic
uncertainties in the \emph{Chandra} cluster mass scale calibration.

\subsubsection{Application of the $\Mtot-\YX$ Relation for Real Data}

In application of the $\YX$-based mass estimates to the real data, we face a
practical problem that $\YX$ should be determined within $\R500$, which is
itself unknown. Moreover, $\YX(r)$ diverges at $r\rightarrow\infty$,
although less quickly than $\Mgas(r)$. The total mass should thus be
estimated with the approach similar to the $\Mgas$-based method
(eq.[\ref{eq:Mgas:Mtot:2}]) --- we find $\R500$ and hence $\M500$ by solving
the following implicit equation
\begin{equation}\label{eq:m-yx:reallife}
  C\,(\TX\,\Mgas(r))^\alpha\, E(z)^{-2/5} = 500\times 4/3\,\pi\, r^3\,
  \rho_c(z), 
\end{equation}
where $C$ and $\alpha$ are the parameters of the power law approximation
to the $\Mtot-\YX$ relation, $\Mtot = C\,\YX^\alpha\,E(z)^{-2/5}$.

\section{Survey Volumes}
\label{sec:volume}

We now need to turn to the next critical component of the cluster mass
function derivation --- determination of the effective survey volume.
Our cluster samples are derived from essentially purely X-ray flux
limited surveys. We can then straightforwardly compute the sample
volumes as a function of X-ray luminosity,
\begin{equation}\label{eq:V(L)}
  V(L) = \int_{z_1}^{z_2} A(f_x,z)\,\frac{dV}{dz}\,dz,
\end{equation}
where $f$ is the X-ray flux corresponding to the object with luminosity
$\LX$ at redshift $z$, $dV/dz$ is the cosmological volume-redshift relation,
and $A(f_x,z)$ is the effective survey area for such objects. A relation
between cluster luminosity and flux,
\begin{equation}
  f = \frac{L}{4\pi\, d_L(z)^2}\, K(z),
\end{equation}
depends on the cosmological background through the bolometric distance
$d_L(z)$ and the $K$-correction factor \citep[see, e.g.,][specifically for
the case of the cluster X-ray spectra]{1998ApJ...495..100J}.  The
$K$-correction depends on the assumed cluster temperature but this
dependence is \emph{very} weak if both fluxes and luminosities are measured
in the soft energy band (0.5--2~keV as we use here). In practice, a
sufficient level of accuracy is achieved by estimating $T$ from the
non-evolving $L_X-T$ relation accurately measured for low-$z$ clusters
\citep{1998ApJ...504...27M,1998PASJ...50..187F}.

Because the objects in our low-redshift sample are all well above the
\emph{RASS} detection threshold, their survey area, $A(f,z)$, is adequately
approximated by a constant value, 8.14~sr, equal to the geometric area of
the sky regions covered \citep[see~\S\,\ref{sec:samle:low-z}
and][]{2002ApJ...567..716R}.  The situation is more complex for our high-$z$
clusters drawn from the \400d{} survey. Sky coverage there is a function of
flux because our distant clusters are generally not much brighter than the
detection thresholds in individual \emph{ROSAT} pointings and because the
detection thresholds also vary widely depending on the exposure time of each
pointing. Formally, the sky coverage is also a function of redshift because
detection efficiency is somewhat sensitive to the cluster angular size.
A~detailed discussion of these effects in application to the 400d survey, as
well as a careful calibration of $A(f,z)$ for the full 400d sample was
presented in \citetalias{2007ApJS..172..561B} (see their \S\,7).

An additional complication arises because we use only a brighter
subsample of the \400d{} sample at $0.35<z<0.473$ (see
\S\,\ref{sec:sample:high-z} and Fig.\,\ref{fig:fmin:ch}). We need,
therefore, to recompute $A(f,z)$ using eq.[2--3] from
\citetalias{2007ApJS..172..561B} with $f_{\rm min}$ in their eq.[2] set
to the actual selection fluxes used in our subsample. This is a
straightforward calculation but the results cannot be conveniently
presented in a paper. We provide machine-readable tables for $A(f,z)$ at
the \400d{} survey WWW
site\footnote{http://hea-www.harvard.edu/400d/CCCP}.

Stability of the \400d\ survey area calculations was extensively
discussed in \citetalias{2007ApJS..172..561B}. The general conclusion is
that the uncertainties in $A(f,z)$ do not exceed 3\%, and therefore they
make a negligible contribution to our overall error budget. A dominant
source of uncertainty in determining the volume as a function of mass is
the details of the $\LX-M$ relation.

\subsection{$\LX-\Mtot$ relations}
\label{sec:lx-mtot}

To fit mass function models to the data, we need to know the survey volume
as a function of mass, not luminosity. The two are trivially related if
there is a well-defined relation between the cluster mass and its
luminosity:
\begin{equation}\label{eq:dV(M)/dz}
  \frac{dV(M)}{dz} = \int_L \frac{dV(L)}{dz}\, P(L|M,z)\, dL,
\end{equation}
where $P(\LX|M,z)$ is the probability for a cluster with mass $M$ to have a
luminosity $\LX$ at redshift $z$. The volume in the given redshift interval
is obtained by integrating this equation,
\begin{equation}\label{eq:V(M)}
  V(M) = \int_{z_1}^{z_2} dz \,\int_L A(f_x,z)\,\frac{dV}{dz}\, P(L|M,z)\, dL,
\end{equation}
where $dV/dz$ is the cosmological volume-redshift relation and
$A(f_x,z)$ is the survey area coverage (cf.\ eq.\ref{eq:V(L)}).

The simplest model that seems to adequately describe the observed
$\Mtot-\LX$ relations can be represented as a power law with approximately
log-normal intrinsic scatter around the mean which is independent of mass
and redshift, and the redshift evolution that changes the normalization but
keeps constant the slope of the power law,
\begin{equation}\label{eq:log:normal:a}
  P(\ln L| M) \propto \exp\left(-\frac{(\ln L - \ln
      L_0)^2}{2\,\sigma^2}\right), 
\end{equation}
where 
\begin{equation}\label{eq:L-M:1}
  L_0 = A(z)\, M^{\alpha}.
\end{equation}
The evolution factor is sometimes approximated as a power law of $(1+z)$
(the simplest model) and sometimes as a power law of $E(z)$
\citep[self-similar evolution inspired models, see
e.g.,][]{1998ApJ...495...80B}:
\begin{equation}\label{eq:L-M:2}
  A(z) = A_0\,(1+z)^\gamma \quad {\rm or} \quad A(z) = A_0\,E(z)^\gamma
\end{equation}
A recent study by \citet[consistent with our results
below]{2007ApJ...668..772M} indicated that the evolution factor,
$E(z)^\gamma$, is in fact close to that expected in the self-similar
model for the ``concordant'' cosmological model.
However, the general consensus has been
\citep[e.g.,][]{2001ApJ...561...13B} that we should not rely on the
simplest theory for the evolution in the $\LX-\Mtot$ relation and
instead should determine it empirically for each background cosmology.
We take this approach in the present study.
\begin{figure*}
\centerline{%
\includegraphics[width=0.99\columnwidth]{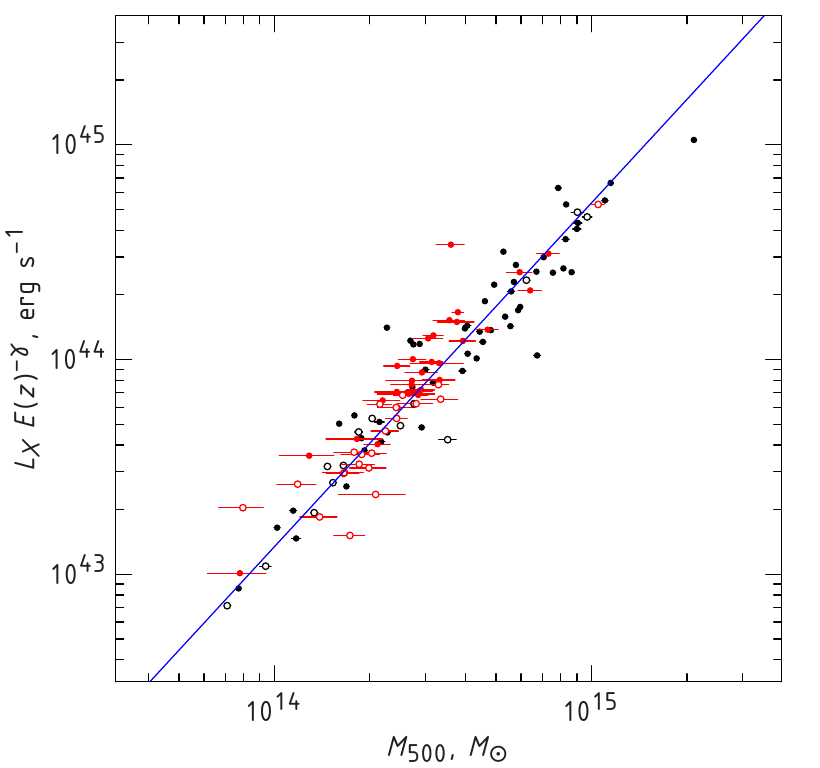}~~%
\includegraphics[width=0.99\columnwidth]{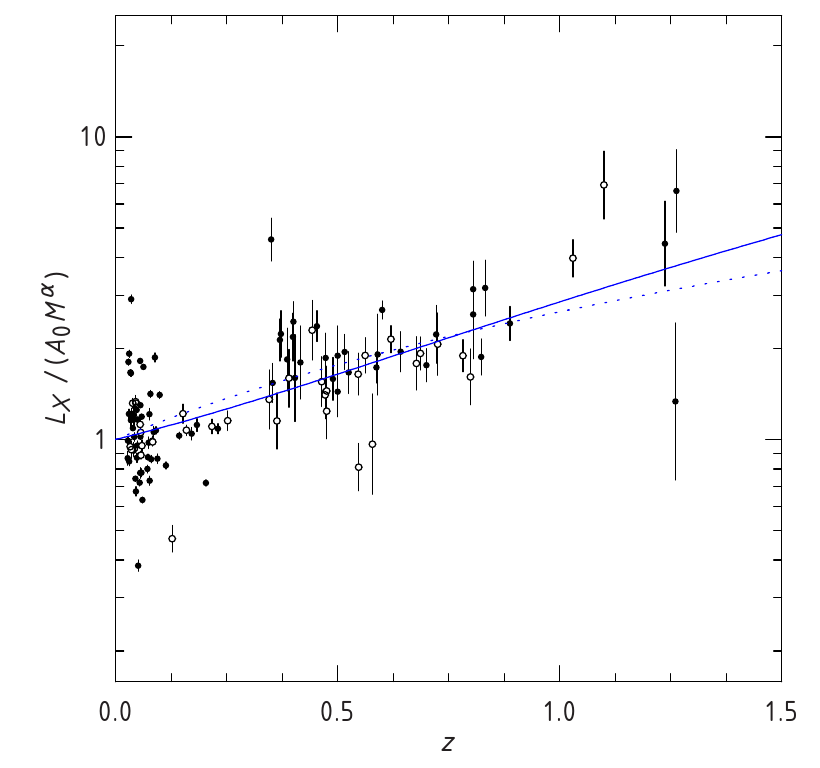}%
}
\caption{Results for the mass-luminosity relation with $\Mtot$ estimated
  from $\YX$. \emph{Left} panel shows the correlation for low-redshift
  clusters (black points) with the best-fit power law relation. The red
  points show the data for high-$z$ clusters with the luminosities
  corrected for the evolution [$E(z)^\gamma$]. All luminosities are
  corrected for the expected Malmqiust bias (see
  Appendix~\ref{sec:malmquist:corr:evol}).  \emph{Right:} Evolution in
  the normalization of the $\LX-M$ relation. Individual measurements
  have been corrected for Malmquist bias and divided by the best-fit
  low-$z$ relation. Solid and dotted lines show the best fit in the form
  $E(z)^\gamma$ and $(1+z)^\gamma$, respectively. In both panels, the
  clusters with large correction ($\Delta \ln L>0.5$) are shown with
  open symbols. The lack of a systematic offset between clusters with
  the estimated strong and weak Malmquist bias proves that the
  correction has been applied correctly. The $z>1$ clusters in this
  panel are from the RDCS survey \citep{2003ApJ...593..705T}; they were
  not used in the fit and are shown only to demonstrate that the
  extrapolation of our best-fit $E(z)^\gamma$ evolution to higher
  redshifts still produces reasonable results.}
\label{fig:L-M}
\end{figure*}

\begin{figure}
\centerline{\includegraphics[width=0.97\linewidth]{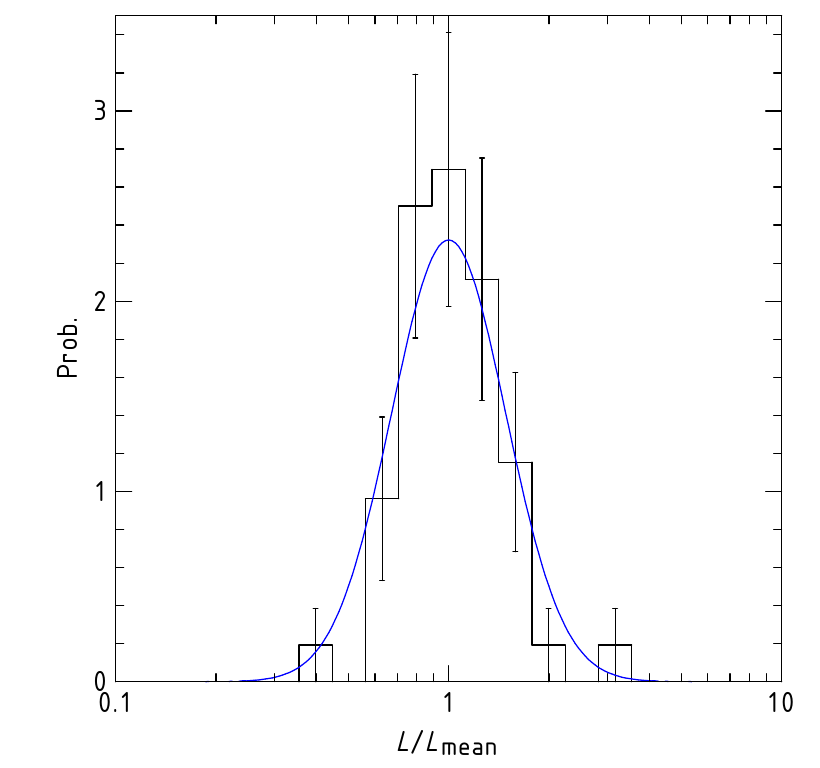}}
\caption{Distribution of the deviations from the mean $\LX-M$ relation
  for the low-$z$ sample (where the contribution of measurement
  uncertainties is negligible). Solid line shows the best-fit log-normal
  distribution with the scatter $\sigma_{\ln L}=0.396$.}
\label{fig:L-M:scatter}
\end{figure}

\subsubsection{Fitting Procedure and Treatment of Malmquist Bias}

In the model specified above, the $\LX-M$ relation is characterized by
four parameters, $A_0$, $\alpha$, $\gamma$, and $\sigma$. These
parameters can be determined using mass estimates for clusters from our
sample. The large size of our sample even allows us to test the basic
assumptions of the model, e.g., that the scatter is log-normal. A major
complication, however, is Malmquist bias.  In a flux limited sample, the
average luminosity of selected clusters is higher than that in the
parent population. The effect becomes strong if the scatter in $\LX$ for
fixed $M$ is large \citep{2006ApJ...648..956S,2007arXiv0706.2189N}, as
observed. Nord et al.\ address the question in which regimes the bias is
unimportant. Stanek et al.\ describe how to estimate the magnitude of
the Malmquist bias by simulating mock catalogs. Stanek et al.\ also
discuss how to derive the mean $\LX-M$ relation and scatter from the
cluster X-ray luminosity function if the cosmological model (including
$\sigma_8$) is assumed known. A similar approach was used by
\cite{2002A&A...383..773I} who consistently modeled the $L-T$ relation
together with the cosmological fit to the cluster temperature function.
However, as we show in Appendix\,\ref{sec:Malmquist:bias}, estimation of
the Malmquist bias can be separated from the cosmological fit to the
mass function, which leads to simpler algorithms than those used by
Stanek et al.\ and Ikebe et al. An approach similar to ours was
independently developed by \citet{2007MNRAS.382.1289P}.

Our algorithm is fully described in Appendix\,\ref{sec:Malmquist:bias}.
Here, we briefly outline the main results and modeling steps and then
proceed to presenting the results.  A typical situation in low-$z$ surveys
with a high flux limit is that the search volume is a power law function of
the object luminosity (e.g., $V\propto L^{3/2}$ in Euclidean space and no
low-$z$ cutoff), and that the evolution can be neglected within the survey's
effective redshift depth. In this case
(Appendix\,\ref{sec:malmquist:noevol}), the Malmquist bias leads to a
constant offset in the normalization of the observed $\LX-M$ relation; the
observed \emph{rms} scatter in $\ln L$ for fixed $M$ equals the standard
deviation of the log-normal distribution ($\sigma$ in
eq.\,[\ref{eq:log:normal:a}]). The true relation is therefore very simple to
recover for typical low-$z$ surveys. For the observed scatter, $\sigma =
0.39$ (see below), the bias is $\Delta \ln L\approx 3/2\,\sigma^2 = 0.23$
(eq.[\ref{eq:malmquist:noevol:2}]), or $\LX$ for fixed $M$ is overestimated
by $\approx 26\%$ (much smaller than the factor of $\sim 2$ bias advocated
by \citeauthor{2006ApJ...648..956S} \citeyear{2006ApJ...648..956S} but
cosistent with the limit from \citet{2006A&A...453L..39R}. This bias is
independent of the actual flux limit of the low-$z$ survey. If uncorrected
for, it leads to overestimation of the volume for fixed mass by $\approx
40\%$ (because $V(M)\propto L_0^{3/2}$ at low $z$, see
eq.~\ref{eq:V(M)}--\ref{eq:log:normal:a}).

The treatment of the Malmquist bias is more complicated if the evolution in
the $\LX-M$ relation cannot be neglected. However, in this case it is still
possible to derive a likelihood function which can be computed numerically
given the survey selection functions and which implicitly depends on the
parameters of the evolving $\LX-M$ relation, $A_0$, $\alpha$, $\gamma$, and
$\sigma$ (eq.[\ref{eq:malmquist:prob:norm}--\ref{eq:malmquist:chi2}]). One
can also compute the average bias for each cluster given $A_0$, $\alpha$,
$\gamma$, and $\sigma$ (Appendix~\ref{sec:malmquist:corr:evol}); using these
corrections we can easily check if the basic model assumptions (e.g., that
the scatter is log-normal and independent of both $M$ and $z$) are
sufficiently accurate. The tests of our fitting procedure using the mock
catalogs show that it recovers the true parameters of the $\LX-M$ relation
without significant biases.

\subsubsection{Results for $\LX-M_Y$ relation}

We independently derive the $\LX-M$ relation for each of our mass
proxies. In this section, we summarize the results obtained with the
$\YX$ proxy (hence the relation is called $\LX-M_Y$); the results for
the $\TX$ and $\Mgas$ proxies are very similar. The best fit (obtained
with our maximum likelihood method) to the evolving relation in the form
eq.[\ref{eq:L-M:1}--\ref{eq:L-M:2}] is
\begin{equation}\label{eq:L-M:fit}
  \begin{split}
    \ln L_X &= (47.392 \pm 0.085) + (1.61\pm0.14)\ln M_{500}\\
    &+(1.850\pm0.42)\ln E(z)-0.39\,\ln(h/0.72)\\
    & \pm (0.396\pm0.039)
  \end{split}
\end{equation}
where the last term on the right hand side indicates the observed
scatter in $\LX$ for fixed $M$. The uncertainties for each parameter are
obtained from the mock catalog simulations described in
Appendix~\ref{sec:malmquist:fitting}. For the median mass in our sample,
the best-fit normalization agrees very well with that from \citep[their
Table 10, after converting the luminosities to the 0.5--2 keV
band]{2002ApJ...567..716R}, even though we do expect some difference due
to corrections for the Malmquist bias applied in our analysis
(Appendix~\ref{sec:malmquist:noevol}). In this regard we note that our
more elaborate procedure for hydrostatic mass estimates should lead to
sistematicallty different results than a simple isothermal $\beta$-model
analysis used in \cite{2002ApJ...567..716R}; the net effect of updated
\Mtot{} measurements and corrections for the Malmquist bias appears to
lead to very small revisions of the normalization of the $\LX-\Mtot$
relation. 

The left panel in Fig.\,\ref{fig:L-M} shows that indeed, the low-$z$ data
are adequately described by a single power law relation. The high-$z$
clusters also follow the same relation with approximately the same
scatter, after correction for the evolution in the overall normalization
[$E(z)^{1.85}$]. The observed evolution in the normalization (right
panel of Fig.\,\ref{fig:L-M}) is consistent with the $E(z)^\gamma$
scaling, but also with a $(1+z)^\gamma$ law. The exact form of the
evolution law is not crucial for our purposes since we use the $\LX-M$
relation only to estimate the survey coverage at each redshift and not
to estimate the cluster masses. The effect of the choice of the
parametrization on the derived $V(M)$ is discussed below.

The observed deviations from the mean relation at low redshifts
(Fig.\,\ref{fig:L-M:scatter}) are consistent with the log-normal
distribution with a scatter of $\sigma_{\ln L}=0.396$ (or $\approx \pm48\%$)
in $\LX$ for fixed $M$. The contribution of the measurement uncertainties to
this scatter is negligible for low-$z$ objects. The expected scatter in the
$\Mtot$ estimates using $\YX$ is also significantly lower. Therefore, it is
reasonable to expect that the observed scatter is a good representation of
that in the relation between $\LX$ and true mass\footnote{Note that we are
  forced to use the total luminosities, including centers and substructures,
  for reasons given in \S\,\ref{sec:substr}. If these components are
  excluded from the flux measurements, the scatter can easily be made lower,
  see, e.g., \cite{2007ApJ...668..772M}}. The current data quality is
insufficient to characterize the shape of the scatter distribution
precisely. For example, we cannot check if the tails of the distribution are
consistent with the log-normal model. The knowledge of tails in the
$P(\LX|M)$ distribution is crucial if one uses $\LX$ as a proxy for cluster
mass \citep{2005PhRvD..72d3006L}. In our case, however, the $\LX-M$ relation
is used only for the survey volume calculations, where the effects of the
$P(\LX|M)$ are minor (see \S\,\ref{sec:volume:model:unc} below).

The observed $48\%$ scatter in the $\LX-M$ relation implies that
Malmquist bias effects are very significant. For example, in a purely
flux-limited low-$z$ sample, the average bias in the luminosity for
fixed $M$ is $\Delta \ln L=0.235$ or $26\%$ (see
eq.[\ref{eq:malmquist:noevol:2}] in
Appendix~\ref{sec:malmquist:noevol}). This is qualitatively similar to
the conclusions of \citet{2006ApJ...648..956S}, although our predicted
bias is lower because Stanek et al.\ have assumed a larger scatter in
the $\LX-M$ relation than that observed in our data.

\subsubsection{Results for $V(M)$}

\begin{figure}[t]
\centerline{\includegraphics[width=0.97\linewidth]{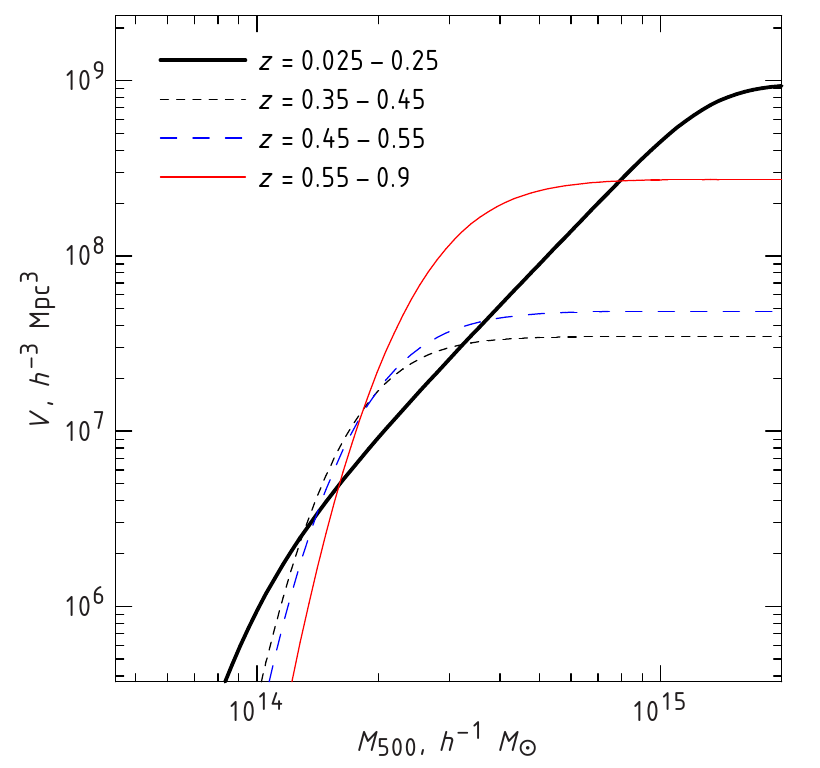}}
\caption{Volumes (comoving) as a function of cluster mass for our
  low redshift sample ($z=0.025-0.25$) and three high-redshift
  subsamples of the \400d{} survey.}
\label{fig:V}
\end{figure}

With the model for the $\LX-M$ relation at hand, we can now compute the
search volumes as a function of cluster mass
(eq.\ref{eq:dV(M)/dz}--\ref{eq:V(L)}). The results for our local sample and
the three redshift bins in the \400d{} sample are shown in
Fig.\,\ref{fig:V}. The volume for the local sample follows a power law
function of $M$ in a broad range of masses, as expected for a flux-limited
sample. A sharp decline of the volume at $M\lesssim1.5\times10^{14}\,\Msun$
is due to a combination of the flux threshold and a lower redshift cutoff of
the sample ($z>0.025$). The sample becomes volume-limited for high masses
because we imposed an upper cutoff ($z<0.25$) in the volume calculation. For
the three high-$z$ subsamples shown in Fig.\,\ref{fig:V}, the dynamic range in
$z$ is smaller and the transition from the volume-limited to strongly
incomplete regimes is much sharper.

\label{sec:volume:model:unc}

\begin{figure}[t]
\centerline{\includegraphics[width=0.97\linewidth]{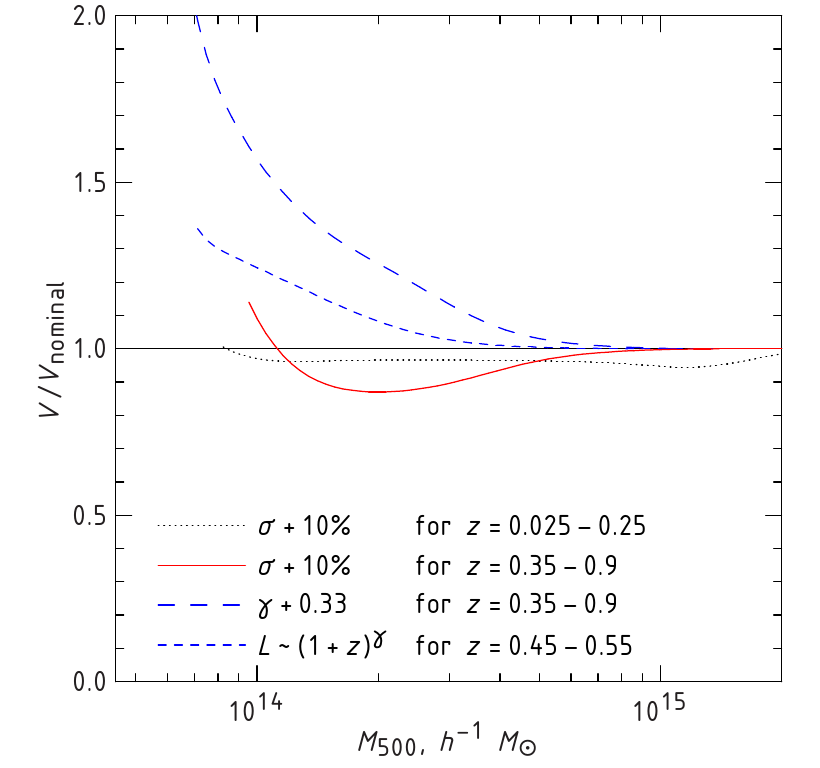}}
\caption{Sensitivity of the volume calculations to various variations of
  the $\LX-M$ relation model.}
\label{fig:V:models}
\end{figure}

We should now discuss how sensitive the survey volume computation is to
the assumptions in the $\LX-M$ relation model. The largest uncertainty
in the volume computation is related to the measurement errors of the
luminosity scale for fixed $M$. The effect is strongest for the high-$z$
data because the normalization of the $\LX-M$ relation is derived using
a smaller number of clusters, with larger measurement uncertainties, and
spanning a range of redshifts (see Fig.\,\ref{fig:L-M}). Overall, the
uncertainty in the high-$z$ relation corresponds to $\pm10.5\%$ in the
$\LX$ scale at $z=0.55$ (Appendix~\ref{sec:malmquist:fitting}). This is
equivalent to varying $\gamma$ by $\pm0.33$ assuming that the low-$z$
normalization is fixed; note that the range $\Delta\gamma=\pm0.33$ is
smaller than that quoted in eq.(\ref{eq:L-M:fit}) because the latter
also includes uncertainties in the low-redshift normalization. The
long-dashed line in Fig.\,\ref{fig:V:models} shows how the volume
calculation for our high redshift sample, $0.35<z<0.9$, is affected by
changing $\gamma$ by $+0.33$. Reassuringly, the relative change of
sample volume is large only for low-mass clusters where it becomes
comparable to the Poisson uncertainty of the derived mass function (see
in \S\,\ref{sec:sys:V-M} below).

By comparison, the sensitivity of $V(M)$ to a particular choice of the
$\LX-M$ evolution model is relatively minor. For example, if we use the
$(1+z)^\gamma$ scaling instead of $E(z)^\gamma$ (eq.[\ref{eq:L-M:2}]),
the largest difference in the best-fit $\LX-M$ relations is near $z=0.5$
(right panel in Fig.\,\ref{fig:L-M}). The corresponding relative change
of $V(M)$ at $z=0.45-0.55$ (dashed line in Fig.\,\ref{fig:V:models}) is
much smaller than the Poisson uncertainties of the mass function in the
same redshift bin (see below). Therefore, the uncertainties related to
the parametrization of the $\LX-M$ evolution can be neglected for our
purposes.

The effects of the scatter uncertainties on the $V(M)$ computations are
comparably small. Note that the situation is crucially different if one
uses $\LX$ to estimate the cluster masses. Consider for example, a case
of volume-limited survey. The $V(M)$ function is unchanged in this case
by variations of $\sigma$, while the estimate of the cluster mass
function is still very strongly affected (see
\citet{2005PhRvD..72d3006L}). Variations of $\sigma$ affect the $V(M)$
computations in two ways. First, there is a positive correlation of $V$
and $\sigma$ because of the scatter term in eq.[\ref{eq:dV(M)/dz}]
(assuming that $V(L)$ increases with $L$).  However, $\sigma$ also
implicitly enters the determination of the $\LX-M$ normalization because
we need to correct for Malmquist bias; the larger the $\sigma$, the
lower the $\LX$ for fixed $M$ inferred from the same data (e.g.,
Appendix~\ref{sec:Malmquist:bias}), and hence the smaller $V(M)$.  We
need to include both these effects to test properly the effect of the
$\sigma$ uncertainties on $V(M)$. This was achieved by fixing the value
of $\sigma$ at the boundaries of its measurement uncertainties ($\pm
10\%$ of the best-fit value, see eq.[\ref{eq:L-M:fit}]), refitting all
other parameters of the $\LX-M$ relation, and computing $V(M)$ for these
new fits. The results are shown in Fig.\,\ref{fig:V:models} by the solid
and dotted line for the high- and low-$z$ bins, respectively. The
variation of volume is negligible for the low-$z$ sample, but is more
substantial for the high-$z$ clusters. It is, however, much smaller than
the effect of uncertainties in the value of $\gamma$ considered above.
Note that increasing the scatter reduces the volume, indicating that the
effect of extra Malmquist bias correction on the $\LX-M$ normalization
outweighs the boost in volume due to an increased scattering kernel in
eq.[\ref{eq:dV(M)/dz}].

\section{Cluster Mass functions in the Concordant $\Lambda$CDM
  cosmology}

\begin{figure}[t]
\centerline{\includegraphics[width=0.97\linewidth]{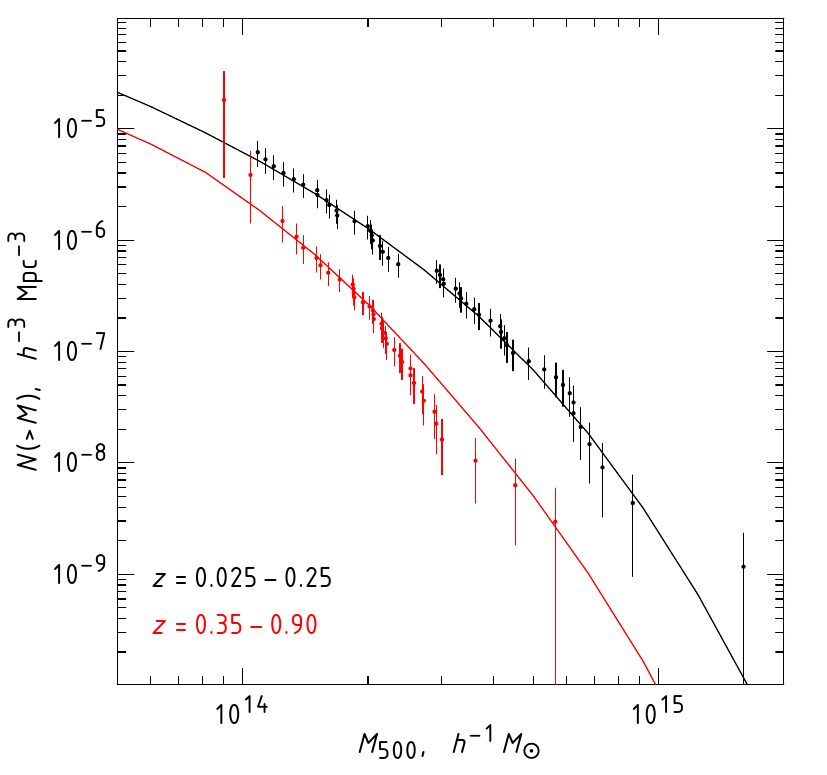}}
\caption{Cluster mass functions for our low and high-$z$ samples. The
  masses were estimated by the $\YX$ method. The errorbars show the
  Poisson uncertainties.
  Solid lines show the model predictions for the adapted cosmological
  model $\Omega_M=0.3$, $\Omega_\Lambda=0.7$, $h=0.72$, with only
  $\sigma_8$ fit to the cluster data (see text). The evolution of the
  mass function is non-negligible within either redshift range. To take
  this into account, the model number densities for each mass were
  weighted with $dV(M)/dz$ (eq.[\ref{eq:dV(M)/dz}]) within the redshift
  bin.  }
\label{fig:Mfun}
\end{figure}

\begin{figure}[t]
\centerline{\includegraphics[width=0.97\linewidth]{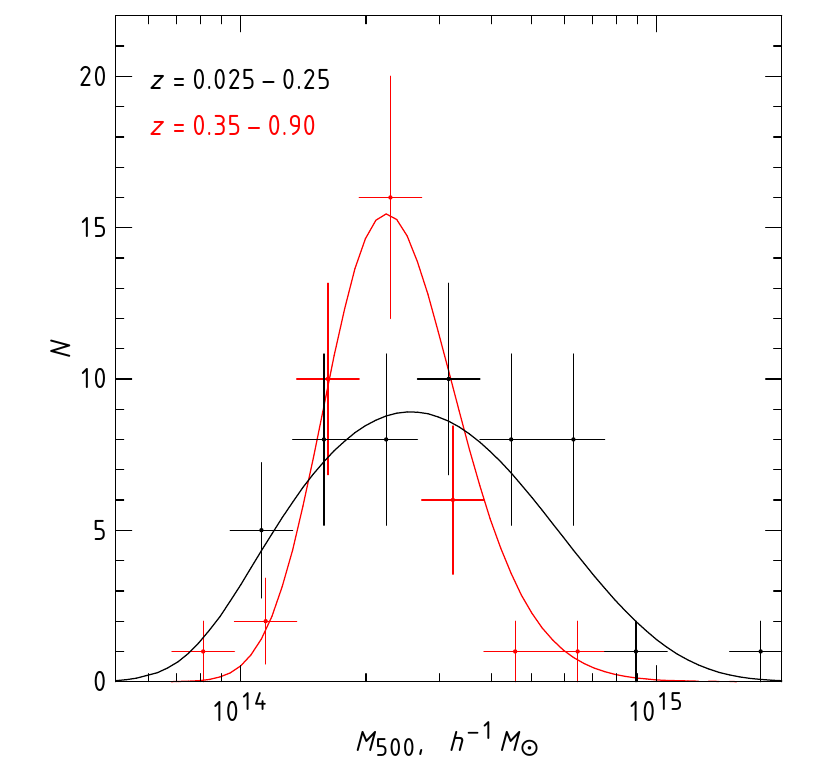}}
\caption{Differential representation of the mass functions shown in
  Fig.\,\ref{fig:Mfun}. The error bars in representation are
  uncorrelated (unlike Fig.\ref{fig:Mfun}), so statistical significance
  of the observed deviations from the best-fit model can be easily assessed. }
\label{fig:Mfun:diff}
\end{figure}

With the survey volume in hand, we can finally compute the mass
functions. Figure~\ref{fig:Mfun} shows the mass function in the
cumulative representation computed as
\begin{equation}
  N(>M) = \sum_{M_i>M} V(M_i)^{-1}.
\end{equation}
%
Our samples span similar mass at low and high redshifts, which is very
important for the robustness of the derived cosmological constraints. A
strong and highly significant decrease in the comoving cluster number
density at a fixed mass is observed between $z=0$ and $z\simeq0.5$, by a
factor of $5.0\pm1.2$ at $M_{500}=2.5\times10^{14}\,h^{-1}\,M_\odot$.  This
reflects the growth of cosmic structure between these redshifts.  Indeed,
the observed evolution of the cluster mass function is in good agreement for
the ``concordance'' cosmological model with the power spectrum normalization
$\sigma_8=0.746$ (solid lines in Fig.\,\ref{fig:Mfun}; we use the mass
function model from \citeauthor{2008arXiv0803.2706T}
\citeyear{2008arXiv0803.2706T} and our approach to the model fitting is
discussed in \S\,\ref{sec:likelihood}). The strongest observed deviation of
the data from the model is a marginal deficit of clusters in the distant
sample near $M_{500}=3\times10^{14}\,h^{-1}\,\Msun$ --- we observe 4
clusters where 9.5 are expected, a $2\sigma$ deviation. The cumulative
function fully recovers by $M_{500}=2\times10^{14}\,h^{-1}\,\Msun$,
approximately the median mass in the distant sample. The differential
representation of the mass function (Fig.\,\ref{fig:Mfun:diff}) also shows
that this deficit is consistent with the Poisson noise expected in the data.

\begin{figure}[t]
\centerline{\includegraphics[width=0.97\linewidth]{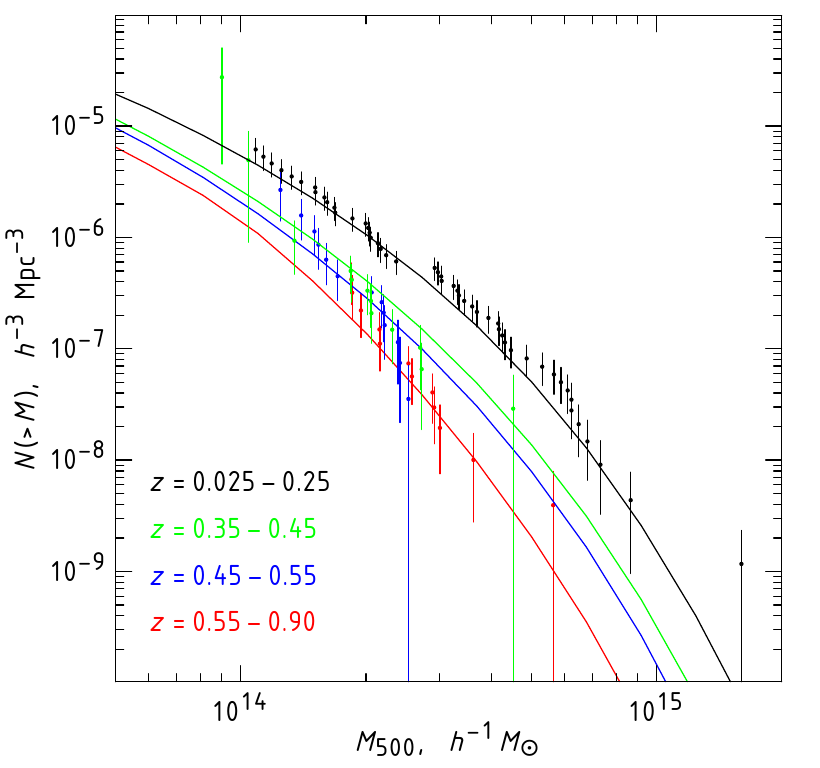}}
\caption{Same as Fig.\,\ref{fig:Mfun} but the high-$z$ sample is split
  into three redshift bins.}
\label{fig:Mfun:bins}
\end{figure}

Our high-$z$ sample can be split into several redshift bins to check if
the observed evolution within the sample is still consistent with the
model. Figure~\ref{fig:Mfun:bins} shows the results for the three bins,
$z=0.35-0.45$, $0.45-0.55$, and $0.55-0.9$, approximately 14 clusters in
each. The data are still in good agreement with the model
predictions. The strongest deviation is a marginal ($\simeq 1\sigma$)
deficit of clusters at $z=0.35-0.45$.

\subsection{Sample Variance}

In addition to the Poisson cluster counting uncertainties, there is
sample variance in the number of clusters in a survey of limited volume
due to large-scale clustering. Depending on the mass scale, the sample
variance can be comparable to, or larger than, the Poisson errors
\citep{2003ApJ...584..702H}. We follow the formalism of
\citeauthor{2003ApJ...584..702H} to assess the importance of sample
errors in the error budget in our case.

We calculate the sample variance for the two geometries. For the local
sample we assume all-sky coverage with an exclusion zone of $\pm 20^\circ$
from the Galactic plane; the variance for this geometry is given by
equation~A7 of \citet{2003ApJ...584..702H}. The second is a pencil-beam
volume with a small circular footprint on the sky, which is appropriate for
the individual \emph{ROSAT} fields included in the \400d{} survey; the
variance for this geometry can be computed using the flat-sky approximation
\citep[eq.~7 in][]{2003ApJ...584..702H}. The variance calculations are done
for our reference cosmology with the power spectrum normalization
$\sigma_8=0.8$, resulting in a slightly higher variance than what would be
predicted for our best-fit cosmological model with slightly lower
$\sigma_8$.
The halo mass function model is from \citet[][their
eq.~B3]{2001MNRAS.321..372J} and the cluster bias model is from
\citet{1999MNRAS.308..119S}. These mass function models use cluster
masses defined within the aperture enclosing an overdensity of 180 with
respect to the \emph{mean} density, and so we need to relate it to the
mass definition adopted here ($\Delta = 500$ with respect to critical
density). We assumed a simple relation, $M_{500}\approx 0.55 M_{180}$,
appropriate for typical concentrations of clusters in our mass range.

The relative importance of the sample variance increases for low-mass
clusters \citep{2003ApJ...584..702H}. Fortunately, it is still
sufficiently small near our mass limit. For example, at the limiting
$M_{500}=1.5\times10^{14}\,h^{-1}\,M_\odot$, the local survey volume
corresponds to the effective redshift depth $z_{\rm max}=0.043$; the
sample variance calculated for this mass limit and redshift range
$z=0.025-0.043$ is $\sigma_{\rm var}\equiv(\langle n^2\rangle/\langle
n\rangle^2-1)^{1/2}\approx 0.16$. This should be compared with the
Poisson errors at this mass limit, $\sigma_{\rm shot}=0.24$
(Fig.\,\ref{fig:Mfun}). Combining these variances in quadrature, we find
that the total uncertainty $\sigma_{\rm tot}=(\sigma_{\rm var}^2 +
\sigma_{\rm shot}^2)^{1/2}$ is only 17\% larger than the Poisson value.
The contribution of sample variance quickly becomes small for higher
masses. For example, at $M_{500}=3\times10^{14}\,h^{-1}\,\Msun$ (the
median mass for the low-$z$ sample), the total uncertainty is only 7\%
larger than the Poisson error; the contribution becomes negligible at
higher masses. Going to lower masses, we predict that $\sigma_{\rm var}$
becomes comparable to $\sigma_{\rm shot}$ for
$M_{500}\approx7\times10^{13}\,h^{-1}\,\Msun$, below our mass limit.

The high-$z$ sample consists of 1600 widely separated (and therefore
independent) pencil-beam pointings. 
For a single pointing of circular radius of $17.5^{\prime}$ and redshift
range $z=0.35-0.45$ (the variances for the higher redshift bins are
similar but somewhat smaller), the sample variance is $\sigma_{\rm
  var,1}\approx 0.65-1.65$ for the samples with $M_{500}$ thresholds
between $10^{14}$ and $10^{15}\ \rm M_{\odot}.$ Assuming that the
individual pointings are uncorrelated (a good assumption for the widely
separated pointings of the 400d survey), the total sample variance is
$\sigma_{\rm var}\approx \sigma_{\rm var,1}\,N^{-1/2}\approx 0.02-0.05$,
where $N=1600$ is the number of \400d{} survey pointings, much smaller
than the Poisson uncertainties. The sample variance can therefore be
safely neglected for our high-$z$ sample.

In principle, sample variance can be included in the calculation of the
likelihood functions for the low-$z$ sample
\citep{2006astro.ph..2251H,2006PhRvD..73f7301H}.  The procedure, however,
would be quite cumbersome in our case and is not worth the effort because
the variance increases the measurement errors by only $17\%$ in the worst
case, and by 7\% or less for the median sample mass. This is considerably
smaller than expected systematic effects and we will therefore neglect the
sample variance hereafter.

\section{Likelihood Function}
\label{sec:likelihood}

Let us now consider the expression for the likelihood function appropriate
for our sample and for our method of deriving the mass functions. The basics
of the likelihood function are very standard and used in large number of
other works. We follow the derivation presented in
\citet{1979APJ...228..939C} for the case of purely Poisson statistics.  We
split the mass intervals into narrow bins, $\Delta M$, so that the
probability to observe a cluster with an estimated mass in this bin is
small, $p(\Mest,z)\,\Delta M\ll 1$, and we have at most one cluster per bin.
The likelihood function in this case can be written as
\citep[c.f.][]{1979APJ...228..939C}
\begin{equation}\label{eq:likelihood:1}
  \begin{split}
    \ln L = \sum_i \ln& \left(p(\Mest_i,z_i)\,\Delta M_i\right) - \\
    & \iint_{M,z}
    p(\Mest,z)\;d\Mest\,dz,
  \end{split}
\end{equation}
where summation is over the clusters in the sample and integration is
over pre-selected $\zmin-\zmax$ and $\Mmin-\Mmax$ intervals. Usually,
the $\Delta M$ terms can be dropped because they are independent of the
model parameters and thus simply add a constant to the likelihood
function. In our case, however, the estimated masses are also a function
of the background cosmology. When $\Mest_i$ is changed because of the
variation in the cosmological parameters, we should correspondingly
stretch the mass interval, $\Delta M = \Delta M^{(0)} M/M^{(0)}$, where
$M^{(0)}$ and $\Delta M^{(0)}$ are the estimated mass and width of the
interval for some fixed reference cosmological model.\footnote{This is
  equivalent to the rule of transformation of the probability density
  function under change of variables, $p(y)\,dy = p(x)\,dx$.} Taking the
logarithm of this expression and dropping constant terms ($M^{(0)}$ and
$\Delta M^{(0)}$), we obtain the likelihood function in the form
\begin{equation}\label{eq:likelihood}
  \begin{split}
  \ln L = \sum_i \ln & p(\Mest_i,z_i) + \sum_i \ln \Mest_i\; - \\
  & -\iint_{M,z} p(\Mest,z)\;d\Mest\,dz.
  \end{split}
\end{equation}
The calculation of individual terms in this expression is discussed in
Appendix\,\ref{sec:likelihood:details}. The likelihood function
implicitly depends on the cosmological parameters through the model of
cluster mass function (reflecting the growth, normalization, and shape
of the density perturbation power spectrum), through the cosmological
volume-redshift relation which determines the survey volume, and through
the distance-redshift and $E(z)$ relations which affect our cluster mass
estimates. The best fit parameters are obtained by maximizing the
likelihood function in eq.[\ref{eq:likelihood}]. We also can use
standard methods \citep[e.g.][]{1979APJ...228..939C} to estimate
uncertainties of the model parameters. The advantage of this approach is
that we do not use any binning in either mass or redshift.

In addition to the best fit parameters and confidence intervals, it is
also useful to be able to characterize the goodness of fit. Even though
the likelihood function cannot be used for this purpose directly, we can
utilize it to obtain an effective $\chi^2$ for every cosmological model.
First, we note that essentially all the cosmological information
provided by the cluster mass function is the normalization and slope of
the linear perturbations power spectrum at $\sim 10$~Mpc scales.
Statistical quality of our sample is sufficient to fit $\sigma_8$
independently in 4 redshift bins (the local sample and the high-$z$
subsamples $z=0.35-0.45$, $0.45-0.55$, and $0.55-0.9$) and tilt to the
entire sample. Individual best-fit values of $\sigma_8$ should be
consistent within the errors if the background cosmology is ``correct'';
similarly, tilt (relative to the best-fit slope constrained by CMB data)
should be consistent with zero if we trust the CDM transfer function
models \citep[see][and references therein]{1998ApJ...496..605E}. To
characterize how close the tilt is to 0 and individual $\sigma_8$'s to a
constant value, we can take advantage of the fact that the deviation of
the quantity $C=-2\ln L$ from the minimum has statistical properties of
the $\chi^2$ distribution \citep{1979APJ...228..939C}. The effective
``tilt'' component of the total $\chi^2$ can be computed as
\begin{equation}\label{eq:tilt:chi2}
  \chi^2_t (\btheta) = \min_{\sigma_8}
  C(0,\sigma_8,\btheta) - \min_{t,\sigma_8} C(t,\sigma_8,\btheta),
\end{equation}
where $\btheta$ is the vector of cosmological parameters
other than tilt and normalization of the power spectrum. Similarly, the
effective $\chi^2$ component for the evolution in the normalization of
the mass function is
\begin{equation}\label{eq:evol:chi2}
  \chi^2_{\rm evol} (\btheta) = \sum\nolimits_j C_j(0,\tilde{\sigma}_8,\btheta) -
  \min_{\sigma_8}C_j(0,\sigma_8,\btheta),
\end{equation}
where summation is over several sufficiently wide redshift bins and
$\tilde{\sigma}_8$ is the best-fit value to the entire sample. Adding
these terms, we obtain the total effective $\chi^2$ for the cluster mass
function data,
\begin{equation}\label{eq:clus:chi2}
  \chi^2_{\rm clus}(\btheta) = \chi^2_{\rm evol} (\btheta) + \chi^2_t
  (\btheta).
\end{equation}
This effective $\chi^2$ can be used to check how consistent are the model
and observed mass functions in terms of general shape and evolution in the
normalization.


A detailed discussion of fitting cosmological parameters to our cluster
data will be presented in Paper~III. Here, we quote only the results of
fitting the power spectrum normalization for our reference cosmology,
$\sigma_8 = 0.746 \pm 0.009$ (purely statistical uncertainties).  The
best fit models are shown by solid lines in
Fig.\,\ref{fig:Mfun}--\ref{fig:Mfun:bins}.

\begin{deluxetable*}{p{6cm}p{4.5cm}ll}
  \tablecaption{Summary of main systematic uncertainties in the cluster mass function\label{tab:systematics}}
  \tablehead{
    \colhead{Source of error} &
    \colhead{Affects} &
    \colhead{Uncertainty}  &
    \colhead{ref} 
  }
  \startdata
  400d selection function\dotfill & $V(M)$ for high-$z$ sample\dotfill &
  $\pm3\%$ & \S~\ref{sec:volume}, \citetalias{2007ApJS..172..561B} \\
  Scatter in $\LX-\Mtot$\dotfill & $V(M)$\dotfill & $\pm10\%$ & \S~\ref{sec:volume} \\
  Evolution in $\LX-\Mtot$\dotfill & $V(M)$ for high-$z$ sample\dotfill
  & $\pm22\%$ for median mass & \S~\ref{sec:volume} \\
  \emph{Chandra} calibration\dotfill & $M_0$\dotfill & $-9\%-0$ & \S~\ref{sec:error:budget:X-ray:cal} \\
  ICM metallicity vs.\ $z$\dotfill & $M_Y(z)$, $M_G(z)$,
  $M_T(z)$\dotfill & $+3\%$, $+1\%$, $+6\%$, respectively & \S~\ref{sec:sys:astro} \\
  Accuracy of X-ray hydrostatic mass estimates \dotfill & $M_0$\dotfill &
  $0-9\%$ & \S~\ref{sec:Xray:vs:WL} \\
  Departures from self-similar evolution\dotfill & $M_T(z)$\dotfill &
  $\pm7\%$ & \S~\ref{sec:M-T:summary} \\
  Departures from self-similar evolution\dotfill & $M_Y(z)$\dotfill &
  $\pm5\%$ & \S~\ref{sec:YX} \\
  $f_g(z)$\dotfill & $M_G(z)$\dotfill &
  $\pm5\%$ & \S~\ref{sec:fgas:corr:evol}
  \enddata
\tablecomments{The volume uncertainties are quoted for the median mass
  in the sample. The $\Mtot$ uncertainties are separated into
  calibration of the low-$z$ mass vs.\ proxy relations ($M_0$) and
  uncertainties in the evolutions of these relation, $M(z)$. $M_T(z)$,
  $M_G(z)$, and $M_Y(z)$ stand for total masses estimated using $\TX$,
  $\Mgas$, and $\YX$, respectively. The evolutionary uncertainties are
  quoted at $z=0.5$.}
\end{deluxetable*}

\section{Systematic Error Budget}
\label{sec:sys:err}

We conclude the analysis with a summary of the systematic error budget
in the mass function measurements. We begin with a discussion of several
sources of observational uncertainties (those affecting measurements of
the basic cluster parameters), and then summarize the modeling
uncertainties --- those related to the mass vs.\ proxy relations and
determination of the survey volumes.

\subsection{Observational Uncertainties}

\subsubsection{Calibration Uncertainties}
\label{sec:error:budget:X-ray:cal}

The accuracy of the basic X-ray observables --- average $T$ and soft
X-ray flux --- is limited by absolute calibration of the \emph{Chandra}
effective area. The calibration parameters most relevant for our study
is the absolute value of the soft-band effective area and the relative
hard- to soft-band calibration. Thanks to the great effort put into
calibration of the \emph{Chandra} telescopes, both on the ground and in
flight, the associated uncertainties are small, but they still need to
be discussed for the sake of completeness.

The absolute soft-band ($\sim 0.5-2$ keV) effective area affects the
measured cluster luminosities and gas masses, $L_X\propto A^{-1}$,
$\Mgas\propto A^{-1/2}$. The largest source of uncertainty in $\Asoft$
is in-flight contamination of the ACIS optical blocking filters by a
hydro-carbon compound.  Fortunately, in our energy band of interest,
this contamination can be accurately measured as a function of time and
position using the on-board calibration source, and so the soft-band
effective area can be brought to its absolute pre-flight calibration,
which is accurate to $\approx3\%$ \citep{Edgar2004}. The effect of such
uncertainties on the derived mass function (through the mass proxies and
the $\LX-M$ relation) is negligible.  The validity of the soft-band
calibration is indirectly confirmed by the excellent agreement in the
\emph{Chandra} and \emph{ROSAT} flux measurements.

The relative hard-to-soft area calibration affects temperatures and
hence hydrostatic total mass measurements. We will characterize this
effect approximately by the relative change of measured temperatures
$\dTcal=\Delta T/T$ for the 5~keV clusters. The hydrostatic $\Mtot$
measurements are affected as $\Delta \M500/\M500 = 3/2\,\dTcal$
\citep[see, e.g., Appendix~A in][]{2006ApJ...640..691V}. This uncertainty
is transferred to our mass function determinations because all mass vs.\
proxy relations are calibrated using hydrostatic $\Mtot$ measurements.

Calibration uncertainties for the cluster temperatures cannot be
characterized exactly.  Approximate estimates can be made from
comparison of the values derived by different telescopes calibrated
independently or by looking at the effect of the most relevant ``fudge''
factors for \emph{Chandra}.  The systematic difference between
\emph{XMM-Newton} and \emph{Chandra} temperatures is approximately 7\%
\citepalias{2005ApJ...628..655V}. The largest remaining \emph{Chandra}
calibration uncertainty is, as of this writing, related to the effect of
the $10-20\,\AA$ hydro-carbon overlayer on the X-ray mirrors.
Experimenting with variations of the overlayer model, we find that the
range of possible temperature variations is $-6\%<\dTcal<0$, and
$\dTcal$ is nearly independent of the cluster temperature and redshift.
This would be equivalent to up to $-9\%$, $z$-independent shift in the
mass scale.

%
%
%

\subsubsection{Astronomical Uncertainties}
\label{sec:sys:astro}

In addition to calibration uncertainties, we checked a number of
``astronomical'' effects which also could affect the measurement of
basic cluster properties. The effects that we checked and determined to
be negligibly small include uncertainties in the Galactic interstellar
absorption measurements \citep[based on neutral hydrogen 21~cm
maps][]{1990ARA&A..28..215D}; absorption by ionized (warm) ISM in the
Galaxy \citep{1993ASPC...35..338R}; difference between plasma spectral
codes; possible variations of the He abundance around the cosmic average
\citep[our conclusion is based on the analysis of][]{pengnagai08}.

The only effect which is marginally significant is the possible
evolution of the ICM metallicity. Since the statistics in the data for
our high-$z$ sample are insufficient for ICM metallicity measurements,
we assumed in each object that the metallicity is equal to $0.3$~Solar,
approximately the mean value for the low-$z$ population. If in fact
there is an evolution in the heavy element metal abundance, our derived
values for $\TX$ and $\Mgas$ are affected slightly. For example, if the
mean abundance for high-$z$ clusters is 0.15~Solar, the derived $\TX$
and $\Mgas$ will be higher by $\approx +4\%$, and +1\%, respectively.
Such a trend \citep[which is probably outside the range allowed by the
data, see][]{2003ApJ...593..705T,2008ApJS..174..117M}, will be
equivalent to changing the mass scale for our high-$z$ sample by $+3\%$,
$+1\%$, and $+6\%$ if masses are estimated through the $\Mtot-\YX$,
$\Mtot-\Mgas$, and $\Mtot-\TX$ relations, respectively.

%

\subsection{Modeling Uncertainties}

Modeling uncertainties in the mass function measurements can be
separated into two components, 1) how accurately we can predict the
survey volume for clusters of a given mass, and 2) how accurately we can
derive cluster masses from the data. The first component mainly depends
on the accuracy of the $\LX-\Mtot$ relation, and the second, on the
$\Mtot$ vs.\ proxy relation. All these uncertainties were discussed in
detail above and so we provide only a summary here.

\subsubsection{Uncertainties in $V(M)$}
\label{sec:sys:V-M}

Uncertainties in the survey volume mainly depend on how accurately we can
recover the $\LX-\Mtot$ relation from the data, assuming that masses are
accurately reconstructed from the $\YX$, $\Mgas$, or $\TX$ proxies.  The
effects of these uncertainties on $V(M)$ are considered in
\S~\ref{sec:volume}. The largest error is related to the measurement of the
evolutionary factor (eq.\,\ref{eq:L-M:2}) and amounts to $\pm22\%$ in volume
for the median mass in our high-$z$ sample, $\sim
2.1\times10^{14}\,h^{-1}\,M_\odot$ (Fig.\,\ref{fig:V:models}). This source
of error is statistical in nature (related to measurement uncertainties in
the $\LX-\Mtot$ parameters). It can therefore be added in quadrature to the
purely Poisson errors ($\pm26\%$ in the cumulative mass function for the
same mass threshold), resulting in a moderate increase in the statistical
errorbars. Although it is possible to include these uncertainties
approximately in the cluster likelihood function or effective $\chi^2$, a
more accurate estimate of their effect on the final results can be obtained
by repeating the entire analysis procedure with the evolution factor varied
within its measurement errors. We take this approach and we will quote the
associated parameter uncertainties in Paper~III.

Other sources of uncertainty from the $\LX-\Mtot$ modeling, such as the
exact scatter in the relation, functional form of the evolution term,
etc., are comparably small. The accuracy of statistical calibration of
the 400d survey selection function also makes a negligible contribution,
$\pm3\%$, to the volume error \citepalias{2007ApJS..172..561B}.

\subsubsection{Uncertainties in derived $\Mtot$}

Separate sources of uncertainties are related to potential biases of the
$\Mtot$ estimates from X-ray proxies. Note that these biases will have
little effect on the volume computations for the given cluster (if we
change the estimated $\Mtot$'s, we need to refit the $\LX-\Mtot$
relation and the net effect will be that the volume for the given
cluster is almost unchanged). In a sense, the $V(M)$ systematics move
the cumulative mass functions in Fig.\,\ref{fig:Mfun} up and down, while
the potential $\Mtot$ biases shift the mass function along the $M$ axis.

The mass biases can be naturally separated into two components. The
first is related to calibration of the $\Mtot$ vs.\ proxy relations for
low-$z$ clusters. Assuming that the evolution in the relation is
nominal, such biases will shift the low and high-$z$ mass function by
the same amount, or, equivalently, will affect the overall
normalization, but not the evolution in the co-moving number density. As
we discussed above (\S~\ref{sec:Xray:vs:WL}), comparison of X-ray and
weak lensing masses provides a good estimate, $\pm9\%$ in mass, for such
biases.

The second source is departures of the evolution in the $\Mtot$ vs.\
proxy relation from the assumed forms. Since evolution is negligible
within the low-$z$ sample, such biases are important only for the
high-$z$ mass function and thus will affect the derived evolution in the
cluster number density, but not the overall normalization of the mass
functions. We estimate that by $z=0.5$, the evolutionary $\Mtot$ biases
can be up to $\pm7\%$ in the $\Mtot-\TX$ relation, and $\pm5\%$ for the
$\Mtot-\Mgas$ and $\Mtot-\YX$ relations (\S\S\,\ref{sec:M-T:unrelaxed},
\ref{sec:fgas:corr:evol}, \ref{sec:YX}, respectively). In the case of
the $\Mtot-\TX$ we also need to add a $\pm6\%$ uncertainty related to
the potential evolution in the ICM metallicity
(\S\,\ref{sec:sys:astro}).

The estimated uncertainties in the $\Mtot$ calibration cannot be easily
included in the likelihood function. Instead, we check (Paper~III) how
they affect the cosmological fit by repeating the entire analysis
procedure with the parameters $\Mtot$ vs.\ proxy relations within the
bounds specified above.  Note also that the use of three different mass
proxies, each with its own bias, provides a good consistency check,
because results obtained with different proxies can be compared to each
other to check for biases.

\section{Summary}

We presented a report on data analysis procedures leading to a measurement
of the galaxy cluster mass functions using \emph{Chandra} observations of
statistically complete samples of low and high-$z$ clusters originally
selected in the X-ray data from \emph{ROSAT}. This measurement relies on a
careful selection of the parent samples, rather detailed \emph{Chandra}
observations of selected objects, and using several robust X-ray proxies for
the total cluster mass ($\YX$, $\Mgas$, $\TX$).

The scaling relations between proxies we use and $\Mtot$ mostly follow the
predictions of the self-similar theory, a very basic and hence reliable
model. We used advanced high-resolution numerical simulations to test the
predictions of this theory with regard to our proxies; these simulations
indicate that only small corrections are necessary, which we use
cautiously. At low redshifts, the $\Mtot$ vs.\ proxy relations were
calibrated by detailed \emph{Chandra} observations of a sample of relaxed
clusters spanning a wide range of mass; our \emph{Chandra} results were
cross-checked against recent weak lensing measurements. 

As a part of this project, we derive a relation between cluster mass and
total X-ray luminosity, using large statistically complete samples and
properly taking into account the Malmquist bias. The relation is adequately
described by a single power law, substantial log-normal scatter, and
evolution of the power law normalization following $E(z)^{1.85}$ ---
assuming that the evolution in the $\Mtot-\YX$ relation is exactly
self-similar as we use $\YX$ to estimate cluster masses.

We present the cluster mass functions estimated assuming a ``concordant''
\LCDM{} cosmology. These data shows a significant evolution in the cluster
comoving number density at a fixed mass threshold, by a factor of $\approx
5$ at $M_{500}=2.5\times10^{14}\,h^{-1}\,M_\odot$ between $z=0$ and $0.5$.

Finally, we provide a summary of estimated systematic uncertainties in our
mass function measurement. Most source of systematics lead to corrections
which are smaller than the Poisson errors in our data. The main exception is
uncertainties in calibration of the absolute mass scale at low redshifts but
it has little impact on the measurment of evolution in the cluster number
density.

The evolution in the cluster mass function reflects the growth of density
perturbations and can be used for the cosmological constraints complementing
those from the distance-redshift relation. The cosmological modeling of
these data will be discussed in a future paper.

\acknowledgements

We thank Wayne Hu, Oleg Gnedin, and Maxim Markevitch for useful
discussions in the course of this work, and D.~Spergel and A.~Loeb for
comments on the manuscript. We also that the referee, T.~Reiptich, for a
very thorough and constructive review. Financial support was provided by
NASA grants and contracts NAG5-9217, GO5-6120A, GO6-7119X, NAS8-39073
(AV, WRF, CJ, SSM), GO5-6120C (HE); NSF grants AST-0239759 and
AST-0507666, NASA grant NAG5-13274 and the Kavli Institute for
Cosmological Physics at the University of Chicago (AK); Sherman
Fairchild Foundation (DN); FONDAP Centro de Astrofisica (HQ); Russian
Foundation for Basic Research grants RFFI 05-02-16540 and RFFI
08-02-00974 and the RAS program OFN-17 (RB and AV).

\bibliography{400d-ch}

\begin{appendix}

  \section{Correction of the Luminosity-Mass relation for the Malmquist
    Bias}
  \label{sec:Malmquist:bias}

  \cite{2006ApJ...648..956S} recently pointed out that observational
  determinations of the mass-luminosity relation can be significantly
  affected by Malmquist bias because of the flux-limited nature of most
  of the available cluster samples. \citeauthor{2006ApJ...648..956S}
  were interested in the relation where $\LX$ is the independent
  variable; such a relation is useful for $\Mtot$ estimates using $\LX$
  as a proxy.  Corrections for Malmquist bias in this case lead to
  complicated computations involving the cluster mass function model.
  We, instead, are interested in computing the survey volume for objects
  of given mass, for which we need to treat $\Mtot$ as independent
  variable and compute luminosity for a given mass (eq.[\ref{eq:V(M)}]).
  The calculations of the Malmquist bias are then much simpler and can
  be done independently of the mass function modeling.

  We assume that the scatter in $L$ for fixed $M$ has a log-normal
  distribution, 
  \begin{equation}
    p(\ln L) = 
    \frac{dN}{d\ln L} \propto 
     \exp\left(-\frac{(\ln L - \ln L_0)^2}{2\,\sigma^2}\right),
  \end{equation}
  where $L_0$ is the average luminosity for the given mass. Typically,
  $L_0$ is a power law of mass, $L_0 \propto M^\alpha$, but we do not
  make this assumption in the calculations below.

  \subsection{Corrections for Malmquist Bias in Non-Evolving $\LX-M$ Relation}
  \label{sec:malmquist:noevol}

  Calculations of the Malmquist bias are particularly simple if the
  evolution in the $\LX-M$ can be neglected, e.g., $L_0(M)$ is the same at
  all redshifts within the sample. This situation is applicable for the
  analysis of the low-$z$ samples and in our case, for establishing the
  low-$z$ reference relations.

  Let us assume that the survey volume as a function of the object
  luminosity can be approximated as a power law
  \begin{equation}
    V(L) \propto L^\delta. 
  \end{equation}
  For example, in the case of Euclidean space and a pure flux-limited
  survey, $V(L)\propto L^{3/2}$ exactly. If there is a lower redshift cutoff
  in the survey, $\delta\ne 3/2$ at the low-$L$ end, even in Euclidean
  space. Likewise, if there is a higher-redshift cutoff, $\delta\rightarrow
  0$ in the high-$L$ end (sample becomes volume-limited).
  
  The $\LX-M$ relation is usually fit in the $\ln M - \ln L$
  coordinates, so we need to compute the bias in $\ln L$ for given
  $M$:
  \begin{equation}\label{eq:malmquist:noevol:1}
    \langle\ln L\rangle - \ln L_0 = 
    \frac
    {\int_{-\infty}^{\infty}(\ln L - \ln L_0)\,p(\ln L)\,V(\ln L)\,d\ln L}
    {\int_{-\infty}^{\infty} p(\ln L)\, V(\ln L)\, d\ln L}
    =
    \frac
    {\int_{-\infty}^\infty x\,\exp(-x^2/2\sigma^2)\, \exp (x\delta)\,dx}
    {\int_{-\infty}^\infty \exp(-x^2/2\sigma^2)\, \exp (x\delta)\,dx},
  \end{equation}
  where we used the substitution $x=\ln L - \ln L_0$. The integrals
  can be worked out analytically,
  \begin{equation}\label{eq:malmquist:noevol:2}
    \langle\ln L\rangle - \ln L_0 = \frac
    {\delta\sqrt{2\pi}\,\sigma^3\,\exp(\delta^2\,\sigma^2/2)}
    {\sqrt{2\pi}\,\sigma\,\exp(\delta^2\,\sigma^2/2)} = \delta\,\sigma^2. 
  \end{equation}
  
  The log-normal scatter in the relation, $\sigma$, is usually unknown
  apriori and thus it should be estimated from the rms scatter around the
  best-fit relation in the $\ln M - \ln L$ plane. Fortunately, the
  flux-limited survey does not introduce bias in the scatter, i.e.
  $\sigma_{\text{obs}} = \sigma$, as we now demonstrate.

  \begin{equation}
    \sigma^2_{\text{obs}} = \langle (\ln L - \ln L_0)^2\rangle - (\ln
    L_0 - \langle\ln L\rangle)^2 = \langle (\ln L - \ln L_0)^2\rangle
    - \delta^2\,\sigma^4
  \end{equation}
  \begin{equation}
    \langle (\ln L - \ln L_0)^2\rangle = \frac
    {\int_{-\infty}^\infty x^2\,\exp(-x^2/2\sigma^2)\, \exp (x\delta)\,dx}
    {\int_{-\infty}^\infty \exp(-x^2/2\sigma^2)\, \exp (x\delta)\,dx} =
    \sigma^2 (1+\delta^2\sigma^2)
  \end{equation}
  (c.f.\ eq.\ref{eq:malmquist:noevol:1}
  and~\ref{eq:malmquist:noevol:2}), and so
  $\sigma_{\text{obs}}^2=\sigma^2$. 

  \subsection{Correction for Individual Clusters}
  \label{sec:malmquist:corr:evol}

  The bias computations from the previous section cannot be applied if
  we aim to model also the evolution in the $\LX-M$ relation, because in
  this case we need to compute the bias in a fixed narrow interval of
  $z$ where $V(L)$ cannot in general be represented with a power law
  (e.g., for an ideal flux-limited survey, $V(L)$ in a narrow interval
  of $z$ is close to a step-function). An alternative (approximate)
  approach is to compute the expected biases in $L$ for individual
  clusters, as considered below. A better approach is to model all
  effects of selection through the likelihood function, as discussed in
  the next section.

  Let us assume that the survey has a single flux threshold, $\fmin$
  (i.e., the cluster is always detected if $f>\fmin$ and not detected if
  $f<\fmin$).  The average luminosity bias of \emph{detected} clusters
  with a given mass is
  \begin{equation}\label{eq:malmquist:evol:1}
    \langle \ln L - \ln L_0\rangle = 
    \langle \ln f - \ln f_0\rangle = \frac
    {\int_{x\Min}^\infty x\,\exp (-x^2/2\sigma^2)\,dx}
    {\int_{x\Min}^\infty \exp (-x^2/2\sigma^2)\,dx} = 
    \frac
    {\exp\left(-x\Min^2/2\sigma^2\right)}
    {\sqrt{\pi/2}\,\,\mathrm{erfc}\left(x\Min\big/\left(\sigma\sqrt{2}\right)\right)}
    \;\sigma,
  \end{equation}
  where $f_0$ is the flux corresponding to the nominal luminosity,
  $L_0$, given by the $\LX-M$ relation, and $x\Min=\ln\fmin-\ln f_0$. 
  For $f_0\gg\fmin$ (very massive clusters), the bias is 0 as expected. 
  For very low mass clusters ($L_0\rightarrow0$, $f_0\rightarrow0$,
  $x\Min\rightarrow\infty$), eq.~\ref{eq:malmquist:evol:1} gives
  $\langle\ln f - \ln f_0\rangle \simeq \ln\fmin-\ln f_0$ (i.e.\
  all detected clusters have fluxes just above the survey threshold). 

  Equation~[\ref{eq:malmquist:evol:1}] is easily generalized for the case
  when the survey selection probability is a smooth function of flux (as is
  the case for the \400d{} sample):
  \begin{equation}\label{eq:malmquist:evol:area}
    \langle \ln L - \ln L_0\rangle = 
    \langle \ln f - \ln f_0\rangle = \frac
    {\int_{-\infty}^\infty x\,\Psel(x+\ln f_0)\,\exp (-x^2/2\sigma^2)\,dx}
    {\int_{-\infty}^\infty \Psel(x+\ln f_0)\,\exp (-x^2/2\sigma^2)\,dx}
  \end{equation}

  \subsection{Likelihood Function and Fitting Procedure}
  
  The best way to treat the Malmquist bias in modeling the relation is
  through a proper definition of the likelihood function. Let $\Psel(\ln
  f)$ be the survey selection efficiency as a function of flux. The
  average luminosity-mass relation gives a ``nominal'' luminosity for
  clusters of given $M$, which corresponds to a ``nominal'' flux
  $f_0$. The probability density function for the cluster to have flux
  $f$ is
  \begin{equation}\label{eq:malmquist:prob:1}
    \frac{dP}{d\ln f} =
    C\,\exp\left(-\frac{(\ln f - \ln
        f_0)^2}{2\sigma^2}\right)\,\Psel(\ln f),
  \end{equation}
  where $C$ is the normalization coefficient defined so that the total
  probability is 1,
  \begin{equation}\label{eq:malmquist:prob:norm}
    C^{-1} = \int_\infty^\infty
    \exp\left(-\frac{(\ln f - \ln
        f_0)^2}{2\sigma^2}\right)\,\Psel(\ln f)\,d\ln f. 
  \end{equation}
  For a survey with a single sharp flux limit [i.e.\ those with $\Psel(f) =
  \theta(f-\fmin)$], \ref{eq:malmquist:prob:norm} becomes $C =
  (\frac{1}{2}\,\mathrm{erfc}[\ln(\fmin/f_0)/(\sigma\sqrt{2})])^{-1}$.

\begin{figure*}
\centerline{%
\includegraphics[width=0.478\linewidth]{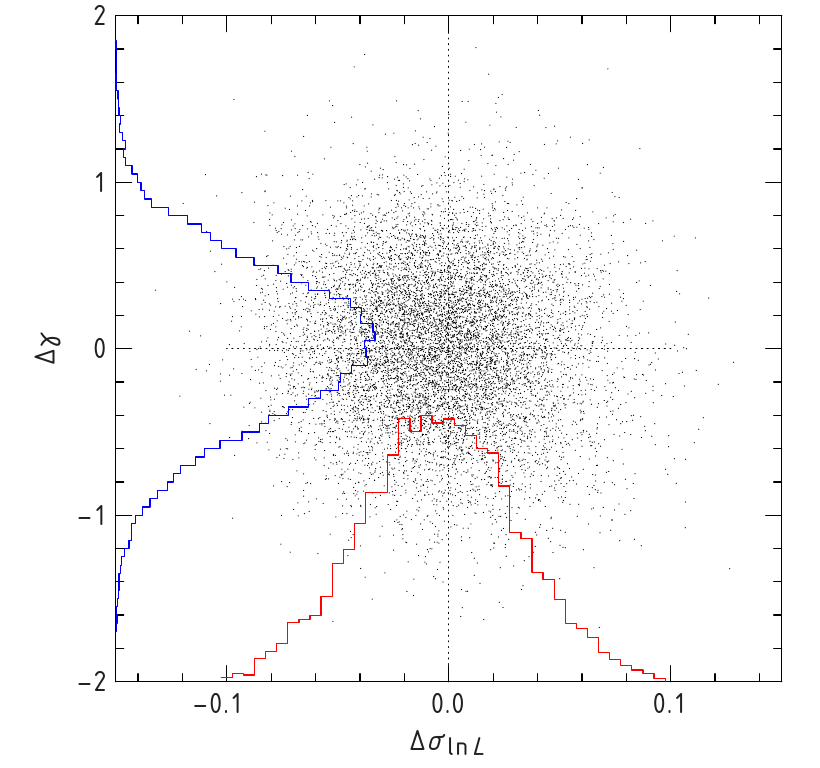}~~%
\includegraphics[width=0.478\linewidth]{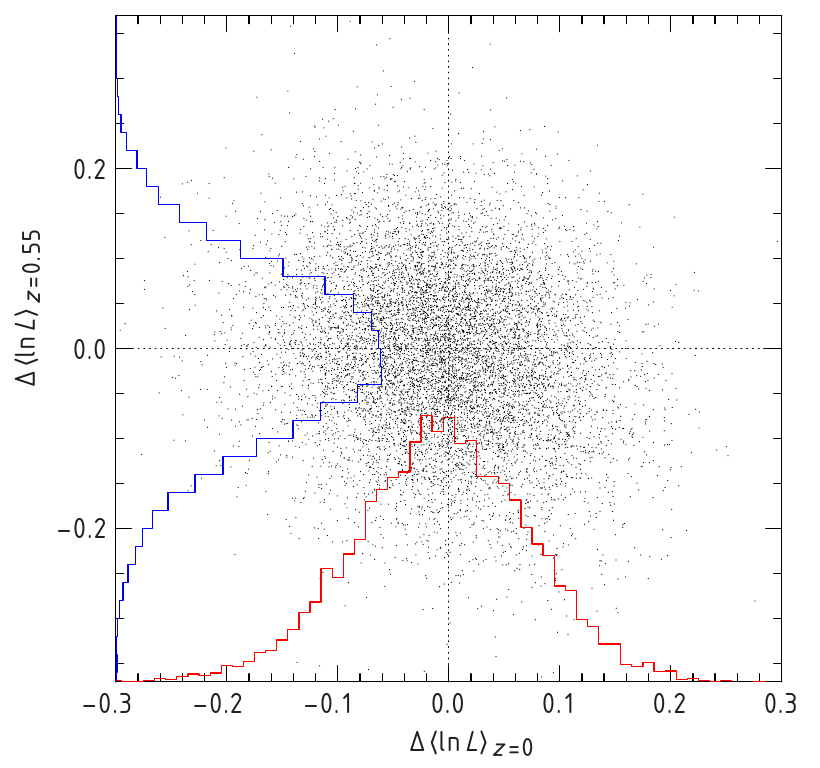}%
}
\caption{Distribution of the $\LX-M$ parameters recovered from mock
  catalogs (see text for details). The points show the deviations of the
  best-fit parameters in each realization from the input values, and
  the histograms show the probability density distribution for each
  parameter. The right panel shows the deviations of best-fit
  luminosities at $z=0$ and $0.55$ for the median mass in the samples
  (the $z=0$ results are equivalent to those for the overall
  normalization, $A_0$) for fits with $\sigma$ fixed at the nominal
  value (see text).}
\label{fig:L-M:mock}
\end{figure*}

  The total likelihood function, $\mathcal{L}$, is the product of
  $dP/d\,\ln f$ for individual clusters. The quantity $-2\ln\mathcal{L}$
  can be used in place of the usual $\chi^2$ for finding the best fit and
  confidence regions \citep{1979APJ...228..939C}. From
  (\ref{eq:malmquist:prob:1}), we have
  \begin{equation}\label{eq:malmquist:chi2}
    -2\ln\mathcal{L} = \sum_i \frac{(\ln f_i - \ln f_{0,i})^2}{\sigma^2}
    - 2\ln C_i - 2\ln\Psel(\ln f_i). 
  \end{equation}
  The first term on the right-hand side of \ref{eq:malmquist:chi2} is the
  usual unweighted $\chi^2$ and the extra two terms are corrections for the
  Malmquist bias. Masses of individual clusters and parameters of the $\LX-M$
  relation enter the likelihood function implicitly, through calculating the
  ``nominal'' luminosities [e.g.\ $\ln L_0 = A \ln \hat{M} + B +
  \mathrm{evol}(z)$] which are then converted to $f_0$'s.

  \label{sec:malmquist:fitting}

  Parameters of the $\LX-M$ relation can be obtained from finding the
  global maximum of the likelihood function~(\ref{eq:malmquist:chi2}).
  In practice, we use a multi-step procedure to fit the $\LX-M$
  parameters.  The scatter, overall normalization, and power law slope
  are determined from the low-$z$ data where the measurement
  uncertainties can be neglected relative to intrinsic scatter. The
  best-fit scatter is corrected by a factor of $(N/(N-1))^{1/2}$, the
  expected bias of the Maximum Likelihood estimate, where $N$ is the
  number of clusters in the low redshift sample. Then, with $A_0$,
  $\alpha$, and $\sigma$ fixed, the evolutionary term ($\gamma$) is
  determined from the fit to the high-$z$ data. The procedure is
  iterated several times until convergence.

  To assess how well our fitting procedure recovers the parameters of
  the $\LX-M$ relation, we applied it to mock cluster samples. The mock
  samples were designed to mimic closely our actual low- and high-$z$
  samples. The cluster masses and redshifts were drawn from the mass
  function model computed in the $\Omega_M=0.28$, $\Lambda=0.72$,
  $\sigma_8=0.79$ cosmology. The luminosities were then simulated
  assuming a mass-luminosity relation with parameters $(\ln A_0, \alpha,
  \gamma, \sigma) = (47.4,1.6,1.8,0.4)$ (c.f.\ our best-fits parameters
  in eq.[\ref{eq:L-M:fit}]), the observed fluxes computed for this
  background cosmology, and finally, the selections appropriate for the
  \emph{ROSAT} All-Sky and \400d{} surveys were applied. The simulated
  lists and the real sample have approximately the same number of
  clusters.

  The distribution of the deviations of best-fit parameters from their
  nominal input values is shown in Fig.\,\ref{fig:L-M:mock}. We are able
  to recover all parameters, normalization, scatter, evolution term
  $\gamma$, and the slope, $\alpha$, (not shown in the figure), without
  significant biases. The widths of the distributions, $\Delta \ln
  A_0=0.085$, $\Delta\alpha=0.14$, $\Delta\gamma=0.42$, and
  $\Delta\sigma=0.039$, correspond to the expected measurement
  uncertainties for each parameter. Note that the uncertainties for
  individual parameters are correlated. For example, the low-$z$
  normalization is obviously anti-correlated with the evolution
  parameter, $\gamma$. The scatter is anti-correlated with both the
  low-$z$ normalization and evolution because the Malmquis bias
  corrections are $\propto \sigma^2$. These correlations have to be kept
  in mind when we estimate the uncertainties in the survey volume
  computations associated with the measurement errors of the $\LX-M$
  relation.  In particular, the most important parameters for $V(M)$ are
  the average $\LX$'s for the median mass of our low and high-$z$
  samples.  For nearby clusters, this corresponds simply to the
  uncertainties in $A_0$, but for high-$z$ clusters, this is a complex
  combination of uncertainties in $A_0$, $\alpha$, and $\gamma$.  The
  results for the average normalizations are shown in the right panel of
  Fig.\,\ref{fig:L-M:mock}.  We are able to recover the true average
  luminosities without a significant bias and with uncertainties of
  $\approx 8.0\%$ and $10.5\%$ at low and high-$z$, respectively.

\section{Likelihood function calculations}
\label{sec:likelihood:details}

\subsection{Calculation of $p(\Mest,z)$}

Generally, the probability density distribution of the observed masses
is given by convolution of the model distribution of the true masses and
the scatter between $\Mest$ and $\Mtrue$. The former is simply the
product of the theoretical mass function $dn/d\Mtrue$ and survey
volume at this redshift, $dV(\Mtrue,z)/dz$ (the calculation of
$dV(M)/dz$ is discussed in \S~\ref{sec:volume}), and so we have 
\begin{equation}\label{eq:conv:model:scatter}
  p(\Mest,z) = \left(\frac{dn}{d\Mtrue}\frac{dV(\Mtrue,z)}{dz}\right)
  \otimes \mathrm{scatter}(\Mest,\Mtrue)
\end{equation}
A log-normal distribution is a good approximation for the scatter in the
mass estimates, and so the convolution in
eq.(\ref{eq:conv:model:scatter}) can be written as
\begin{equation}\label{eq:dndm:convolved}
  p(\Mest,z) =  \frac{1}{\Mest}\frac{1}{\sqrt{2\pi}\,\sigmaest}
  \int_{-\infty}^\infty \frac{dn}{d\ln\Mtrue}\frac{dV(\Mtrue,z)}{dz}
  \exp\left(-\frac{(\ln\Mest-\ln\Mtrue)^2}{2(\sigmaest)^2}\right)
  \;d\ln\Mtrue
\end{equation}
The function $p(\Mest,z)$ enters the expression for likelihood in
summation over observed clusters (first term in
eq.[\ref{eq:likelihood}]) and in the integral over the observed range
(second term in the same equation).  Equation~\ref{eq:dndm:convolved}
should be evaluated numerically, but the calculation of all the terms is
straightforward. The term $dn/d\ln\Mtrue$ is the differential cluster
mass function at the given redshift. Cosmological parameters enter the
calculation of $dV(\Mtrue,z)$ through the volume-redshift relation and
the evolving cluster $\LX-M$ relation which is derived
(\S\ref{sec:lx-mtot}) using $\LX$ and $\Mtot$ estimated in this
cosmology.

\subsection{Integration of $p(\Mest,z)$}

We now need to evaluate the second term in eq.[\ref{eq:likelihood}],
\begin{equation}
  I = \int\limits_{\Mmin}^{\Mmax}\int\limits_{\zmin}^{\zmax}
  p(\Mest,z)\;d\Mest\,dz.
\end{equation}
Using \ref{eq:dndm:convolved}, we have
\begin{equation}
  I  =\frac{1}{\sqrt{2\pi}}
  \int\limits_{\Mmin}^{\Mmax}d\Mest\int\limits_{\zmin}^{\zmax} dz
  \int\limits_{-\infty}^\infty
  \frac{1}{\Mest}\frac{1}{\sigmaest}
  \frac{dn}{d\ln\Mtrue}\frac{dV(\Mtrue,z)}{dz}
  \exp\left(-\frac{(\ln\Mest-\ln\Mtrue)^2}{2(\sigmaest)^2}\right)
  \;d\ln\Mtrue
\end{equation}
Changing the order of integration, we have
\begin{equation}
  I =
  \int\limits_{\zmin}^{\zmax}
  dz
  \int\limits_{-\infty}^\infty d\ln\Mtrue
  \frac{dn}{d\ln\Mtrue}\,\frac{dV(\Mtrue,z)}{dz}\,
  \int\limits_{\ln\Mmin}^{\ln\Mmax}
  \frac{1}{\sqrt{2\pi}\,\sigmaest}
  \exp\left(-\frac{(\ln\Mest-\ln\Mtrue)^2}{2(\sigmaest)^2}\right)
  \;d\ln\Mest
\end{equation}
The last term in this equation is the integral of the normal
distribution (can be computed numerically using the library error
function):
\begin{equation}
   \int\limits_{\ln\Mmin}^{\ln\Mmax}
  \frac{1}{\sqrt{2\pi}\,\sigmaest}
  \exp\left(-\frac{(\ln\Mest-\ln\Mtrue)^2}{2(\sigmaest)^2}\right)
  \;d\ln\Mest =
  \mathfrak{N}\left(\frac{\ln\Mmin-\ln\Mtrue}{\sigmaest},\frac{\ln\Mmax-\ln\Mtrue}{\sigmaest}\right),
\end{equation}
where
\begin{equation}
\mathfrak{N}(x_1,x_2) \equiv
\frac{1}{\sqrt{2\pi}}\int_{x_1}^{x_2} \exp(-x^2)\,dx,
\end{equation}
so finally,
\begin{equation}\label{eq:int:prob:1}
  I = 
  \int\limits_{\zmin}^{\zmax}
  dz
  \int\limits_{-\infty}^\infty
  \frac{dn}{d\ln\Mtrue}\,\frac{dV(\Mtrue,z)}{dz}\,
  \mathfrak{N}\left(\frac{\ln\Mmin-\ln\Mtrue}{\sigmaest},\frac{\ln\Mmax-\ln\Mtrue}{\sigmaest}\right)\;
  d\ln\Mtrue
\end{equation}
The quantities $\sigmaest$ are the total uncertainties of the mass
estimates, including intrinsic scatter and measurement errors,
\begin{equation}
  \sigmaest_i =
  \left(\sigma_{\text{intr}}^2+\sigma_{\text{meas}, i}^2\right)^{1/2}.
\end{equation}
In practice, $\sigmaest_i$ are not the same because at least
$\sigma_{\text{meas}, i}$ varies from cluster to cluster. A reasonable
strategy to include these variations is to replace
$\mathfrak{N}(\ldots,\ldots)$ with an average over all sample members,
\begin{equation}
  \label{eq:int:prob}
  I = 
  \int\limits_{\zmin}^{\zmax}
  dz
  \int\limits_{-\infty}^\infty
  \frac{dn}{d\ln\Mtrue}\,\frac{dV(\Mtrue,z)}{dz}\,\,\left[\frac{1}{N}\sum_{i=1}^N
  \mathfrak{N}\left(\frac{\ln\Mmin-\ln\Mtrue}{\sigmaest_i},\frac{\ln\Mmax-\ln\Mtrue}{\sigmaest_i}\right)\right]\;
  d\ln\Mtrue
\end{equation}
We use this equation to evaluate the second term in the expression for
likelihood function (eq.[\ref{eq:likelihood}]).

\end{appendix}

\end{document}